\documentclass[a4paper,fleqn,usenatbib,useAMS]{mnras}
\usepackage{pdflscape}
\usepackage[T1]{fontenc}
\usepackage{ae,aecompl}
\usepackage{graphicx}   
\usepackage{amsmath}    
\usepackage{amssymb}    
\usepackage{multirow}

\title[Nova LMC 2009a with XMM-Newton]{Nova LMC 2009a as observed with
 XMM-Newton, compared with other novae}

\author[M. Orio et al.]
{Marina Orio,$^{1,2}$\thanks{E-mail: orio@astro.wisc.edu}
Andrej Dobrotka,$^3$
Ciro Pinto,$^4$
Martin Henze,$^5$
Jan-Uwe Ness,$^6$
\newauthor
Nataly Ospina,$^{7,8}$
Songpeng Pei,$^9$ 
Ehud Behar,$^{10}$
Michael F. Bode,$^{11}$
\newauthor
Sou Her,$^1$
Margarita Hernanz,$^{12,13}$
and Gloria Sala$^{13,14}$\\
$^1$ Department of Astronomy, University of Wisconsin, 475 N. Charter Str., Madison WI 53704\\
$^2$ INAF--Osservatorio di Padova, vicolo dell' Osservatorio 5,
   I-35122 Padova, Italy \\
$^3$ Advanced Technologies Research Institute,
 Faculty of Materials Science and Technology in Trnava,\\
 Slovak University of Technology in Bratislava, Bottova 25, 917 24 Trnava, Slovakia
 \\
$^4$ INAF-IASF Palermo, via Ugo la Malfa, 153, 90146 Palermo, Italy \\
$^5$ Department of Astronomy, San Diego State University, San Diego, CA 92182, USA \\ 
$^6$  European Space Astronomy Agency (ESA), European Space Astronomy Center (ESAC),
 Camino Bajo del Castillo s/n, \\
 28692 Villanueva de la Ca\~nada, Madrid, Spain \\
$^7$ Department of Physics and Astronomy, Padova University, via Marzolo, 3, 35131 Padova \\
$^8$INFN Sezione di Padova, Via Marzolo, 8, 35131 Padova, Italy \\
$^{9}$ Department of Physics and Astronomy, Padova University, vicolo Osservatorio, 3, 
35122 Padova, Italy \\
$^10$ Department of Physics, Technion, Haifa, Israel \\
$^{11}$Astrophysics Research Institute, Liverpool John Moores University,
IC2, Brownlow Hill, Liverpool, L3 5RF, UK\\
$^{12}$Institut de Ciencies de l' Espai (ICE-CSIC). Campus UAB. c/ Can Magrans s/n, 08193, Bellaterra, Spain \\
$^{13}$ Institut d' Estudis Espacials de Catalunya, c/Gran Capita 2-4,
 Ed. Nexus-201, 08034, Barcelona, Spain \\
$^{14}$ Departament de F\`isica, EEBE, Universitat Politecnica de Catalunya.
 BarcelonaTech., Av. d' Eduard Maristany 10-14,\\ 08019,
Barcelona, Spain 
}
\date{Accepted XXX. Received YYY; in original form ZZZ}
\pubyear{2019}
\begin{document}
\label{firstpage}
\pagerange{\pageref{firstpage}--\pageref{lastpage}}
\maketitle

\begin{abstract}
 We examine four 
 high resolution reflection grating spectrometers (RGS) spectra
 of the February 2009 outburst of the luminous recurrent nova LMC 2009a. 
 They were very complex and  rich in intricate absorption
 and emission features. The continuum was consistent
 with a dominant component originating in the atmosphere of 
 a shell burning white dwarf (WD) with peak effective temperature
 between 810,000 K and a million K, 
 and mass in the 1.2-1.4 M$_\odot$ range.
 A moderate blue shift of the absorption features of a few
 hundred km s$^{-1}$ can be explained with a residual nova
 wind depleting the WD surface at a rate of 
about 10$^{-8}$ M$_\odot$ yr$^{-1}$. 
 The emission spectrum seems to be due to both photoionization
 and shock ionization in the ejecta. 
 The supersoft X-ray flux was irregularly variable
 on time scales of hours, with decreasing amplitude of the variability.
 We find that 
 both the period and the amplitude of another, already known 33.3 s modulation, 
varied within timescales of hours. We compared N LMC 2009a with 
other Magellanic Clouds novae, including  4 serendipitously discovered 
as supersoft X-ray sources (SSS) among 13 observed 
 within 16 years after the eruption. The new detected targets
 were much less luminous than expected: we suggest that 
 they were partially obscured by the accretion disk.
  Lack of SSS detections in
 the Magellanic Clouds novae more than 5.5 years after the eruption 
 constrains the average duration of the nuclear burning phase.
\end{abstract}



\section{Introduction}
Classical and recurrent novae (CNe, RNe) are now routinely discovered in other
 galaxies of the Local Group, offering useful terms of
 comparison of the nova phenomenon in ambients with different
 metallicity and star formation history. 
Known novae in the Magellanic
 Clouds are not numerous, due to the small mass of the two galaxies,
 but they   occur in an environment of much
 lower metallicity than the Galaxy, and at relatively close
 distance, only about 5 times as high as the farthest luminous
 Galactic novae well studied in recent years \citep[for instance,
 the RN U Sco is at 12$\pm$2 kpc distance, see][]{Schaefer2010}.
\subsection{Nova LMC 2009A}
Nova LMC 2009a (also N LMC 2009-02 in the notation
 including the outburst month) was discovered on 2009 February
05.067 UT by \citet[][]{Liller2009} and spectroscopically
 confirmed by \citet[][]{Bond2009}. It was later identified as a RN, 
 coinciding with N LMC 1971b \citep[][]{Bode2016}.
RN are the ones recurring
 on human time scales, although the models indicate 
 that all outburst recur (i.e. Prialnik 1986).
 An optical spectrum obtained in outburst by \citet[][]{Orio2009} 
showed prominent emission lines of H, He and N, so this 
 nova is  a He/N one in the classification scheme of \citet{Williams1992},
like the other RNe we know.
 \citet{Orio2009} reported a large expansion velocity with the
 full width at half maximum of the H$\alpha$ line 
 slightly above 4200 km s$^{-1}$. 
Additional spectra
 obtained by \citet[][]{Bode2016} showed expansion velocities
derived from different lines and at different phases between 1000 and 4000
 km s$^{-1}$. Coronal line emission before day 9 indicated shocks in the ejecta.
The initial decay was fast, and the time t$_3$ for a decay by 3 magnitudes
 lasted from 10.4 days in the V band to 22.7 days in the infrared K.
 The time t$_2$ for a decay by 2 magnitudes 
 ranged from 5 days (V filter)
 to 12.8 days (K filter). These parameters are pertinent to a classification as
 a ``very fast'' nova \citep{PayneG1964}.

 By comparison with the grid of nova models by \citet{Yaron2005},
the characteristic parameters of the outburst (recurrence time of 38
 years, velocity reaching $\simeq$4000 km s$^{-1}$,
 t$_3$=11.4 days in the V band, amplitude of about 9 mag in V),
 place the nova the highest WD mass range (1.4 M$_\odot$), with a
 rather young and hot WD at the start of accretion, and
 mass accretion rate $\dot m$  of a few 10$^{-8}$ M$_\odot$.
 However, the models include only a constant mass
 accretion rate $\dot m$, which
 may not be the case in some novae, and probably not for RN, whose
 recurrence time has been observed to vary (while the envelope mass
 accreted to trigger the burning outburst is expected to remain the same).

\begin{figure*}
\includegraphics[width=120mm]{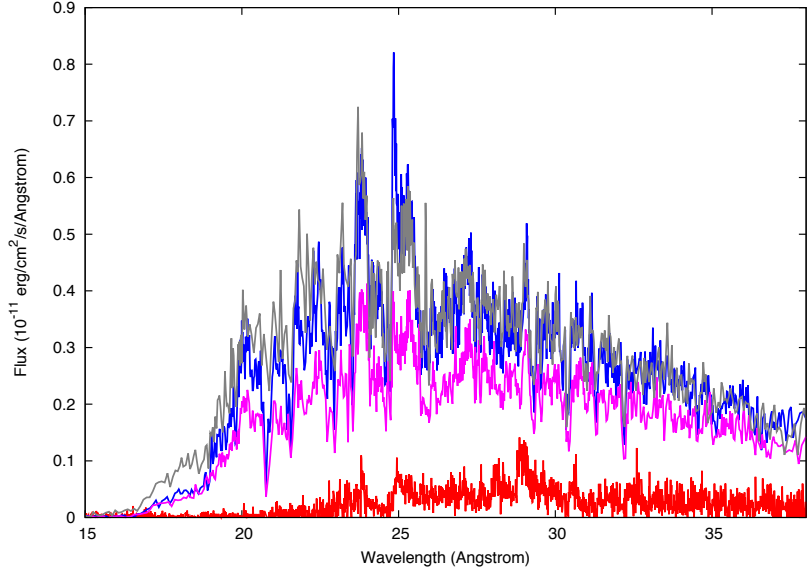}
\caption{The RGS spectra of Nova LMC 2009 in units of measured flux versus
 wavelength, observed on days 90 (2009 May, in red), 165
 (2009 July, in blue), 197 (2009 August, grey) and 
 229 (2009 September, purple).}
\end{figure*}

\citet[][]{Bode2016} identified the progenitor system; the optical
 and infrared magnitude in different filters and the colour indexes 
 are best interpreted with the presence of a sub-giant feeding a luminous
 accretion disk. 
Modulations with
 a period P=1.2 days, most probably orbital in nature, were evident in
the UV and optical flux since day 43 \citep{Bode2016}.
 Two other RNe with orbital periods of the order of a day and sub-giant
 evolved secondaries are the Galactic novae 
U Sco and V394 CrA.
There is also evidence that also a third RN, V2487 Oph, hosts
 a sub-giant, although its 
 orbital period has not been measured yet \citep{Strope2010}.
 Other nova systems with suspected subgiant secondaries
 and day-long orbital periods are: KT Eri \citep[also a candidate RN, but so far 
 without previous known outbursts,][]{Bode2016}, 
  HV Cet \citep[][]{Beardmore2012}
and V1324 Sco \citep[][]{Finzell2015}.
\begin{figure*}
\includegraphics[width=120mm]{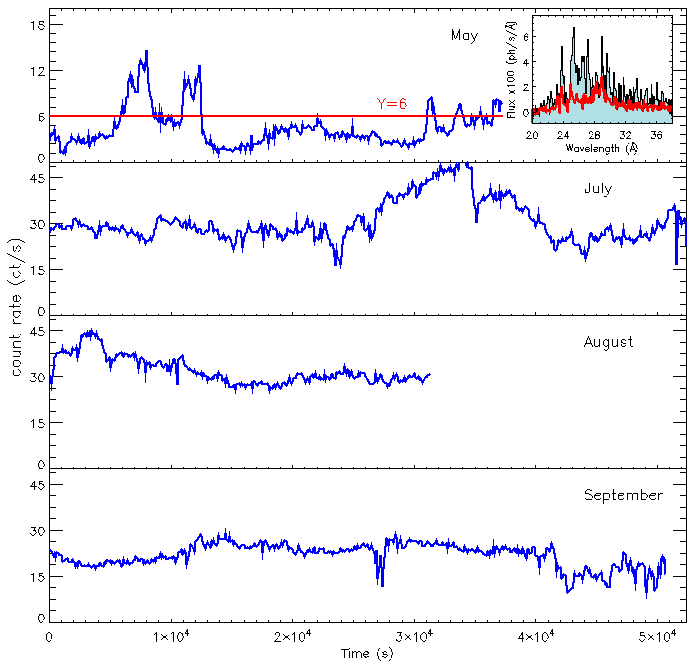}
\caption{The EPIC-pn light curve, from top to bottom, as measured on days 90, 165, 197, 229.
 The red horizontal line in the top panel
 shows the pn
cutoff for the averaged ``low count rate'' and ``high count rate'' spectra,
 shown in the inset in red and black, respectively
(note that the RGS count rate varied proportionally to
 the pn count rate variation). }
\end{figure*}

 Luminous  CNe and RNe  
 are monitored regularly with {\sl Swift} in UV and X-rays. 
 It is known that all nova shells emit X-rays in outburst
 \citep[e.g.][]{Orio2012}, although 
 they are not usually luminous enough to be detected 
 at LMC distance. However, when the ejecta become
 optically thin to soft X-rays, the photosphere of the WD contracts and
 shrinks to close to pre-outburst dimension 
\citep{Starrfield2012, Wolf2013} while 
 CNO burning still occurs close to the surface, with only
 a thin atmosphere on top, for a period of time
 ranging from days to years \citep{Orio2001, Schwarz2011, Page2020}.
 Because the WD effective temperature T$_{\rm eff}$ is
 in the 150,000 K to a million K range, the WD atmosphere
 peaks in the X-ray range or very close to it, and the WD detected as a
 luminous supersoft X-ray
 source (SSS), observable at the distance of the Clouds,
 also thanks to the low column density.

Strengthening of the He II 4686\AA \ line in the N LMC 2009a spectrum
preceded the emergence of the central WD
 as a supersoft
 X-ray source (hereafter, SSS) observed in X-rays with
 the {\sl Swift} X-Ray Telescope (XRT) since day 63.
 The SSS initially was at lower luminosity, but became much
 more luminous around day 140.
 The following X-ray observations indicated
 an approximate constant average luminosity (albeit with
 large fluctuations from day to day),
 until around day 240 of the outburst. The SSS
 was always variable, periodically and aperiodically.
 and clear oscillations with the 1.2 days period were
 observed \citep[][]{Bode2016},
 with a delay of 0.28P with respect to the optical modulations.
\begin{figure*}
\begin{center}
\includegraphics[width=170mm]{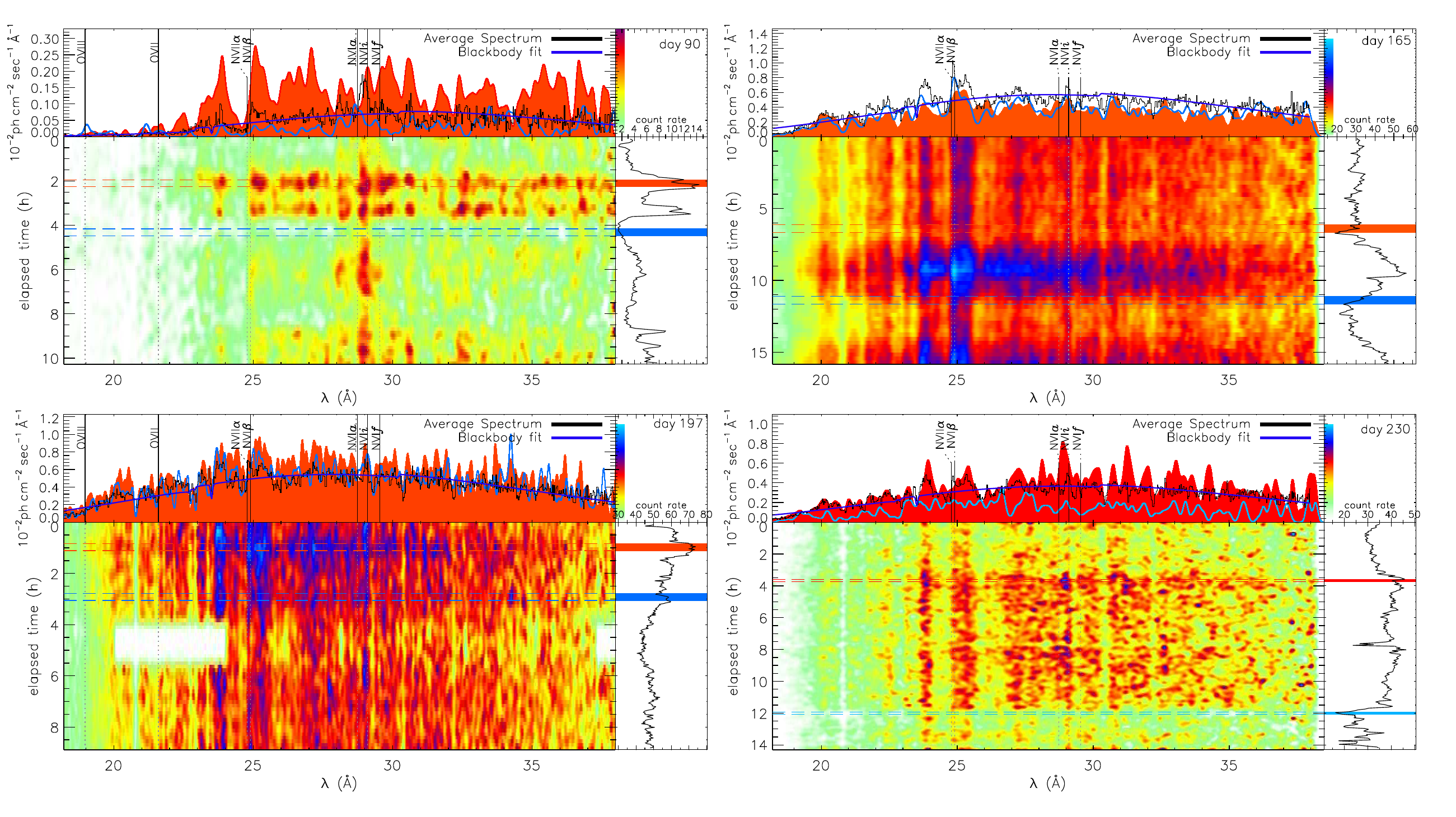}
\end{center}
\caption{Visualization of the spectral evolution within the observations.
 The four
panels show the fluxed spectra as function of wavelength (top left),
colour-coded intensity map as function of time and wavelength (bottom
left), rotated light curve with time on vertical and count rate on
horizontal axes (bottom right). In the top right panel, a vertical
bar along the flux axis indicates the colours in the bottom left panel.
In the top left panel, the red (highest) and blue (lowest) spectra
have been extracted during the very short sub-intervals marked in the light curve
(bottom right) with shaded areas and bordered by dashed horizontal lines
in the bottom left panel. The average spectrum is shown in black, and the dark blue 
 light curve is a blackbody fit obtained assuming depleted oxygen abundance
 in the intervening medium (see text).}
\end{figure*}

Not all novae are sufficiently X-ray luminous to be studied with
 the gratings in detail, especially
 if they are as far as the Magellanic Clouds,
 and we did not want to miss the occasion of the X-ray luminous
 Nova LMC 2009, so in addition to {\sl Swift} X-Ray Telescope (XRT)
 \citep{Bode2016}, {\sl XMM-Newton} was used for longer exposures and
 high spectral resolution.

%
%
\begin{table*}
\caption{XMM-Newton observations of Nova LMC 2009.}
\label{table:obs}
\begin{center}
\begin{tabular}{rrrrrrrr}\hline\hline \noalign{\smallskip}
        ObsID & Exp. time$^a$ & Date$^b$ & MJD$^b$ & Day$^c$ & pn & MOS1 & MOS2 \\
         & (ks) & (UT) & (d)\\ \hline \noalign{\smallskip}
  0610000301 & 37.7 & 2009-05-06.43 & 2454957.43 & 90.4 & thin/small &
 medium/small & medium/small \\ \hline \noalign{\smallskip}
  0610000501 & 58.1 (40.0) & 2009-07-20.04 & 2455032.04 & 164.9 & thin/small & medium/small &medium/small
\\ \hline \noalign{\smallskip}
  0604590301 & 31.9  & 2009-08-20.59  &  2455063.59 & 196.17   & thin/small & thin/small & thin/full 
\\ \hline \noalign{\smallskip}
  0604590401 & 51.1 & 2009-09-23.02 &  2455097.02 & 228.93   & thin/small & thin/small & thin/full 
\\ \hline \noalign{\smallskip} 
\end{tabular}
\end{center}
\noindent
Notes:\hspace{0.1cm} $^a $: Exposure time
 (cleaned of high flaring background
 intervals and dead time
 corrected) of the observation; $^b $: Start date of the observation; $^c $: Time in days after the discovery of Nova
 LMC 2009 in the optical  on 2009 February 05.067 UT \citep[MJD 54867.067, see][]{Liller2009}. \\
\end{table*}
\begin{figure*}
\includegraphics[width=0.41\textwidth]{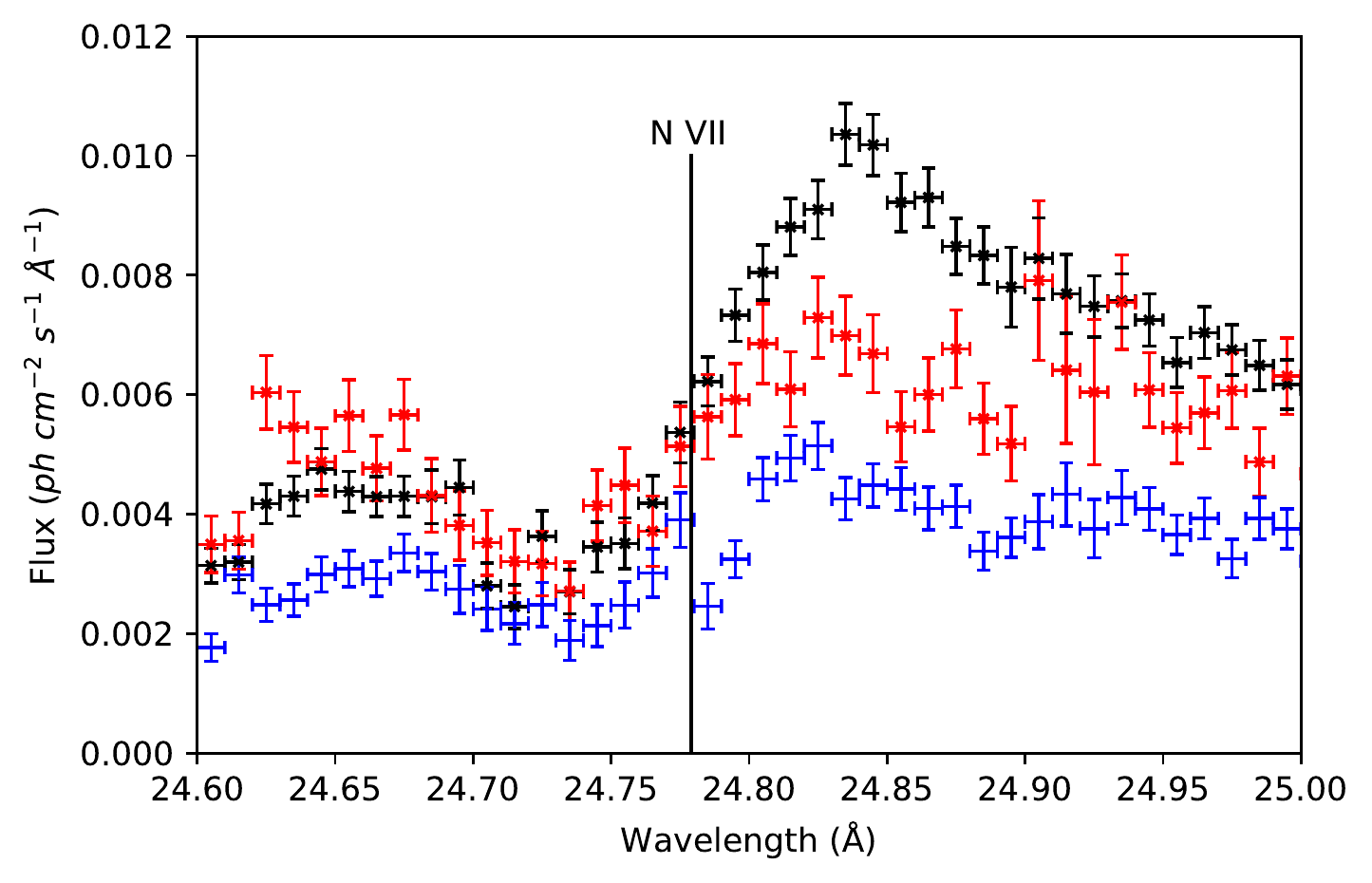}
\includegraphics[width=0.41\textwidth]{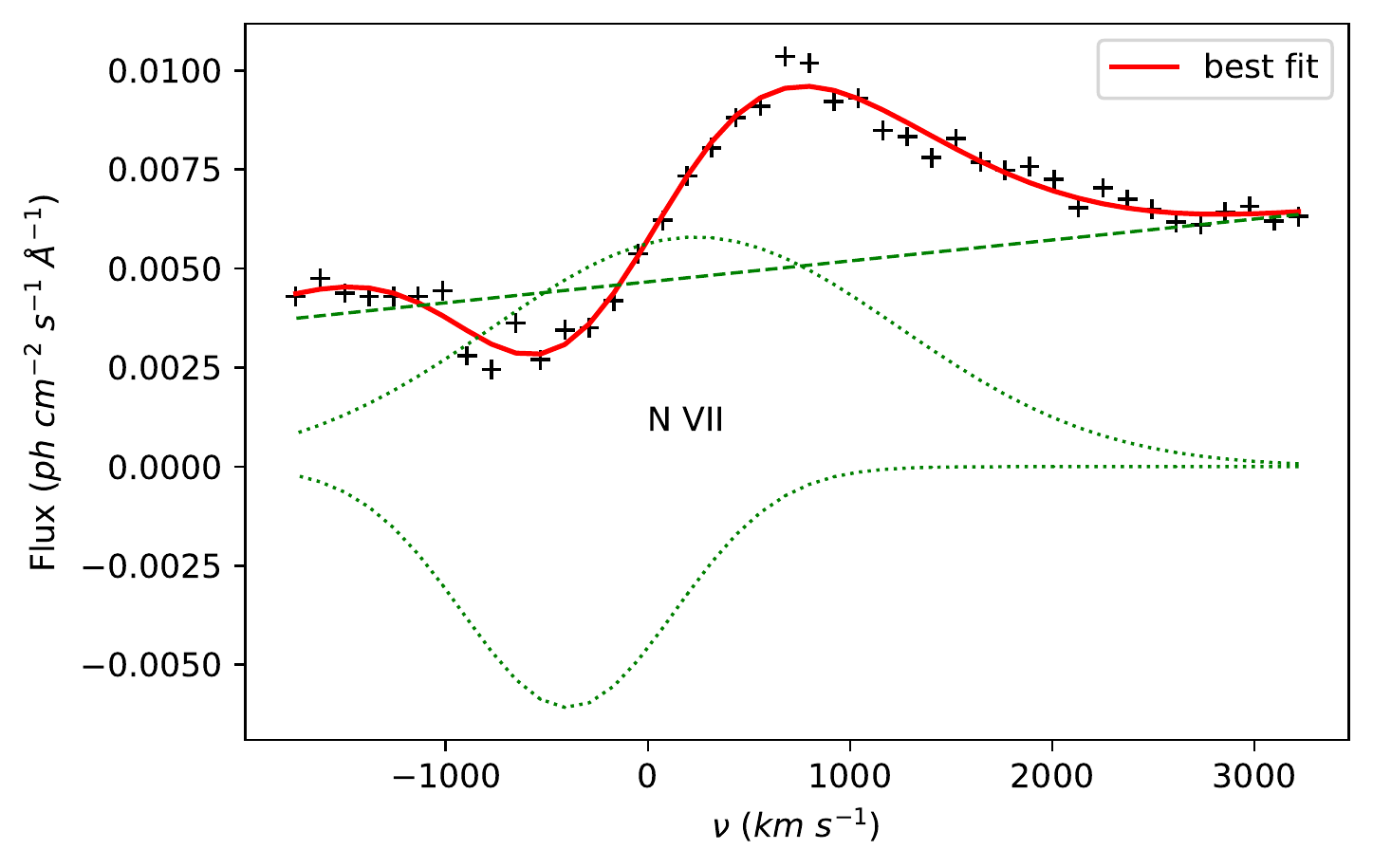}
\includegraphics[width=0.41\textwidth]{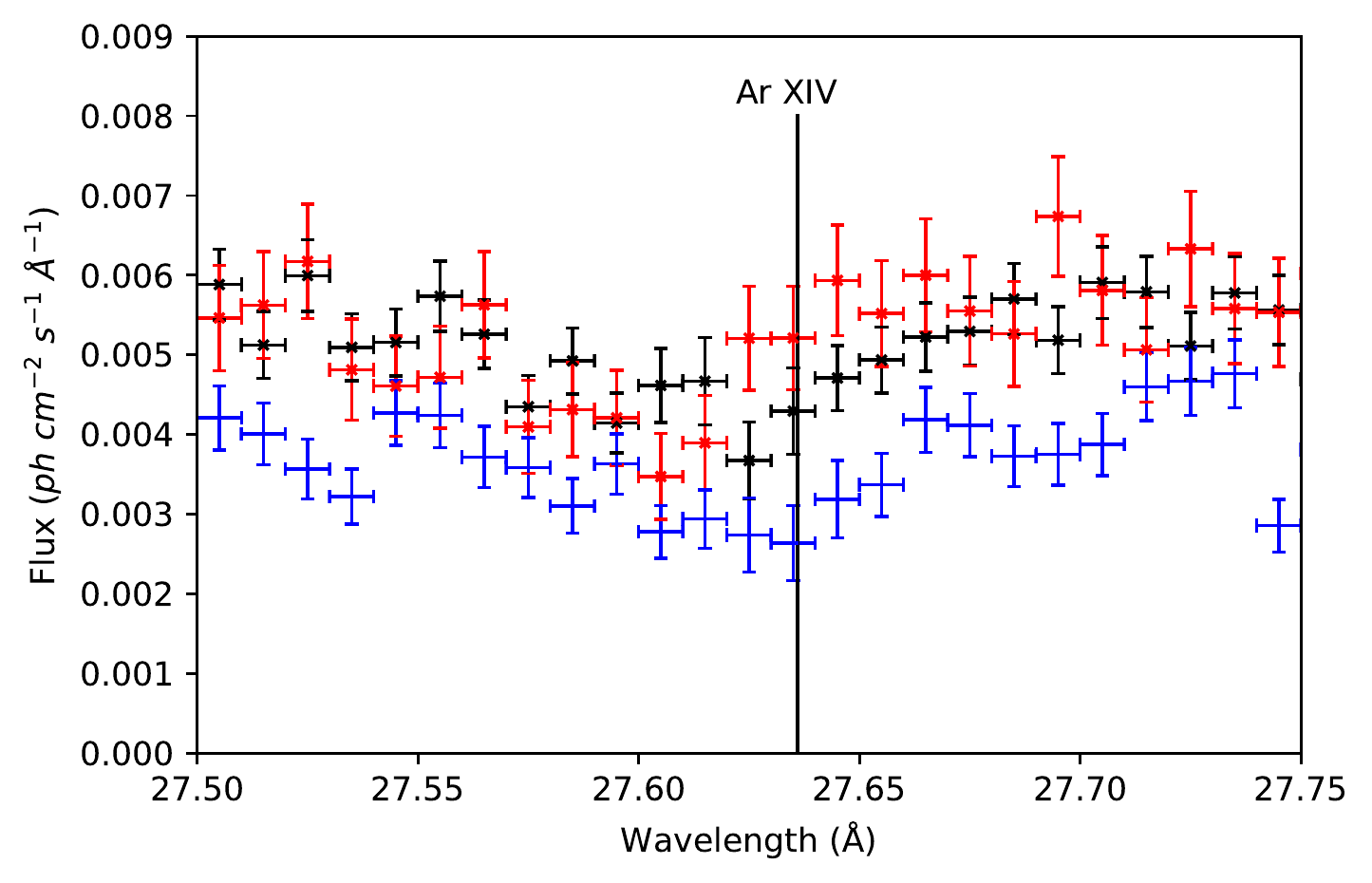}
\includegraphics[width=0.41\textwidth]{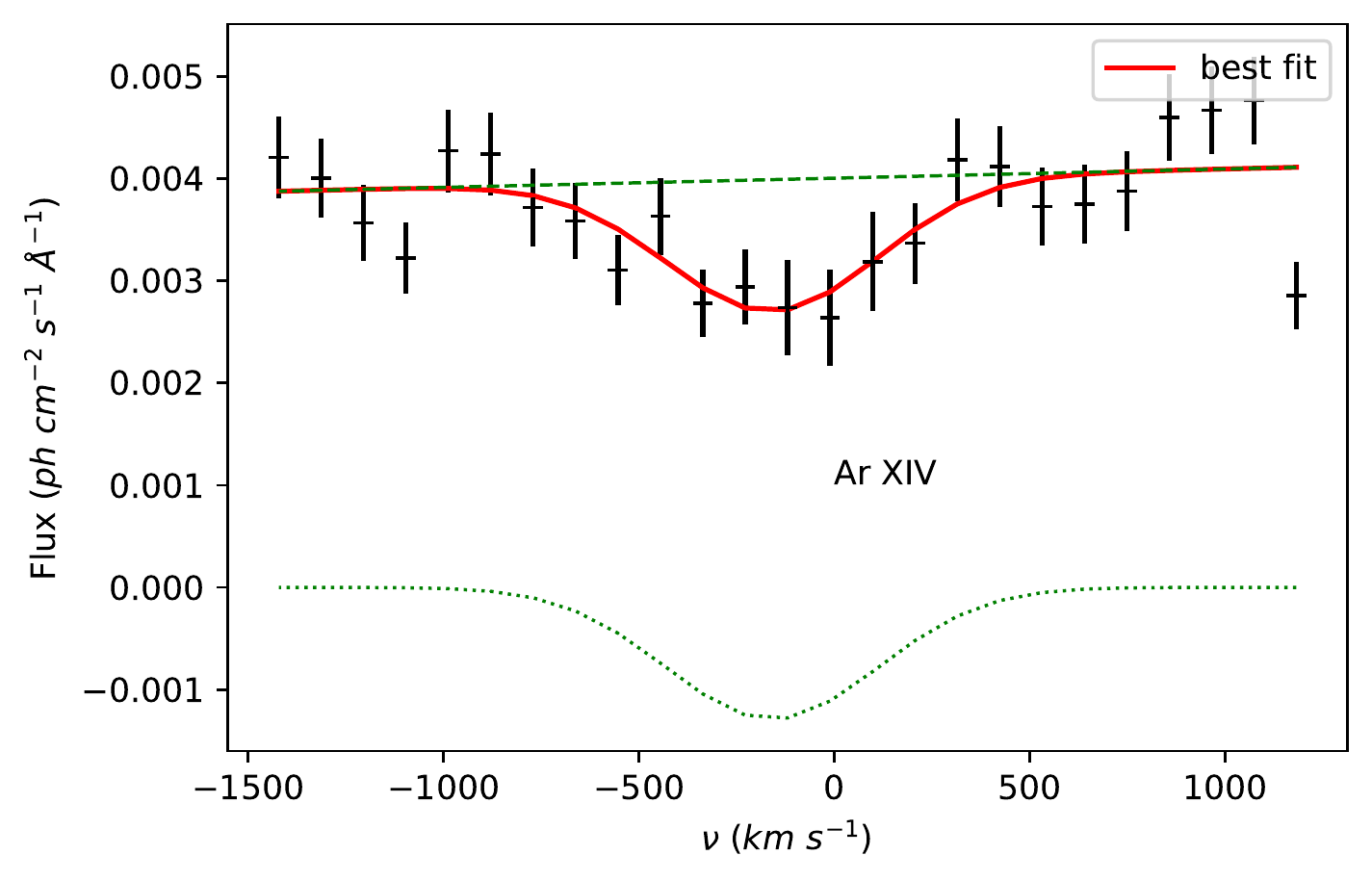}
\includegraphics[width=0.41\textwidth]{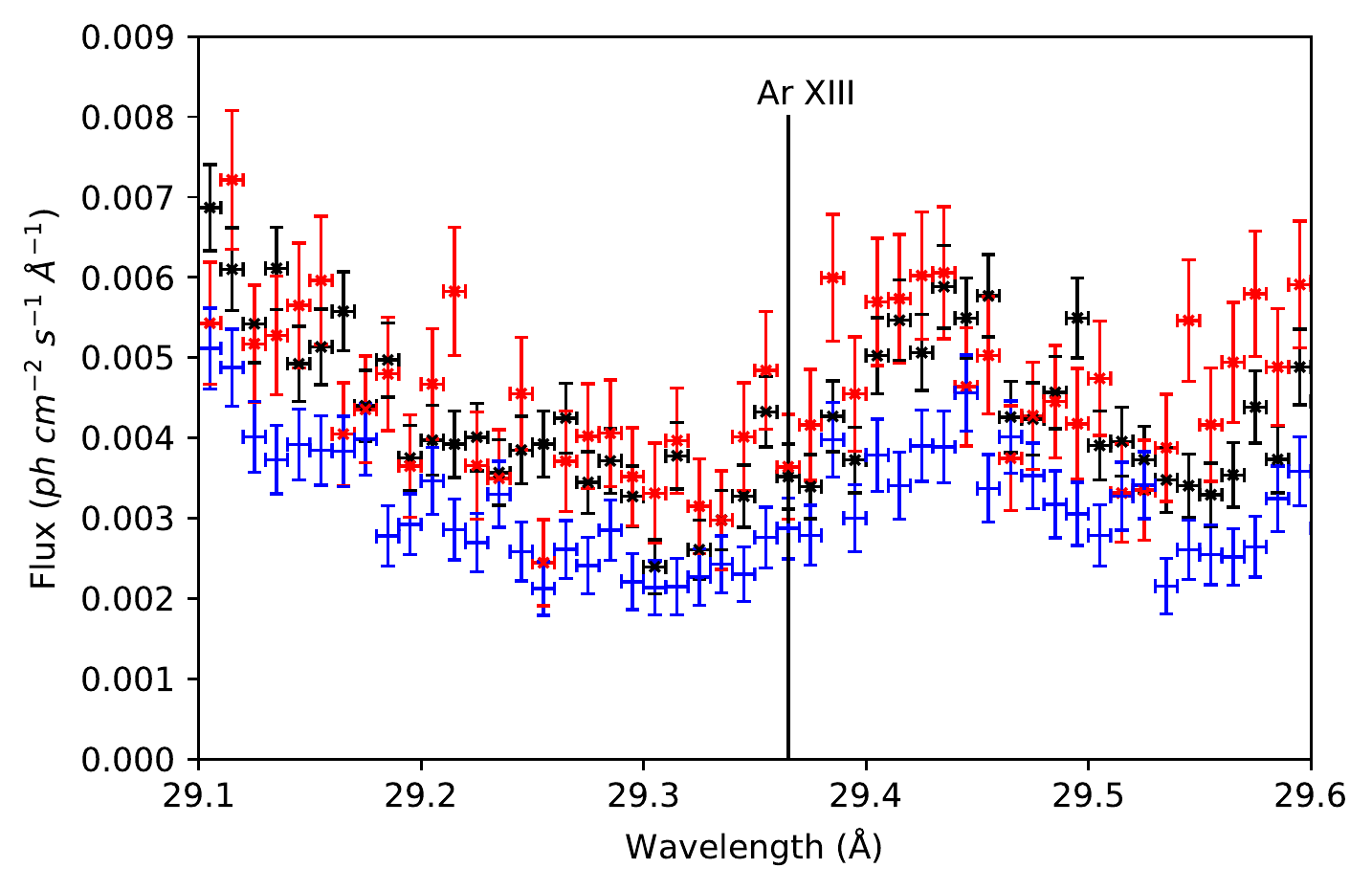}
\includegraphics[width=0.41\textwidth]{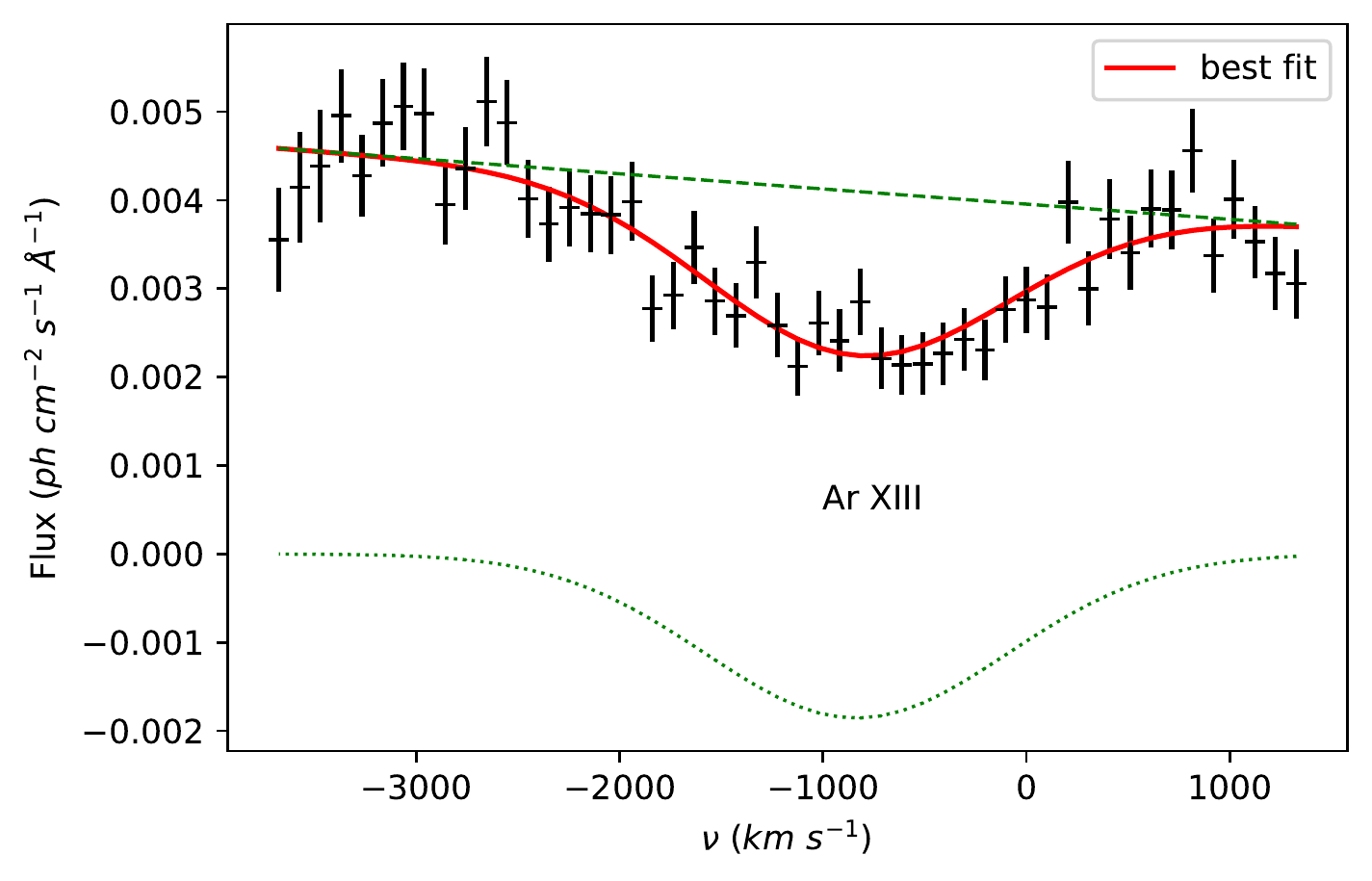}
\includegraphics[width=0.41\textwidth]{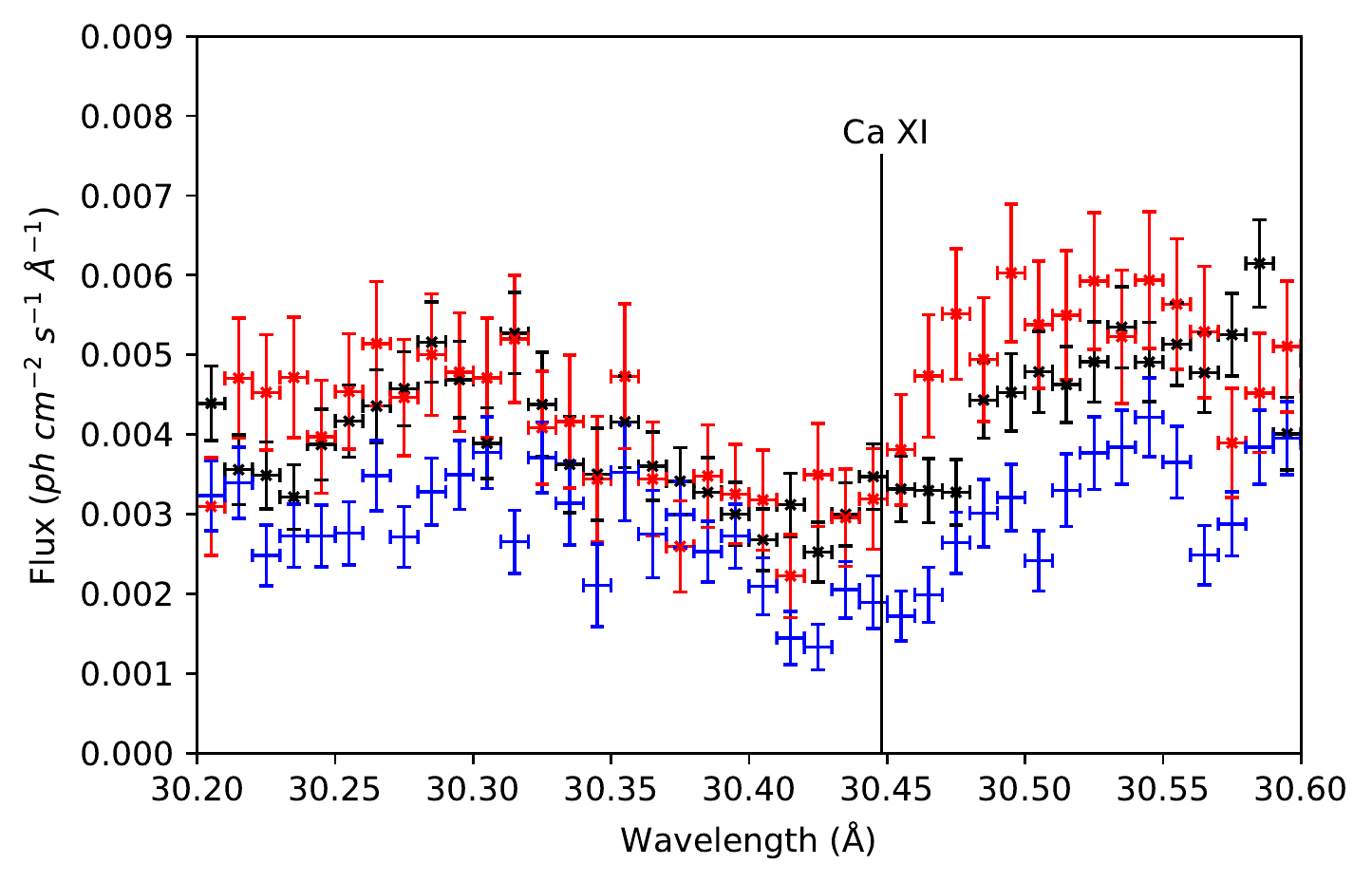}
\includegraphics[width=0.41\textwidth]{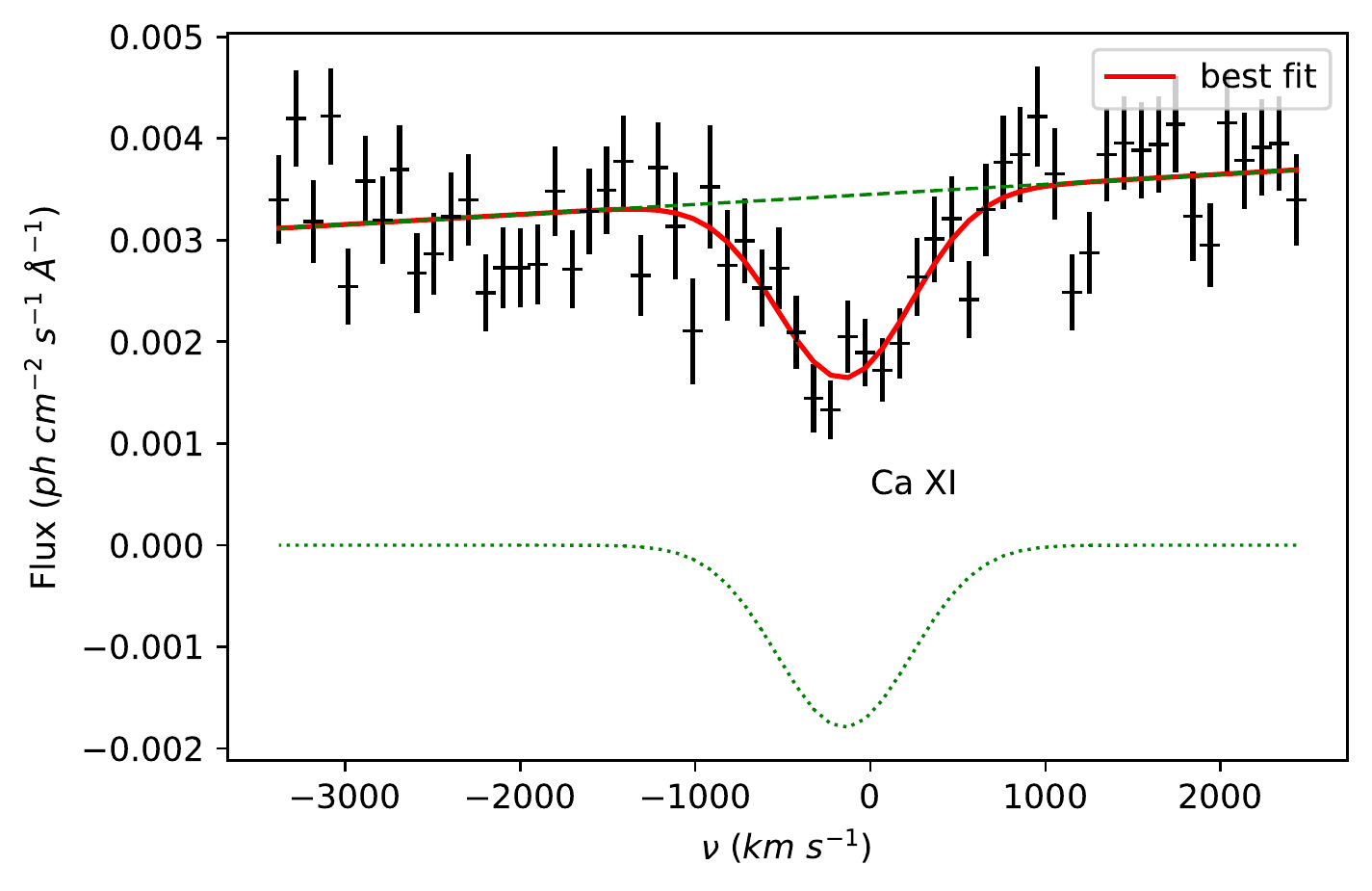}
\includegraphics[width=0.41\textwidth]{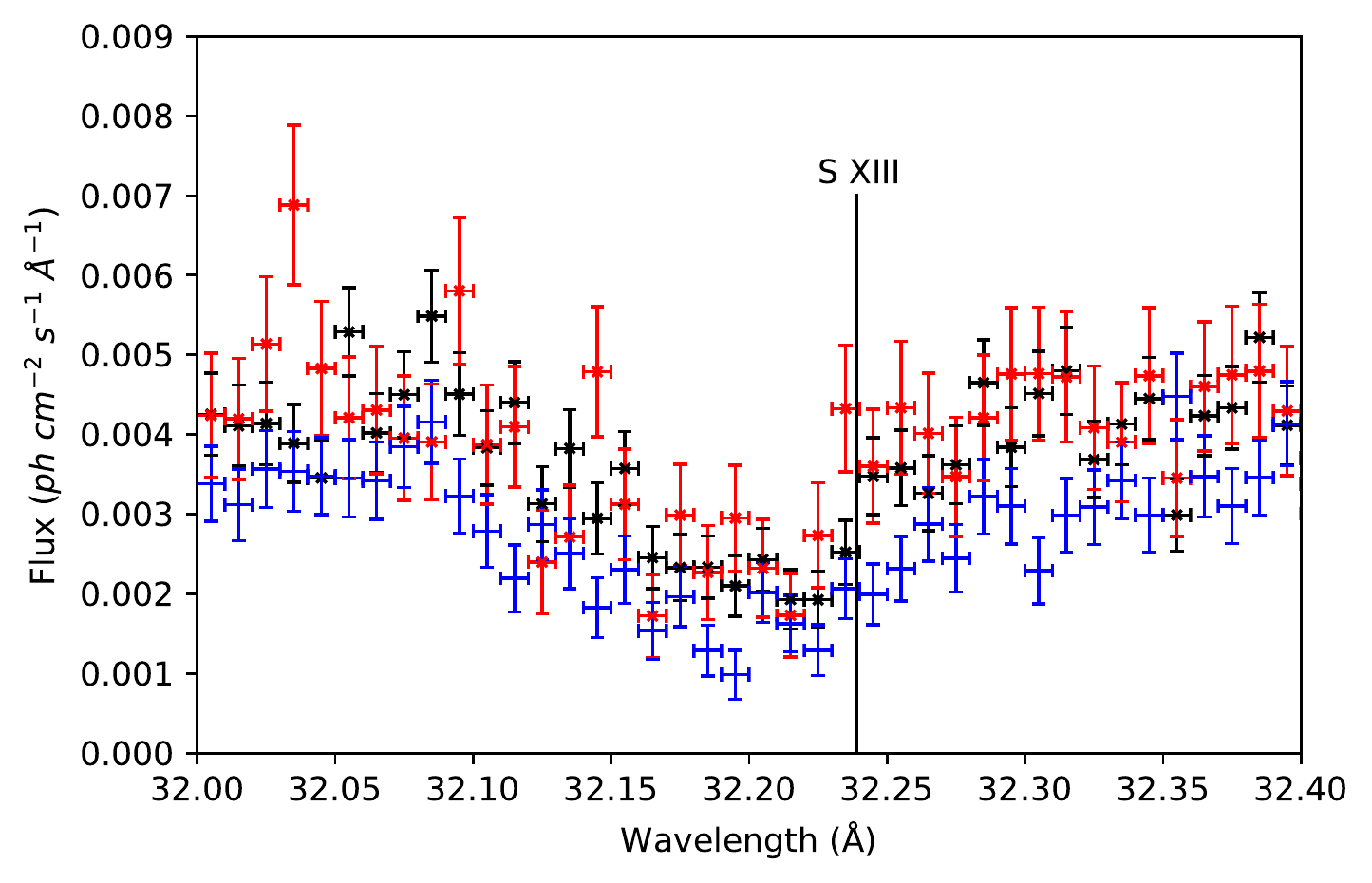}
\includegraphics[width=0.41\textwidth]{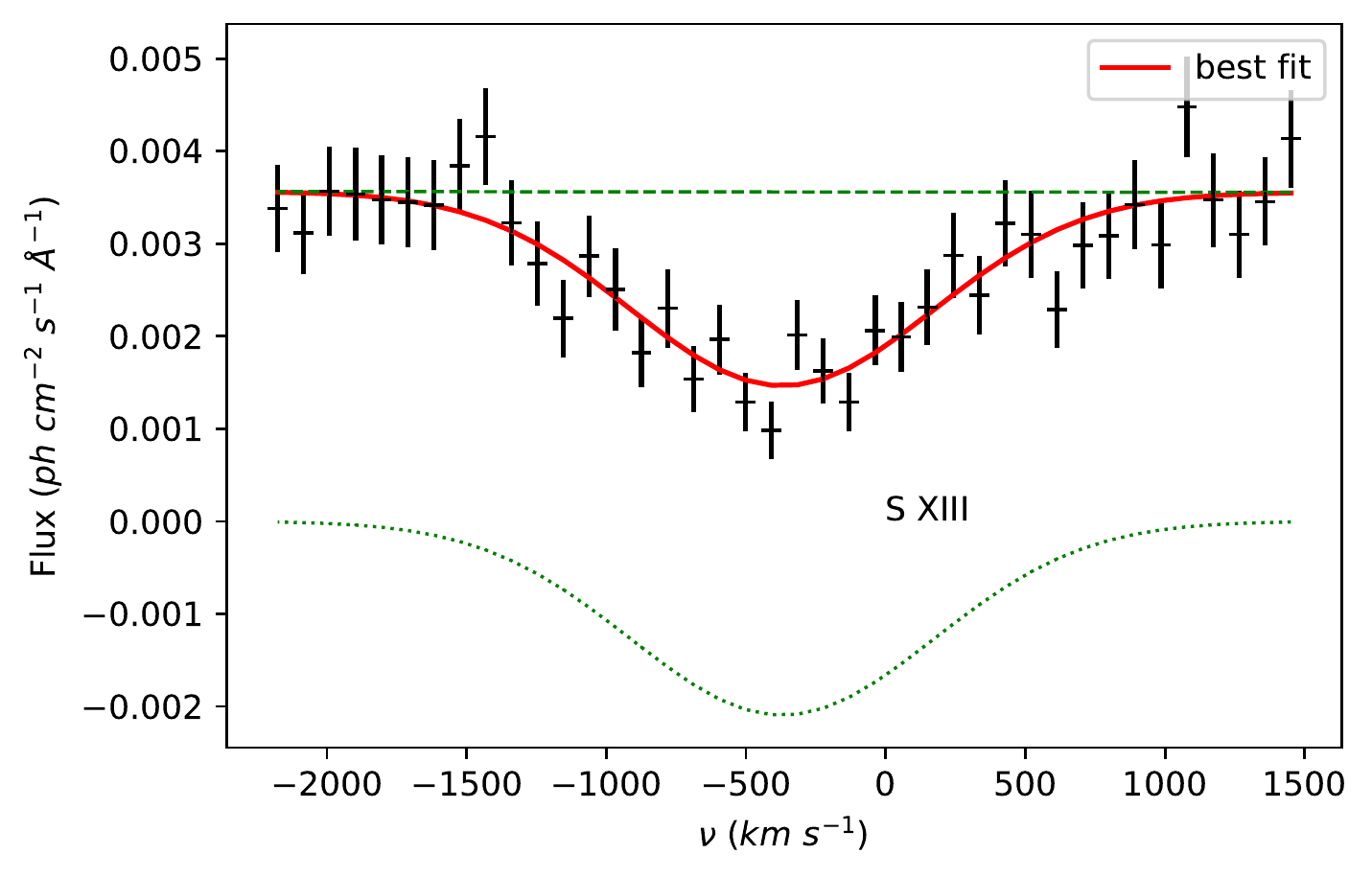}
\caption{On the left, the profiles of common features observed 
 with the RGS (averaged RGS1 and RGS2) at days 165 (red), 197 (black), and 229
 (blue). On the right, in velocity space the same line and the fit for day 229
(N VII) and for day 197 (other plots). 
}
\end{figure*}
\section{The Observations}  
 The rise observed with the {\sl Swift} XRT
 prompted two observations in the Director Discretionary Time (DDT),
 requested by W. Pietsch, 90 and 165 days after the optical
 maximum, observed on 2009 February 6 \citep{Liller2009}.  In July of 2009,
 the nova was sufficiently X-ray bright
 to trigger also two pre-approved target of Opportunity (TOO)
 observations awarded to PI M. Orio. These exposures were done
respectively days 197 and 229 after the optical discovery 
 at the observed optical maximum.
All four {\sl XMM-Newton} observations are listed in Table 1.
 While partial results were presented in \citet{OrioIAU, Orio2017}, this
 paper contains the first comprehensive analysis of all the data.

The {\sl XMM-Newton} observatory
consists of five different instruments behind three X-ray mirrors,
plus an optical monitor (OM), and all observe simultaneously.
 For this paper, we used the spectra from the Reflection Grating
Spectrometers \citep[RGS;][]{denHerder2001}, and the light curves
of the EPIC pn and MOS cameras.
 The calibrated energy range of the EPIC cameras is 0.15-12 keV for the pn, and 0.3-12 keV
 for the MOS. The RGS wavelength range is 6-38 \AA, corresponding to
 the 0.33-2.1 keV energy range.
 Table 1 shows that in the first two observations the EPIC 
 cameras were used in imaging mode, the pn detector was  ``small window mode''
 to mitigate pile up and the medium filter was used,
 while the MOS was used with the ``small'' frame and the medium filter. 
 The set up of the third and fourth observation was the same,
 with larger MOS2 window size and with the thin (instead of
 the medium) filter for both MOS. 

The EPIC pn and MOS detectors
 were used to extract light curves  with the {\sl XMMSAS}  ({\sl XMM} Science
 Analysis System)  task XMMSELECT after applying barycentric corrections
 to the event files, choosing only single-photon events (PATTERN=0) in the
 events' files.  A reference for this and the other {\sl XMMSAS}
 tasks mentioned below is  
\url{https://xmm-tools.cosmos.esa.int/external/xmm_user_support/documentation/sas_usg/USG.pdf}.
 The resulting source light curves were corrected for 
 background variations using the XMMSAS task \texttt{epiclccorr}. 
 Because there was essentially no emission above 0.8 keV, we used only the RGS
 with their high spectral resolution for the spectral analysis. We
 extracted the RGS spectra with the {\sl XMMSAS task} \texttt{rgsproc}. 
 Periods of high background were rejected.
\section{Irregular variability and spectral variability}
The averaged, fluxed RGS spectra for each of
 the four exposures  are shown in the same plot in Fig. 1,
 as fluxed spectra and in units of erg cm$^{-2}$ s$^{-1}$ \AA$^{-1}$ .
All spectra show a luminous continuum and emission lines.
While in last three observations the count rate level was comparable
 and the spectrum did not change very significantly,
 the first RGS spectrum obtained in
 May of 2009 (on day 90 of the outburst)
 still had a much lower average continuum, consistently with the rise phase
 observed with {\sl Swift} \citep{Bode2016}.

In Fig. 2 we present the EPIC-pn light curves during the four exposures,
because the pn is the instrument with the largest count rate and the
 best time resolution.
 In all four exposures there was large aperiodic variability:
 the count rate varied by an order of magnitude on day 90,  
 by a little more of a factor of 3 on days 165, and by about 
60\% on days 197 and 229. 
 The MOS and RGS light curves of each exposure
are modulated exactly like the pn one. Despite
 a moderate amount of pile-up in the pn spectrum,  the
 irregular variability of the source appeared the same in all
 instruments. We corrected
 for pile-up by excluding an inner region and leaving
only the PSF wing, and found that 
 pile-up does not affect the proportional amplitude of the irregular variations
 and light curve trend. 
 Irregular variability over time scales
 of hours has been observed
 at different epochs in the X-ray light curve
in several other SSS-novae, most notably in N SMC 2016 \citep{Orio2018},
V1494 Aql \citep{Drake2003}, and in one of
 the  early RS Oph exposures \citep{Nelson2008}.
 However, RS Oph was observed many times for hours, and we know that 
 the supersoft X-ray flux stabilized in the later exposures.
Before proceeding with a spectral analysis, we examined the spectral
 variability to assess whether it is connected
 with flux variability.  

 In the top panel of Fig. 2 a red line indicates the 
 count rate limit of 6 counts s$^{-1}$
 measured with the pn, chosen for the intervals in which we extracted
  high and low count rate high resolution RGS spectra on
 day 90. This is 
 the only exposure for which we found strong variability
 in the strength of the spectral features,
as shown in the inset in the first panel of the figure.
 There is especially a striking difference in the 25-28 \AA\ range,
 where some emission lines were almost only present when the flux
 increased significantly for short periods.
 In the other observations (days 165, 197, and 229)
 the  count rate variability
 was not matched by a change in the spectral shape of the continuum, 
 to which most of the flux variation is due. Also the 
strength of most emission lines varied less, and the variations 
 mostly followed with the change in the continuum. Fig. 3 shows
 ``time maps'', that is variations of each line during the exposures,
 as well as spectra extracted during two very short intervals of high and low
 count rate for each observation.  
 We found a correlation of the count rate with the
 strength of the  N VI and N VII lines. 
  The absorption edge of N VII, that abruptly cuts the 
 flux below 18.587 \AA, did not vary with the flux level,  
 and for this reason we rule out that the variability was caused by 
 changes in intrinsic absorption.
 A likely 
 interpretation of the variability of the continuum flux is 
 that along the line of sight 
 a partially covering, ``opaque'' absorber appeared and disappeared.
 In this scenario, the WD surface is 
 for some time partially obscured by the accretion disk that was not disrupted
 in the outburst, or by large, optically thick and asymmetric
 regions of the ejecta. Because of the irregular variability over
 short time scales, in Nova LMC 2009 we favor the second hypothesis, and  
 elaborate  this idea in the Discussion and Conclusions sections.
We rule out intrinsic variations of the WD flux, 
 because the nova models indicate that  burning 
 occurs at a constant rate, thus the thin atmosphere above the burning
 layer can hardly change temperature.
\section{Identifying the spectral features}
Identifying the spectral features for this nova appears much more
 challenging than it has been for other novae observed in the last
 15 years \citep[e.g.][]{Rauch2010, Ness2011, Orio2018, Orio2020}.
 We started by examining the strongest features.
In the spectrum of day 90, we clearly identify two strong 
 emission features: a redshifted  Lyman-$\alpha$
 H-like line of N VII (rest wavelength 24.78 \AA,
 possibly partially blended with a weaker line
 of N VI He$\beta$ at 24.889 \AA),
 and the N VI He-like resonance line at 28.78 \AA.
  We fitted the two stronger
lines with Gaussian functions and obtained a
 redshift velocity of 1983$\pm 190$ km s$^{-1}$ for the N VII
 line, and, consistently,  a velocity of 1926$^{+220}_{-250}$ km s$^{-1}$ 
 for N VI. The integrated
flux in these two lines is 1.23$\pm0.18 \times 10^{-13}$ erg
 cm$^{-2}$ s$^{-1}$ and 2.37$\pm0.18 \times 10^{-13}$ erg cm$^{-2}$ s$^{-1}$,
 respectively.

 The spectra of the three following dates, when the SSS was at plateau luminosity,
 show a forest of features in absorption, and also
 some in emission. The continua of these spectra appear remarkably
 similar, but many features varied from one exposure to
 the next, unlike  in
other novae observed at the peak of the SSS emission,
where the absorption features were not
 found to change significantly in different exposures
 done within weeks \citep[e.g. RS Oph, V4374 Sgr,
 V2491 Cyg, see ][]{Nelson2008, Rauch2010, Ness2011}.  
 We focused only on the
 absorption lines that are clearly common to all the last three spectra. 
 In addition to the interstellar
 absorption lines of O I at 23.51 \AA\ and N I
 at 31.3 \AA, we found only five strong common features, whose profiles
 we show in Fig. 3. The y-axis shows the averaged RGS1 and RGS2 count rate.

 The line of
 N VII at 24.78 \AA \ appears to have a P-Cyg
 profile in at least two exposures. Rather
 than being a ``true'' P-Cyg, it may be due to the superposition of absorption
 and emission lines produced in different regions \citep[e.g. absorption
 in the WD atmosphere, emission farther out in the shell, like
 in U Sco, see][]{Orio2013}. 
 The other lines in Fig.3 are only in absorption.
The first three
 were already observed at almost the same wavelength in
 other novae \citep{Ness2011}, and were marked as yet ``unidentified''.
We suggest identification of two of these lines as argon:
 Ar XIII at rest wavelength  29.365 \AA,
 and Ar XIV at 27.631-27.636 \AA.
 We also propose a tentative identification of the 
 Ca XI (30.448 \AA) and S XIII (32.239 \AA).  

In Table 2 we report blueshift velocity, broadening velocity and
optical depth obtained with these identifications. Fits for
 one of the spectra are shown in
 Fig. 3 in the panels on the left.
 To calculate the blueshift velocities, we followed 
the method described by \citet{Ness2010} to determine the line shifts, widths,
and optical depths at the line center of the absorption lines for the spectra for the
 four exposures. Following \citet{Ness2011}, we did
 not include the absorption correction, because it is not important in determining
 velocity and optical depth.
 The narrow spectral region around each line was fitted with a function
 $$C(\lambda) \times e^{-\tau(\lambda)}$$
 where $\tau(\lambda)$ is a result of the fit
 and is the opacity for each line.
We assumed that $C(\lambda)$ is a linear function for each line in modelling
 the continuum. 
 We also fitted the N VII emission component with a Gaussian function.

The blueshift
 velocity is modest compared to other novae (RS Oph, see \citet{Nelson2008},
 V4743 Sgr, see \citet{Rauch2010}, V2491 Cyg see \citet{Ness2011},
 N SMC 2016, described in \citet{Orio2018}), but this nova was observed
 as a SSS at a much later post-outburst time. The spread of blueshift
 velocity was also evident in V2491 Cyg \citep[see][]{Ness2011}.
\begin{figure*}
\includegraphics[width=87mm]{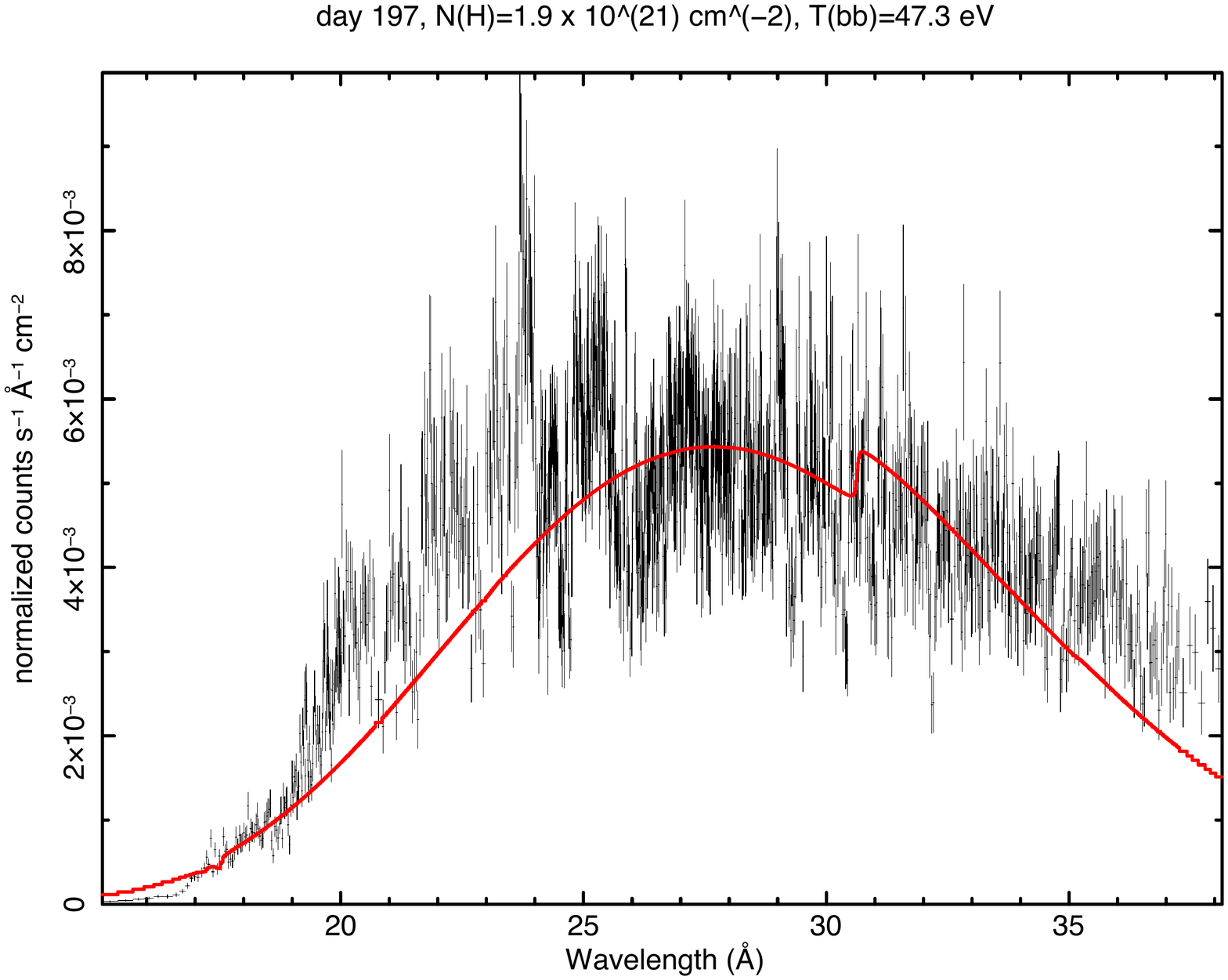}
\includegraphics[width=87mm]{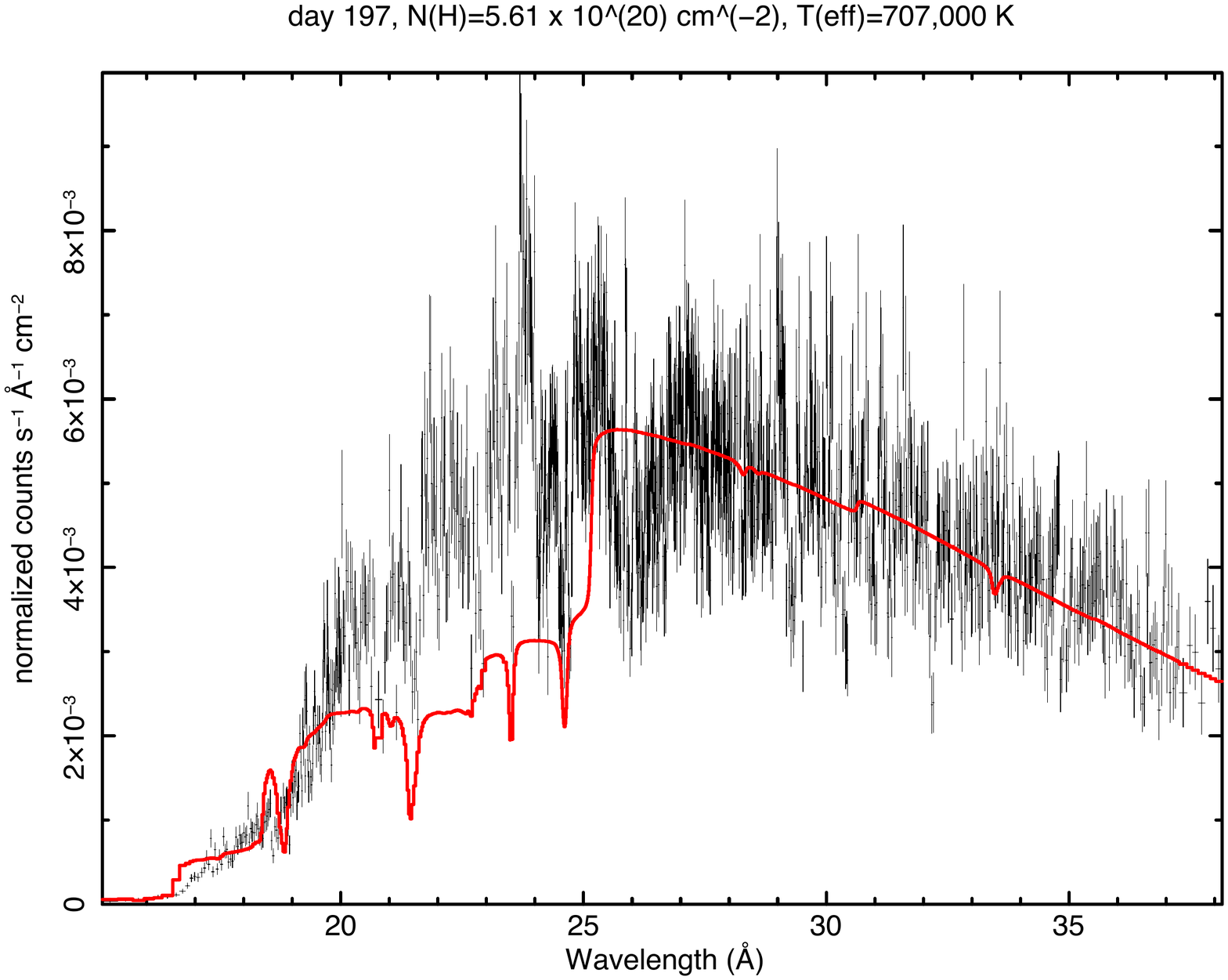}
\includegraphics[width=87mm]{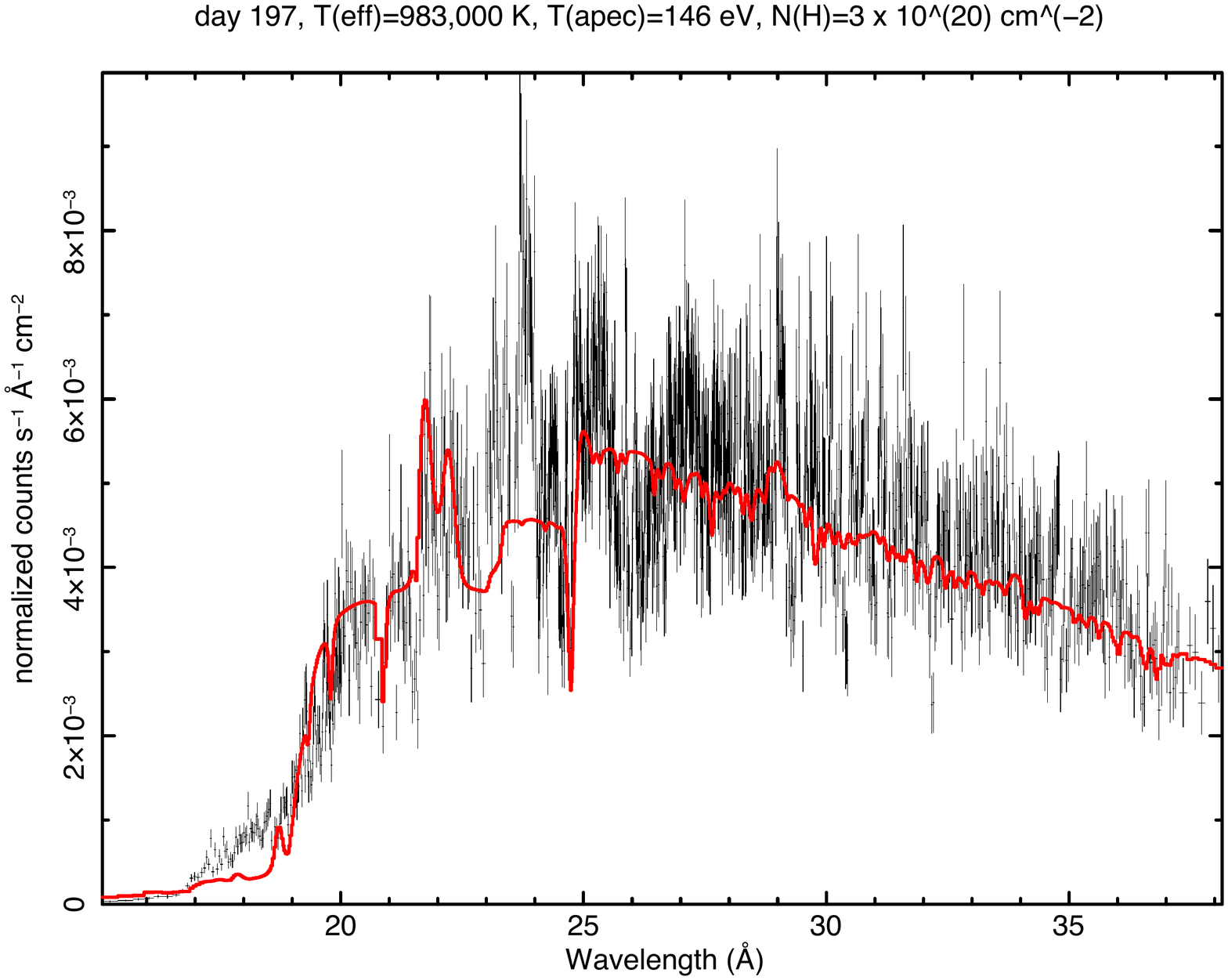}
\includegraphics[width=87mm]{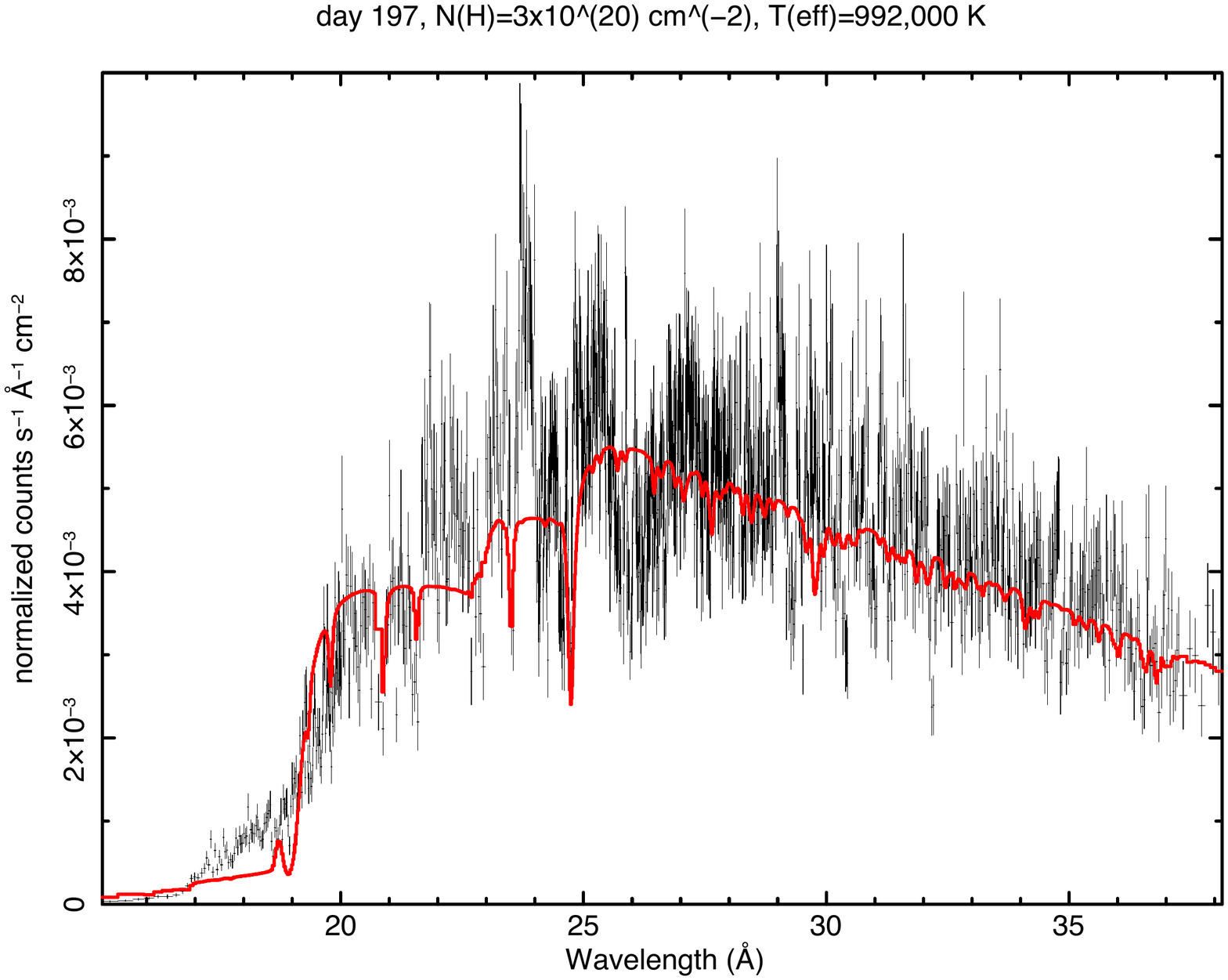}
\caption{Clockwise, from the top left,
 comparison of the RGS fluxed spectrum on day 197 with different
 models, respectively:
 a blackbody with oxygen depleted absorbing
interstellar medium, a metal poor model TMAP atmosphere, 
  a metal rich TMAP atmosphere, and the
 metal rich atmosphere with a superimposed thermal plasma component
 (single temperature) in collisional ionization equilibrium. 
 See Table 3 and text for the details.}
\end{figure*}
\begin{figure*}
\includegraphics[width=87mm]{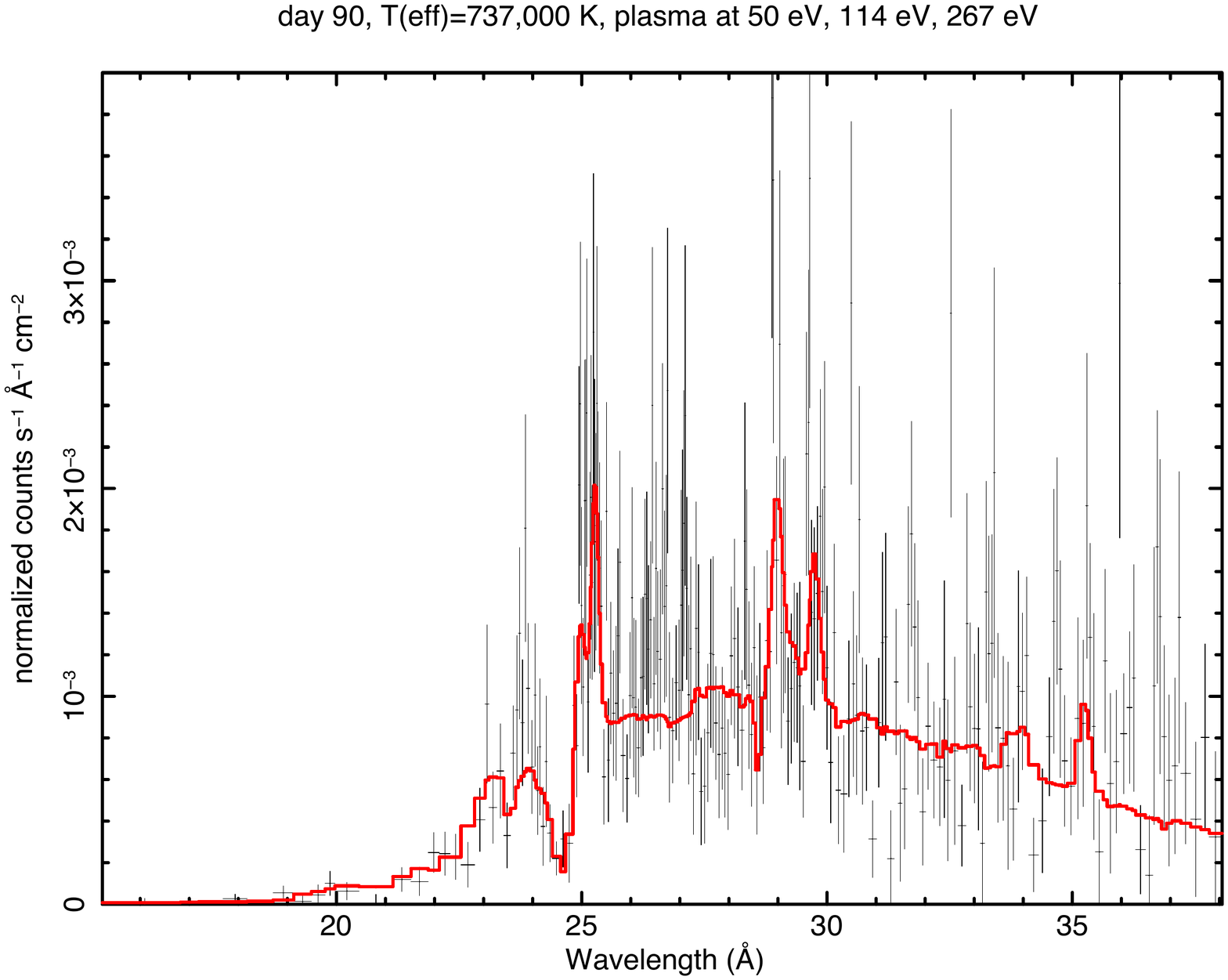}
\includegraphics[width=87mm]{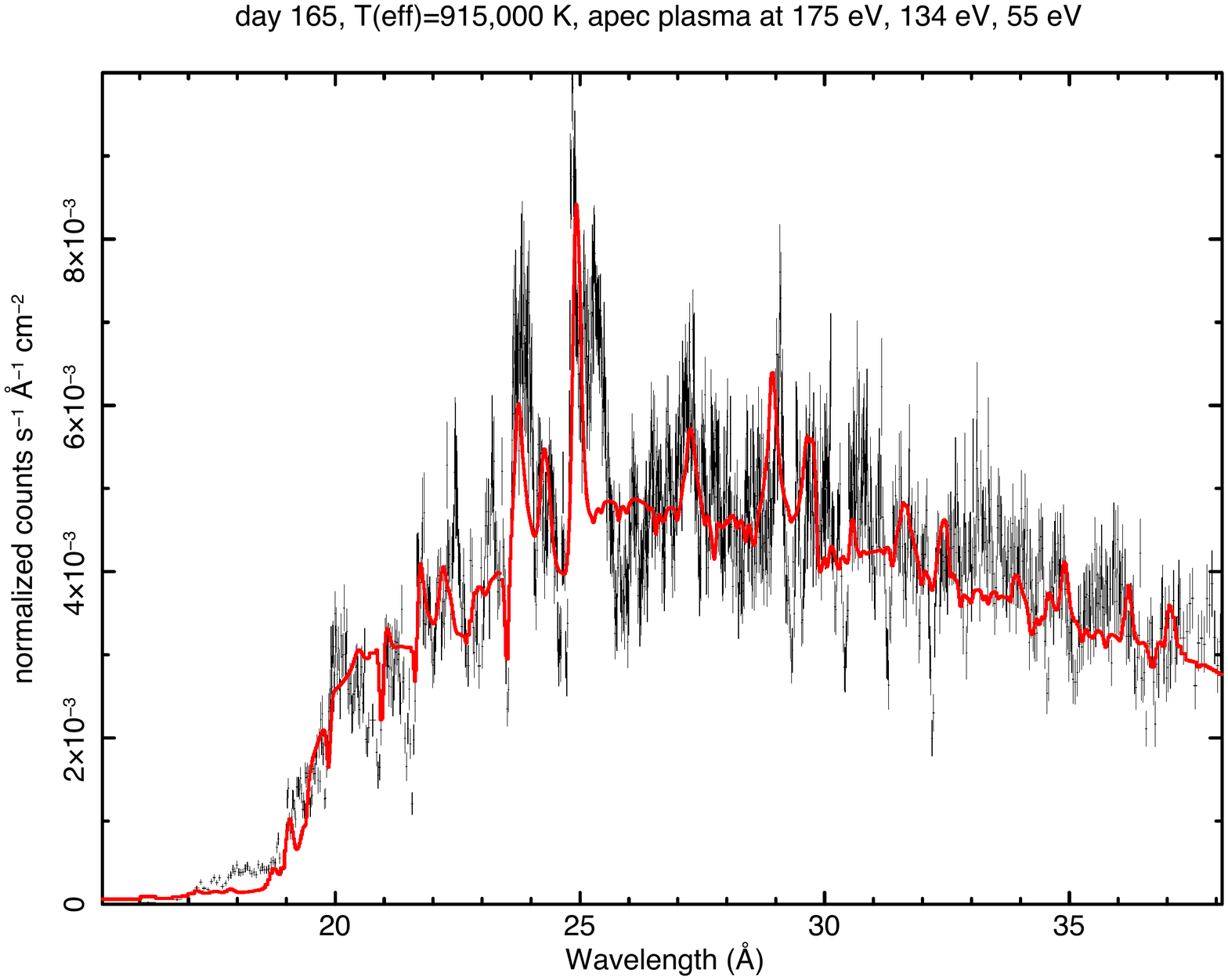}
\includegraphics[width=87mm]{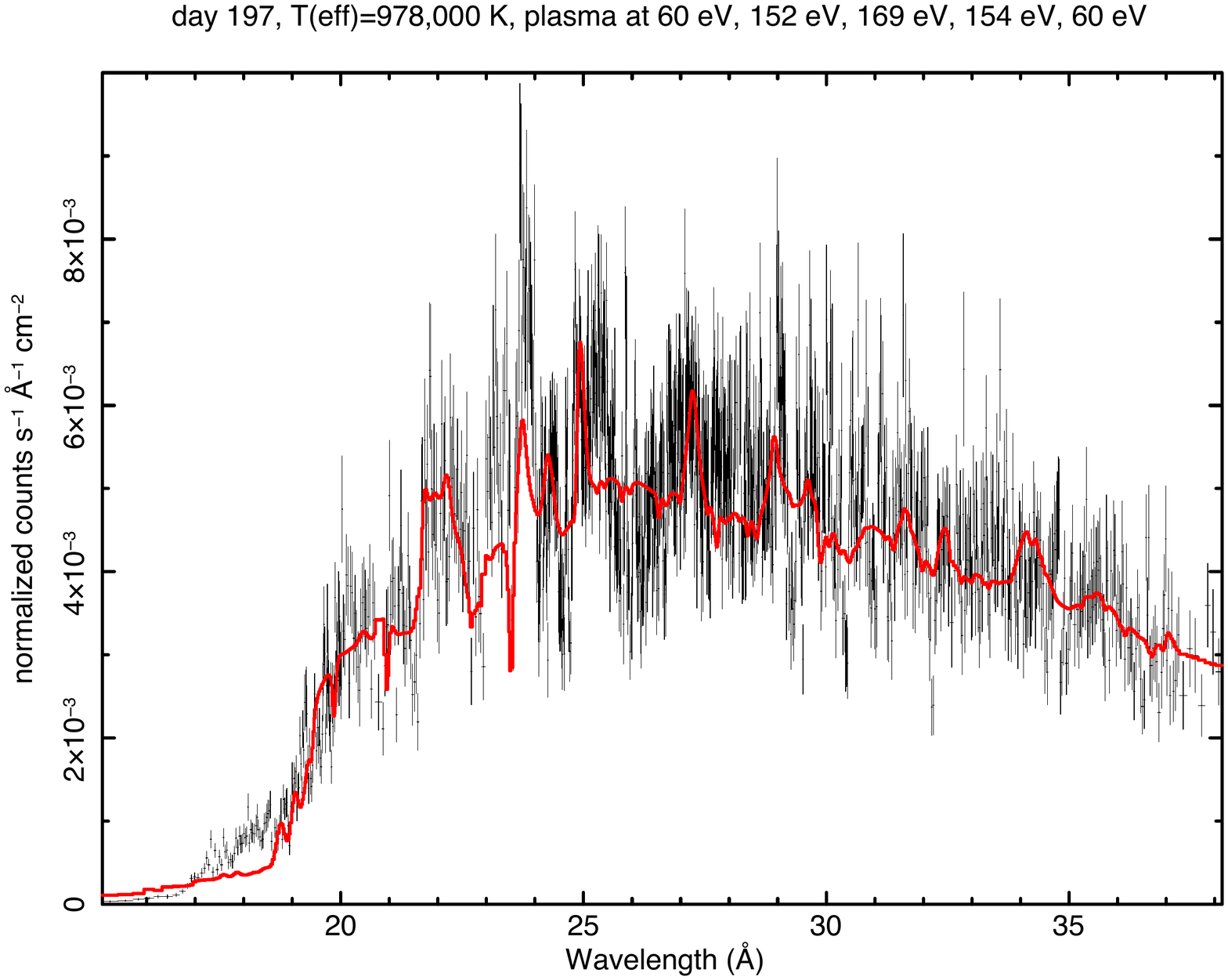}
\includegraphics[width=87mm]{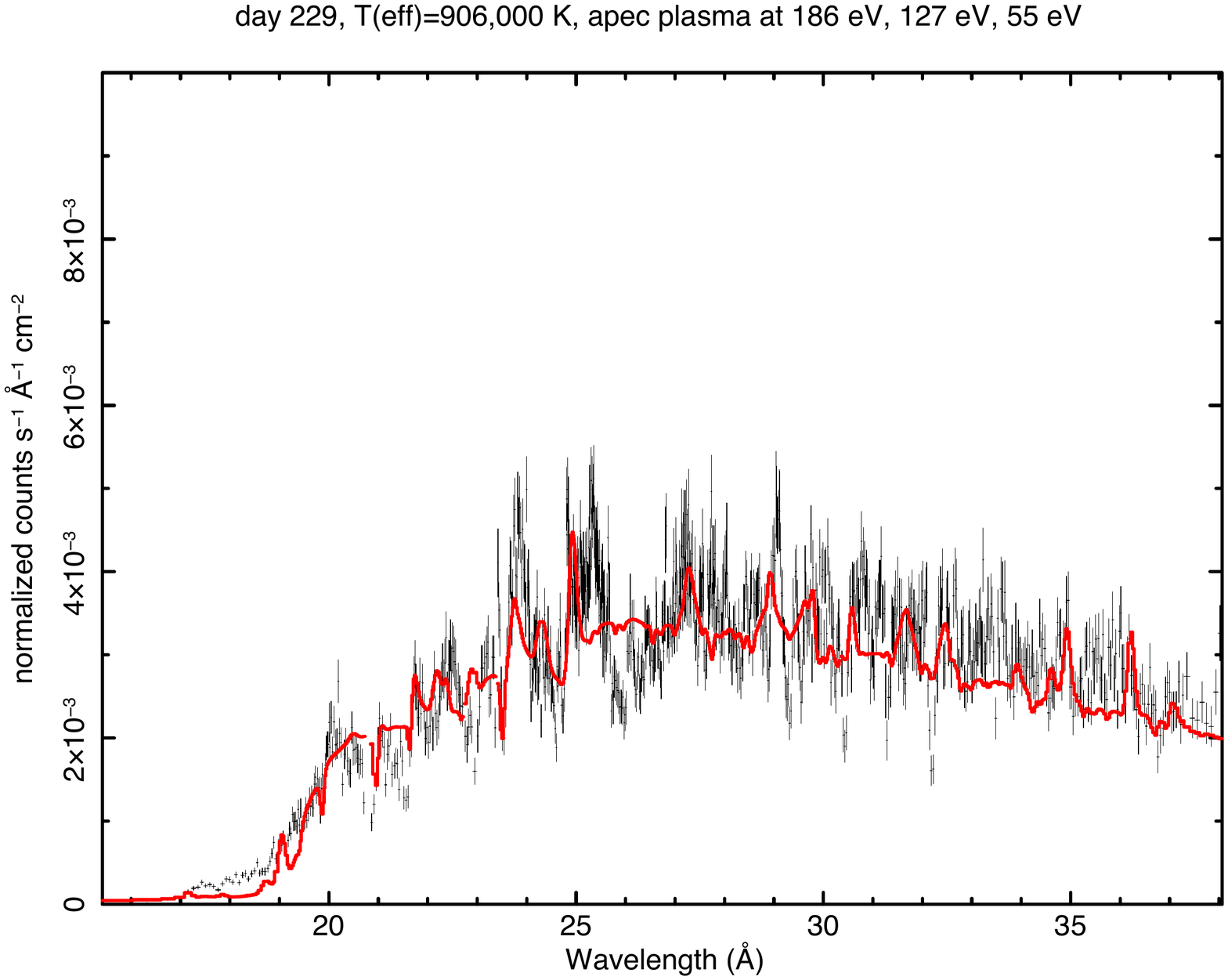}
\caption{ Fit to the RGS spectra of days 90 (of the ``high'' period as
 shown in Fig.2), 165, 197 and 229 with
 the composite model with parameters in Table 4.}
\end{figure*}

\section{Spectral evolution and the viable spectral models}
With the gratings, we measure the integrated flux 
 without having to rely on spectral fits like
 with broad band data.  The  flux in
 the 0.33-1.24 keV (10-38 \AA) range increased by only 10\%
 between days 165 and 197,  as shown in Table 2. 
 On day 197, the source was also somewhat ``harder'' and hotter.
 A decrease  by a factor of 1.6 was registered 
 on day 229, when the sources had ``softened'' again.
 This evolution is consistent with the maximum temperature estimated 
 with the {\it Swift} XRT around day 180 and the following
 plateau, with final decline of the SSS occurring between days 257 and 279 \citep{Bode2016}.
\begin{table*}
\begin{flushleft}
\caption{Rest wavelength, and observed 
 wavelength, velocity, broadening velocity, and optical depth resulting
 from the fit of spectral lines with
 proposed identification.}
\label{table:absorption 2009a}
\begin{center}
\begin{tabular}{lccccc}
\hline
Ion & $\lambda_0$ & $\lambda_m$ & $v_{\rm shift}$ & $v_{\rm width}$ & $\tau_{\rm c}$  \\
& (\AA) & (\AA) & (km\,s$^{-1}$) & (km\,s$^{-1}$) &  \\
\hline
\multicolumn{6}{c|}{\bf 0610000501}\\
\hline
N VII  & 24.779 & $24.746 \pm 0.005$ & \mbox{ $-394 \pm 66$} & \mbox{ $737 \pm 154$} & \mbox{ $0.0061 \pm 0.0034$}  \\
 &   & $ 24.799\pm 0.049 $  & \mbox{ $+242 \pm 94$} &   &    \\
Ar XIV   & 27.636 & $27.615 \pm 0.006$ & \mbox{$-223 \pm 64$ } & \mbox{$523 \pm 110$ }  & \mbox{$0.0014 \pm 0.0002$ }  \\
Ar XIII  & 29.365 & $29.288 \pm 0.010$ & \mbox{$-789 \pm 104$ } & \mbox{$1216 \pm 202$ }  &  \mbox{ $0.0027 \pm 0.0003$}  \\
Ca XI    & 30.448 & $30.413 \pm 0.008$ & \mbox{$-344 \pm 75$ }  & \mbox{$746 \pm 117$ }  & \mbox{$0.0021 \pm 0.0003$ }  \\
S XIII   & 32.239 & $32.200 \pm 0.005$ & \mbox{$-361 \pm 49$ }  & \mbox{$539 \pm 77$ }  & \mbox{$0.0023 \pm 0.0003$ }  \\
\hline
\multicolumn{6}{c}{\bf 0604590301}\\
\hline

N VII  & 24.779 &  $24.728 \pm 0.006$ & $-616 \pm 67$ &  $470 \pm 106$ &  $0.0025 \pm 0.0004$  \\
Ar XIV   & 27.636 & $27.599 \pm 0.005$ & $-406 \pm 52$ &  $323 \pm 83$ &  $0.0018 \pm 0.0004$  \\
Ar XIII  & 29.365 & $29.285 \pm 0.021$ & $-813 \pm 215$ &  $1370 \pm 596$ &  $0.0030 \pm 0.0013$  \\
Ca XI    & 30.448 & $30.403 \pm 0.006$ & $-435 \pm 63$ &  $504 \pm 95$ &  $0.0021 \pm 0.0003$  \\
S XIII   & 32.239 & $32.184 \pm 0.009$ & $-513 \pm 80$ &  $630 \pm 132$ &  $0.0022 \pm 0.0004$  \\

\hline
\multicolumn{6}{c}{\bf 0604590401}\\
\hline
N VII  & 24.779 &   $24.732 \pm 0.006$ & \mbox{ $-569 \pm 71$} & \mbox{ $413 \pm 119$} & \mbox{ $0.0012 \pm 0.0004$} \\
&   & $ 24.829 \pm 0.004 $  & \mbox{ $+604 \pm 54$} &   &  \\
Ar XIV   & 27.636 &   $27.621 \pm 0.006$ & \mbox{ $-159 \pm 70$} & \mbox{ $383 \pm 111$} & \mbox{ $0.0013 \pm 0.0003$} \\
Ar XIII  & 29.365 &  $29.283 \pm 0.008$ & \mbox{ $-835 \pm 83$} & \mbox{ $1051 \pm 149$} & \mbox{ $0.0019 \pm 0.0002$} \\
Ca XI    & 30.448 &   $30.433 \pm 0.006$ & \mbox{ $-149 \pm 57$} & \mbox{ $539 \pm 87$} & \mbox{ $0.0018 \pm 0.0002$} \\
S XIII   & 32.239 &  $32.199 \pm 0.006$ & \mbox{ $-370 \pm 55$} & \mbox{ $767 \pm 96$} & \mbox{ $0.0021 \pm 0.0002$} \\
\hline
\end{tabular}
\end{center}
\end{flushleft}
\end{table*}
\begin{table*}
\caption{Integrated flux measured in the RGS range 10-38 \AA,
effective temperature, column density and unabsorbed flux in the
 WD continuum resulting from the
 fit with two atmospheric models with different metallicities, as described
 in the text: M1 is metal enriched and M2 is metal poor.}
\begin{center}
\begin{tabular}{cccccccc}
\hline
 Day & RGS Flux & M1 T$_e$ & M1 Unabs. flux & M1 N(H) & M2 T$_e$ & M2 Unabs. flux & M2 N(H) \\
     & erg cm$^{-2}$ s$^{-1}$ & K & erg cm$^{-2}$ s$^{-1}$ & 10$^{20}$ cm$^{-2}$ &  K & erg cm$^{-2}$ s$^{-1}$ & 10$^{20}$ cm$^{-2}$ \\
\hline
\hline
 90.4 &  5.38 $\times 10^{-12}$ & 738,000 & 1.29  $\times 10^{-11}$ & 6.5 &
 624,000 & 1.36 $\times 10^{-11}$  & 5.7 \\
164.9 & 8.08 $\times 10^{-11}$  & 910,000 & 9.08 $\times 10^{-11}$ &  3.5 & 784,000 &
 9.08 $\times 10^{-10}$ & 6.1 \\
 196.17 & 1.20 $\times 10^{-10}$ & 992,000 & 9.06 $\times 10^{-11}$  & 4.5 & 707,000 &
 1.21 $\times 10^{-10}$ & 5.6 \\
 228.93 & 3.91 $\times 10^{-11}$ & 947,000 & 5.89 $\times 10^{-11}$ & 3.0 &
701,000 & 7.29 $\times 10^{-11}$ &  4.4 \\
\hline
\hline
\end{tabular}
\end{center}
\label{flux}
\end{table*}
 We tried global fits of the spectra, with several steps. Although
 we did not obtain a complete, statistically significant and comprehensive
 fit, we were able to test several models and reached
 a few conclusions in the physical mechanisms from which the
 spectra originated. We focused mostly on
 the maximum (day 197), for which we show the result with all all models.
 The first steps were done by using XSPEC \citep{Dorman2001} to fit the models.
 
 $\bullet$ {\bf Step 1.}
 The first panel of Fig. 5 shows the fit with a blackbody for day 197. 
 We first fitted the spectrum with the TBABS model in 
 XSPEC \citep{Wilms2000}, however a better fit,
 albeit  not statistically significant yet, was
 with obtained with lower blackbody
 temperature 47.3 eV (almost 550,000 K), with the TBVARABS formulation by the same authors for the 
 intervening absorbing column N(H). This prescription allows to
 vary the abundances of the absorbing medium, and in the best
 fit we could obtain we 
 varied the oxygen abundance, allowing it to decrease to almost zero,
 because an absorption edge of O I at 22.8 \AA\ in the ISM makes a significant
difference when fitting a smooth continuum like a blackbody.
 However, it can be seen in Fig. 5 even the best blackbody
fit is not very satisfactory in the hard portion of the spectrum.

 $\bullet$ {\bf Step 2.} Assuming, as the models predict \citep[e.g.][]{Yaron2005,Starrfield2012,Wolf2013},
 and several observations have confirmed 
 \citep[e.g.][]{Ness2011,Orio2018}, that most of the X-ray flux of
 the SSS originates in the atmosphere of the post-nova WD, 
 we experimented by fitting atmospheric models for
 hot WDs burning in shell. We used the grid of TMAP models
 in Non Local Thermodynamical Equilibrium (NLTE) by \citet{Rauch2010},
available  in the web site
 \url{http://astro.uni-tuebingen.de/#rauch/TMAP/TMAP.html}. We
 wanted to evaluate whether the 
 significant X-ray brightening between May and July (day 90 to
 day 165)  was due to the WD shrinking and becoming hotter
 in the constant bolometric luminosity phase predicted
 by the nova models \citep[e.g.][]{Starrfield2012}, or to decreasing
 column density (N(H)) of absorbing material in the shell.
In several novae, and most notably 
 in V2491 Cyg \citep{Ness2011} and N SMC 2016 \citep{Orio2018},
 despite blueshifted absorption lines that indicate a residual,
 fast wind from
 the photosphere, a static model gives a good first order fit and
 predicts most of the absorption features.
 We moved towards the
 blue the center of all absorption lines by a given amount for all
 lines, leaving the blueshift as a free parameter in the 100-1500 km s$^{-1}$ range,
 compatible with the values found in Table 2.

 Table 3 (for all four exposures) and Fig. 5 (for day 197) show the fit with two different  grids of models available in the web site
 \url{http://astro.uni-tuebingen.de/#rauch/TMAP/TMAP.html}.
 In the attempt to
 optimize the fit, we calculated both the CSTAT parameter \citep{Cash1979, Kaastra2017}
 and the reduced $\chi^2$, but 
 we cannot fit the spectrum very well with only the atmospheric
 model (thus we do not
 show the statistical errors in Table 3),
 since an additional component appears to be superimposed on the WD spectrum
 and many fine details are due to it.
 \citet{Bode2016} attributed an
 unabsorbed luminosity 3-8 $\times 10^{34}$ erg s$^{-1}$ to the nova ejecta.
 Another
  shortcoming  is that the TMAP model does not include argon. Argon L-shell ions
 are important between 20 and 40 \AA \ and argon may be even enhanced in
 some novae \citep[the oxygen-neon ones, see][]{Jose2006}. Above, we have identified
 indeed two  of these argon features, that appear strong and quite constant 
 in the different epochs. 

 Model 1 (M1) is model S3 of the 
 ``metal rich'' grid (used by \citet{Bode2016} for the
 broad band Swift spectra), including all elements up to nickel, 
 and Model 2 (M2) is from the  metal poor ``halo'' composition
 grid,  studied for
 non-nova SSS sources in the halo or Magellanic Clouds, which
 are  assumed  
 to have accreted metal poor material and may have 
 only undergone very little mixing above the burning layer.
 The
nitrogen and oxygen mass abundance in model S3 are as follows:
 nitrogen is 64 times the solar value, oxygen is 34 times the solar value.
 In contrast, carbon is depleted because of the CNO cycle, and is
 only 3\% the solar value. 
 
Novae in the Magellanic Clouds may not be similar
 to the non-ejecting SSS. They may be metal-rich and not have retained
 the composition of the accreted material, since convective mixing
 is fundamental in causing the explosion, heating the envelope and bringing towards
 the surface $\beta^+$ decaying nuclei \citep{Prialnik1986}. 
 The first interesting fact is that adopting the TBVARABS formulation
 like for a blackbody does not make a significant difference
 in obtaining the best fit. In fact, the atmospheric models have
 strong absorption edges and features that ``peg'' the model at a certain
 temperature and are much more important in the fit than any
 variation in the N(H) formulation, especially with
 low absorbing column like we have towards the
 LMC. In the following context, we show only models obtained
 with TBABS (fixed solar abundances).

 Both the metal poor
 and metal enriched models imply a very compact and massive WD, with logarithm of
 the effective gravity log(g)=9. In fact, the resulting effective temperature 
 is too high for a less compact configuration, the SSS would have
 largely super-Eddington luminosity.
 Table 4 does not include the blue-shift velocity, which was a variable
 parameter but was fixed for all lines: it is in the 300-600 km s$^{-1}$ 
 for each best fit, consistently with the measurements in
 Table 2.

 In Fig. 5 we show the atmospheric fits for day 197. 
 The metal rich model, with higher effective temperature,  appears more suitable. 
 We note that in the last three spectra, whose best fit parameters are shown 
 in Table 3,
 the value of the column density N(H) for this model was limited to a minimum
 value of 3 $\times 10^{20}$ cm$^{-2}$, consistently with
 LMC membership. The  fact
 that the metal rich model fits the spectrum better indicates that, even 
 in the low metallicity environment of the LMC, the 
 material in the nova outer atmospheric layers during a mass-ejecting 
  outburst mixes up  with ashes of the burning and with
 WD core material. The WD atmosphere is
 thus  expected to be metal rich and especially rich in nitrogen. 
However, the metal rich model overpredicts an absorption edge of N VII at 18.587 \AA, which is instead underpredicted by the ``halo'' model. 
 This is a likely indication that the abundances may be a little lower than in
 Galactic novae, for which the ``enhanced'' TMAP better fits the absorption edges 
\citep[e.g.][]{Rauch2010}.
The metal-rich model indicates that the continuum is consistent with
 a WD at effective temperature around 740,000$\pm$50,000 K on day 90,
 and hotter in the following exposures, reaching
 almost a million K on day 197. Although the fit indicates
 a decrease in intrinsic absorption after the first observation, the increase in
 apparent luminosity is mostly due to the increase
 in effective temperature and we conclude that 
 the WD radius was contracting until at least day 197.

\begin{table*}
\caption{Main parameters of the best fits obtained by adding three BVAPEC components to the 
 enhanced abundances TMAP model, and allowing the nitrogen
 abundance to vary. The flux is the unabsorbed one. 
 The  N(H) minimum value 3 $\times 10^{20}$ cm$^{-2}$ and the fit converged
to the minimum value in the second and fourth observation. We assumed
 also a minimum BVAPEC temperature of 50 eV.}
\begin{center}
\begin{tabular}{ccccc}
\hline
  Parameter  & Day 90.4 ``high spectrum''  & Day 164.9 & Day 196.17 & Day 228.93 \\
\hline
\hline
  & & & & \\
 N(H) $\times 10^{20}$ (cm$^{-2}$) &     8.7$^{+3.8}_{-1.8}$ &      3 & 6.8$_{-1.9}^{+4.2}$    & 3 \\
 & & & & \\
 T$_{\rm eff}$ (K)                    & 736,000$^{+11,000}_{-21,000}$ & 915,000$\pm 5,000$  &
  978,000$\pm 5,000$ & 907,000$\pm 5,000$ \\ 
 & & & & \\
 F$_{\rm SSS,un} \times 10^{-11}$ (erg s$^{-1}$ cm$^{-2}$) 
            & 2.64$_{-0.71}^{+1.52}$ & 7.69$_{-0.04}^{+0.01}$ &  9.49$_{-0.17}^{+0.20}$ &
  5.58$_{-0.08}^{+0.16}$ \\
  & & & & \\
 kT$_1$ (eV)& 50$^{+8}_{-0}$ & 55$_{-5}^{+2}$  & 68$\pm 8$ & 55$_{-9}^{+2}$  \\
  & & & & \\
 F$_{\rm un,1} \times 10^{-11}$ (erg s$^{-1}$ cm$^{-2}$) &
             0.71$_{-0.69}^{+0.70}$ & 0.12$_{-0.02}^{+0.05}$ & 5.83$_{-2.50}^{+11.00}$ &
     0.17$_{-0.16}^{+0.26}$ \\ 
  & & & & \\
 kT$_2$ (eV) & 114$_{-18}^{+22}$    & 134$\pm 2$ & 152$\pm 6$ & 127$\pm 8$ \\
  & & & & \\
 F$_{\rm un,2} \times 10^{-12}$ (erg s$^{-1}$ cm$^{-2}$)
           & 1.74$^{+1.00}_{-0.70}$ & 3.29$\pm$0.03 & 4.45$^{+6.00}_{-4.00}$ &
    1.93$_{-0.29}^{+0.46}$ \\
  & & & & \\
 kT$_3$ (eV) & 267$_{-165}^{+288}$  & 176$\pm 3$      & 174$\pm 10$       & 177$\pm 6$  \\  
  & & & & \\
  F$_{\rm un,3} \times 10^{-12}$ (erg s$^{-1}$ cm$^{-2}$) 
          & 0.77$^{+1.00}_{+17.57}$ &  5.56$\pm0.50$ & 4.46$_{-4.00}^{+5.21}$  & 3.29$\pm0.39$ \\
  & & & & \\
 F$_{\rm un,total} \times 10^{-11}$ (erg s$^{-1}$ cm$^{-2}$)
    & 3.60 & 8.69 & 16.21  & 6.28 \\
 & & & & \\
\hline
\hline
\end{tabular}
\end{center}
\end{table*}
\begin{table}
\caption{Main physical parameters of the SPEX Black Body(BB)+2-PION model fit for day 197.
 ISM indicates values for the intervening interstellar medium column density.
 The errors (67\% confidence
 level) were calculated only for the PION in absorption, assuming fixed parameters for
 PION in emission. The abundance values marked with (*) are parameters that reached
 the lower and upper limit.}
\begin{center}
\begin{tabular}{cc}
\hline
\hline
N(H)$_{\rm ISM}$ &  9.6$\pm 0.4 \times 10^{20}$ cm$^{-2}$ \\
0/0$_\odot$ (ISM)  & 0.06 (*) \\
T$_{\rm BB}$     & 812,000$\pm 3020$ K \\
R$_{\rm BB}$     & 8  $\times 10^8$ cm \\
L$_{\rm BB}$      & 7 $\pm 0.2 \times 10^{37}$ erg s$^{-1}$ \\ 
N(H)$_1$ &  6.4$^{+2.5}_{-1.9} \times 10^{21}$ cm$^{-2}$ \\
v$_{\rm width}$ (1) & 161$\pm 4$ km s$^{-1}$ \\
v$_{\rm blue shift}$ (1) & 327$^{+16}_{-21}$  km s$^{-1}$ \\
N/N$_\odot$ & 100$^{-7}$ (*) \\
0/0$_\odot$ & 13$\pm 2$ \\
$\xi$ (1) & 489.7$^{+50}_{-30}$ erg cm \\
$\dot m$  & 1.84  $\times 10^{-8}$ M$_\odot$ yr$^{-1}$ \\
n$_{\rm e}$ (1) & 2.85 $\times 10^{8}$ cm$^{-3}$ \\
N(H)$_2$ &  2.6 $\times 10^{19}$ cm$^{-2}$ \\
v$_{\rm width}$ (2) & 520 km s$^{-1}$ \\
v$_{\rm red shift}$ (2) & 84  km s$^{-1}$ \\
$\xi$ (2) & 30.2  erg cm \\
n$_{\rm e}$ (2) & 2.87 $\times 10^{4}$ cm$^{-3}$ \\
L$_{\rm X}$ (2) & 1.1 $\times 10^{36}$ erg s$^{-1}$ \\ 
\hline
\hline
\end{tabular}
\end{center}
\label{flux}
\end{table}

 The ``enriched'' model atmosphere seems to be the most suitable
 in order to trace the continuum, except for the  excess flux above 19 \AA.
 Because the continuum shape of the atmospheric model does not change
 with the temperature smoothly or ``incrementally'' like a blackbody,
 and is extremely dependent on the absorption edges, we cannot obtain
 a better fit by assuming an absorbing medium
 depleted of oxygen (or other element). This is the same also for
 the ``halo'' model.  Therefore, the fits we
 show were obtained all only with TBABS, assuming solar abundances
 in the intervening column density between us and the source.

$\bullet$ {\bf Step 3.}
 In these spectra, we do not detect He-like triplet lines sufficiently
 well in order to use line ratios as diagnostics \citep[see, for instance,][and references
therein]{Orio2020}. 
In the last panel in Fig. 5, we show the fit with TMAP and a component of
 thermal plasma in collisional ionization equilibrium \citep[BVAPEC in XSPEC,see][]{Smith2001}.
 The fit improvement is 
 an indication that many emission lines from the nova outflow
are superimposed on the SSS emission, but clearly a single thermal
 component is not sufficient to fit the whole
 spectrum.  Nova shells may have luminous emission lines in
 the supersoft range after a few months from
the peak of the outburst \citep[V382 Vel, V1494 Aql][]{Ness2005,Rohrbach2009}.
 Such emission lines in the high resolution X-ray spectra of novae are still
 measured when the SSS is eclipsed or obscured \citep[][]{Ness2013},
 indicating an  origin far from the central source.
 Shocked ejected plasma in collisional ionization equilibrium has
 been found to originate the X-ray emission of
 several novae, producing emission features in different
 ranges, from the ``hard'' spectrum of V959 Mon \citep{Peretz2016} to
 the much ``softer'' spectrum of T Pyx \citep{Tofflemi2013}.
 
 Adding only one such additional component with plasma temperature
 keV improves the fit by modeling oxygen emission lines,
 but it does not explain all the spectrum in a rigorous way. 
 The fit improved by adding one more BVAPEC component, and improved 
 incrementally, when a third one was added. The fits shown in Fig. 5
 yield a reduced $\chi^2$ parameter still of about 3, due to many
 features that are  are still unexplained. Also, 
the N VII K-edge remains too strong and this is not due to an overestimate
 of N(H), since the soft portion of the flux is well modeled.
In Fig. 6 and in Table 4 we show fits with 
 three BVAPEC components of shocked plasma
 in collisional ionization equilibrium. To limit the
 number of free parameters, the results we present
 in Table 4 were obtained with variable nitrogen
 abundances as free parameter, that turned out to reach even values
 around $\simeq$500 for one of the components. We obtain a better
 fit if the nitrogen abundance is different in the the three
 components, 
 but different ``combinations'' of temperature and nitrogen abundance give
 the same goodness of the fit. We also tried fits leaving free also the 
oxygen and  carbon abundances, leaving the other abundances
 at  solar values: although the fit always converged with 
 al least one of these elements enhanced in at least one  
 plasma component,  there was no clear further improvement.
The best composite fit with ``free'' nitrogen is shown in Fig. 6 for all four spectra. 
 Some emission
 lines are still unexplained, indicating a complex origin,
probably in many more regions of different temperatures and densities. 
The fits are not statistically acceptable yet, with a 
 reduced value of $\chi^2$ of 1.7 for the first spectrum, about 3
 for the second and third and 4.7 for the fourth  (we note that adopting cash statistics
 did not result in very different or clearly improved fits),  however, the figure shows that
 the continuum is well modeled and many of the emission lines are also explained.

$\bullet$ {\bf Step 4.} Since we could not model all the emission lines  
 with the collisional ionization code, the next step was to explore 
 a photoionization code. We used the PION model \citep{Mehdipour2016}
 in the spectral fitting package SPEX \citep{Kaastra1996}, having also
 a second important aim: exploring how the abundances may change the  
 continuum and absorption spectrum of the central
 source. In fact, SPEX allows to use PION also
 for the absorption spectrum. The photoionizing source is assumed to be
 a blackbody, but the resulting continuum is different from that of a simple
 blackbody model, because the absorbing layers above it remove emission at the 
 short wavelengths, especially with ionization edges (like in the static
 atmosphere).  An important difference between PION and 
 the other photoionized plasma  model in SPEX, XABS,
 previously used for nova V2491 Cyg \citep{Pinto2012}
 is that the photoionisation equilibrium is calculated self-consistently using
 available plasma routines of SPEX (in XABS instead the photoionization
 equilibrium was pre-calculated with an external code).
 The atmospheric codes like TMAP include
 the detailed microphysics of an atmosphere in non-local
 thermodynamic equilibrium, with the detailed radiative transport
 processes,  that are not calculated in PION, 
 however with PION we were able to observe how the line profile varies 
 with the wind velocity and, most important, to vary the abundances of
 the absorbing material.  This step is  
 thus an important experiment, whose results may be used to
 calculate new, ad-hoc atmospheric calculations in the future.
 The two most important elements to vary 
 in this case are nitrogen (enhanced with respect to carbon and to its
 solar value because mixing with the ashes in the burning layer) and
 oxygen (also usually enhanced with respect to the solar value in novae).  
 We limited the nitrogen abundance to a value of 100 times the solar value
 and found that this maximum value is the most suitable to explain several
 spectral features. Oxygen in this model fit is less enhanced than in all TMAP models 
 with ``enhanced'' abundances, resulting to be 13 times the solar value.  
 With such abundances, and  with depleted oxygen abundances in the intervening
 ISM (the column density of the blackbody),
 we were able to reproduce the absorption edge of nitrogen
 and to fit the continuum well. It is remarkable that this simple
 photoionization model, with the possibility of ad-hoc abundances as
 parameters, fits the continuum and many absorption lines better than 
 the TMAP atmospheric model.

 To model some of the emission lines, we had to
 include a second PION component, suggesting that the emission spectrum
 does not originate in the same region as the absorption one,
 as we suggested above.  
 Table 5 and Fig. 7 show the best fit obtained for day 197.
 Due to the better continuum fit, even if we seem to have modeled fewer emission
 lines, $\chi^2$ here was about 2 (compared to a value of 3 in Fig. 6).
 The blackbody temperature of the ionizing source is 70 keV, or about 
 812,000 K, and its luminosity turns out to be L$_{\rm bol}$=7.03 $\times 10^{37}$ erg s$^{-1}$, and by making the blackbody assumption
 we know these are only a lower and upper limit, respectively, for the
 effective temperature and bolometric luminosity of the central source
 \citep[see discussion by][]{Heise1994}.
 The X-ray luminosity of the ejecta is 10$^{36}$ erg s$^{-1}$,
 or  1.5\% of the total luminosity. 

 The value of the mass outflow rate $\dot m$ in the table is not  a free
 parameter, but it is obtained from the other free parameters of the
 layer in which the absorption features originate (region 1), following
 \citet{Pinto2012} as $$\dot m= (\Omega/4 \pi) 4 \pi  \mu m_H v_{\rm blue shift} L/\xi$$ 

 and the electron density of each layer is obtained as:
$$ n_H = (\xi/L) (N_H/f_c/\beta)^2$$

  where $m_H$ is the proton mass, $\mu$ is the mean atomic weight
 (we assumed 11.5 given the enhanced composition),
 $\Omega$ is the fraction of a sphere that
 is occupied by the outflowing material, $f_c$ is a clumpiness factor,
 and $\beta$ is a scale length. For a first order calculation, we assumed
 that $\Omega$,  $f_c$ and $\beta$  are equal to 1. 
 The small radius of the emitting blackbody is compatible
 only with a very massive WD, and/or with an emitting region that is smaller
 than the whole surface.  
 The value of $\dot m$ obtained in  the best fit is of course orders
 of magnitude lower
 than the mass loss during the early phase of a nova, but 
 the evolutionary and nova wind models predict even a complete halt to mass
 loss by the time the supersoft source emerges \citep[e.g.]{Starrfield2012, Wolf2013}.

 An interesting fact is that we were indeed able to explain several
 emission features, but not all. It is thus likely that there is a superposition
 of photoionization by the central source and ionization due to shocks in colliding 
 winds in the ejecta. However, we did not try any further composite 
 fit, because too many components were needed and we found
 that the number of free parameters
 is too large for a rigorous fit. In addition, the variability of 
some emission features
 within hours make a precise fit an almost impossible task. 

  Assuming LMC distance of about 49.6 kpc \citep{Pietrzynski2019},
the unabsorbed flux in the atmospheric models at LMC distance implies only
 X-ray  luminosity 
 close to 4 $ \times 10^{36}$ erg s$^{-1}$ on the first date, peaking close to 
 $\simeq 2.7 \times 10^{37}$ erg s$^{-1}$ on day 197 (although the  model
 for that date includes 
 a $\simeq$60\% addition to the X-ray luminosity due to a very bright plasma component
 at low temperature, 70 eV, close to the 
 the value obtained with PION for the blackbody-like
 ionizing source). We note that the X-ray luminosity with
 the high T$_{\rm eff}$ we observed, represents over 98\% of
 the bolometric luminosity of the WD. Thus, the values obtained with the fits  
  are a few times lower than the post-nova WD bolometric luminosity
 exceeding 10$^{38}$ erg s$^{-1}$ , predicted by the models \citep[e.g.][]{Yaron2005}. 
 Even if the luminosity may be higher in an expanding atmosphere \citep{vanRossum2012}.
 it will not exceed the blackbody luminosity of the PION fit, 
 7 $\times 10^{37}$ erg s$^{-1}$, which is still a factor of 2 to 3 
 lower than Eddington level.
 Although several novae have become as luminous SSS as predicted by the
 models \citep[see N SMC 2016][]{Orio2018}, in other post-novae much lower
 SSS luminosity has been measured, and 
 in several cases,  this has been attributed to 
 an undisrupted, high inclination accretion disk
 acting as a partially covering absorber \citep{Ness2015},
 although also the ejecta can cause this
 phenomenon, being opaque to the soft X-rays while having
 a low filling factor.
 In U Sco instead, the interpretation for the low observed
 and inferred flux was different. It is in fact very likely that 
 only observed Thomson scattered radiation was observed, while the central
 source was always obscured by the disk \citep{Ness2012, Orio2013}.  
 However, also the reflected radiation of
 this  nova was thought to be partially obscured
 by large clumps in the ejecta in at least one observation \citep{Ness2012}. 
  
 In the X-ray evolution of N LMC 2009, one intriguing fact is that not only
 the X-ray flux continuum, but also that the emission features
 clearly increased in strength after day 90. This seems to imply that 
 these features are either associated with, or originate very close to, the WD 
 to which we attribute the SSS continuum flux. 
\begin{figure*}
\includegraphics[width=120mm]{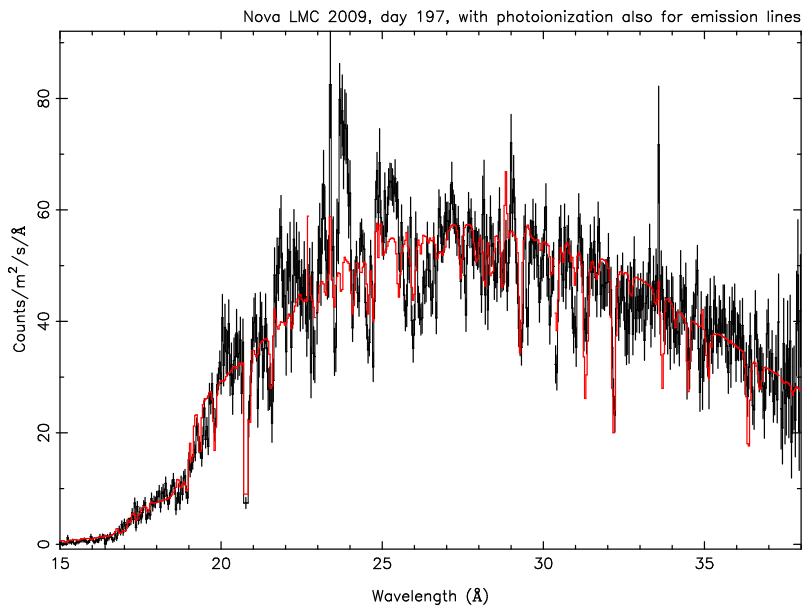}
\caption{Fit to the spectrum of day 197 with two PION regions as in Table 5.}
\end{figure*}

   $\bullet$ {\bf Step 5.} An additional experiment we did with spectral
 fitting was with the ``expanding atmosphere'' 
 ``wind-type'' WT model of \citet{vanRossum2012}, which
 predicts shallower absorption edges and may explain some
 emission features with the emission wing of atmospheric
 P-Cyg profiles. One of the parameters is the wind velocity at infinite,
 which does not translate in the observed blue shift velocity, 
 The P-Cyg profiles become ``smeared out'' and smoother as 
 the wind velocity and the mass loss rate increase.
 Examples of the comparison of the best model with
 the observed spectra, plotted with IDL and obtained by imposing the condition
 that the flux is emitted from the whole WD surface at 50 kpc distance,
 are shown for days 90 and 197 in Fig. 8 in the top panel. The models
 shown are those in the calculated
 grid that best match the spectral continuum, however, 
 that there are major differences between model and observations.
 Next, we assumed that only a quarter of the WD surface is observed,
 thus choosing models at lower luminosity, but we obtained
 only a marginally better match. The interesting facts are:
 the mass outflow rate is about the same as estimated
 with SPEX and the PION model, b) this model predicts
 a lower effective temperature for the same luminosity, and c) 
 most emission features do not seem to be possibly
associated with the ``wind atmosphere''
 as calculated in the model. It is not unrealistic to 
 assume that they are produced farther out in the ejected nebula,
 an implicit assumption in the composite model of Table 4 and Fig. 6.
\begin{figure*}
\includegraphics[width=87mm]{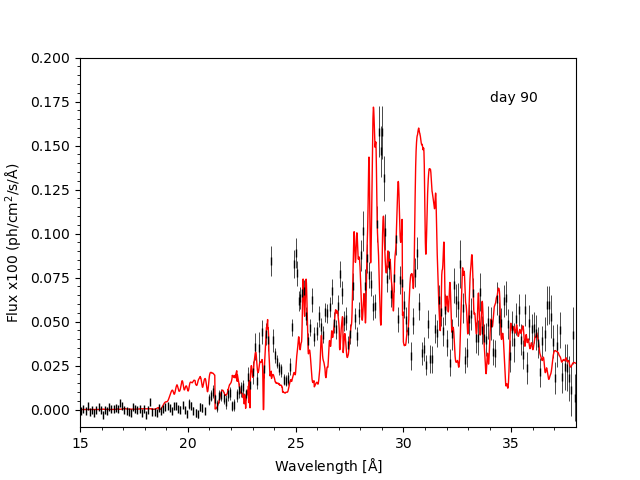}
\includegraphics[width=87mm]{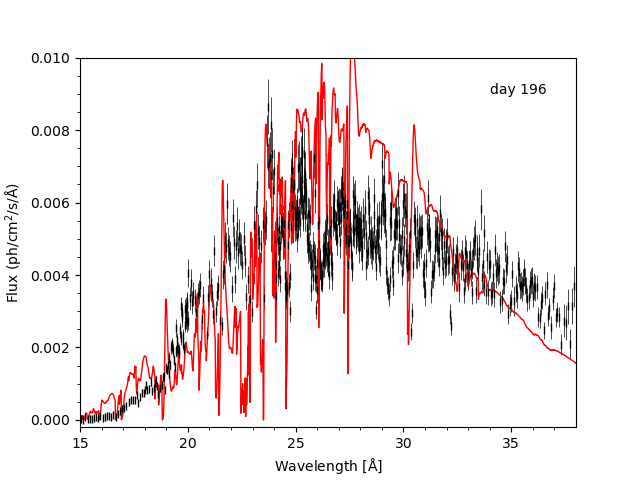}
\includegraphics[width=87mm]{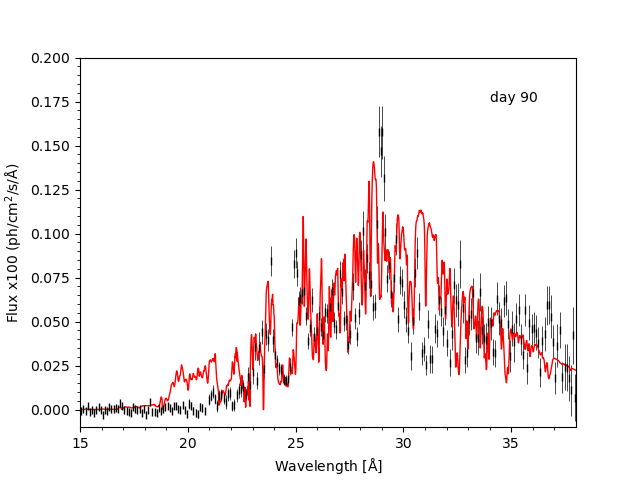}
\includegraphics[width=87mm]{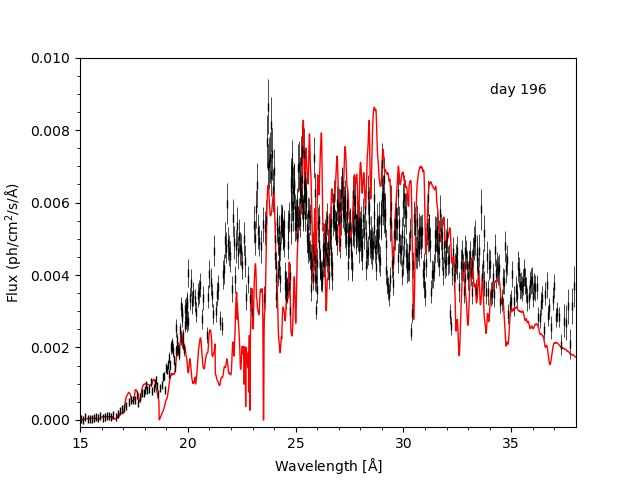}
\caption{Comparison of the ``WT model'' with the observed 
 spectra on day 90 (averaged spectrum)
 and day 197.  In the upper panels,
 the condition that the luminosity matched that observed
 at 50 kpc is imposed, assuming respectively for
 day 90 and 197: T$_{\rm eff}$=500,000 K,
 log(g$_{\rm eff}$)=7.83, N(H)=2 $\times 10^{21}$ cm$^{-2}$,
$\dot m=2 \times 10^{-8}$ M$_\odot$ yr$^{-1}$,
 v$_{\infty}$=4800 km s$^{-1}$ and
 T$_{\rm eff}$-550,000 K, log(g$_{\rm eff})$=8.30, $\dot m=10^{-7}$ M$_\odot$ yr$^{-1}$, N(H)=1.5 $\times 10^{21}$ cm$^{-2}$,
 v$_{\infty}$=2400  km s$^{-1}$. In the lower panels, the assumption was
 made that only a quarter of the WD surface is observed, and 
 the parameters are  T$_{\rm eff}$=550,000 K,
 log(g$_{\rm eff}$)=8.08, $\dot m=10^{-8}$ M$_\odot$ yr$^{-1}$, N(H)=1.8 $\times
 10^{21}$ cm$^{-2}$
 v$_{\infty}$=2400 km s$^{-1}$ and
 T$_{\rm eff}$-650,000 K, log(g$_{\rm eff})$=8.90, $\dot m= 2 \times
10^{-8}$ M$_\odot$ yr$^{-1}$, N(H)=2 $\times 10^{21}$ cm$^{-2}$,
 v$_{\infty}$=4800  km s$^{-1}$ respectively for day 90 and 197.}
\end{figure*}
\begin{figure*}
\includegraphics[width=130mm]{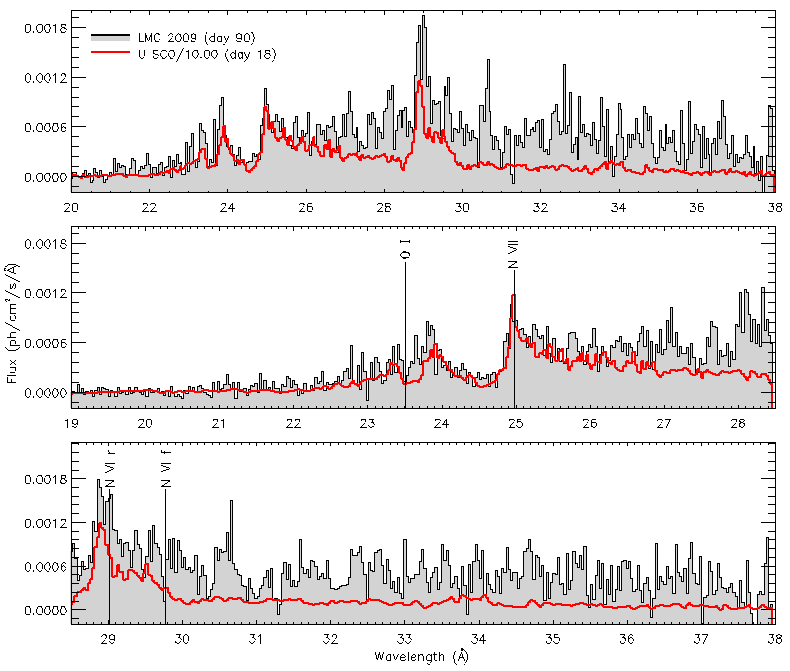}
\caption{The {\sl XMM-Newton} RGS X-ray spectrum of of Nova LMC 2009a on
 day 90 of the outburst (grey), compared in the left panel
 with the {\sl Chandra} Low Energy Transmission
 Gratings of U Sco on day 18, with the photon
 flux multiplied by a factor of 10 for N LMC 2009a (red). 
 The distance to U Sco is at least 10 kpc \citep{Schaefer2010} 
 or 5  times larger than
 to the LMC), so the absolute X-ray luminosity of N LMC 2009a
 at this stage was  about 2.5 times that of U Sco. 
 H-like and He-like resonance and forbidden
 line of nitrogen are marked, with
 a redshift 0.008 (corresponding to 2400 km s$^{-1}$, measured
 for the U Sco emission features).
 These features in U Sco were also measured
 in absorption with a  blueshift corresponding to
 2000 km  s$^{-1}$, producing an apparent
 P-Cyg profile. We also indicate the O I interstellar
 absorption at rest wavelength.
 }
\end{figure*}
\section{Comparison with the spectra of other novae}
 To date, 25 novae and supersoft X-ray sources have been observed
 with X-ray gratings, often multiple times, so  
 a comparison with other novae in the supersoft X-ray
 phase is useful and can
 be correlated with other nova parameters.
 While the last part of this paper illustrates an archival search
 in X-ray data of other MC novae, in this Section we analyse some 
 high resolution data of Galactic and Magellanic Clouds (MC) novae bearing some
 similarity with N LMC 2009.

 {\bf 1. Comparison with U Sco.}
The first spectrum (day 90) can be compared with one of 
  U Sco,   another RN that has
 approximately the same orbital period \citep[][]{Orio2013}.
 In this section all the figures illustrate the fluxed spectra.
 In U Sco the SSS emerged
 much sooner and the turn-off time was much more rapid. 
 Generally, the turn-off and turn-on time are both inversely
 proportional to the mass of the accreted envelope, and
 \citet{Bode2016} note that this short time is consistent 
 with the models' predictions for a WD mass 1.1$<$m(WD)$<1.3$ M$_\odot$,
 which is also consistent with the high effective temperature. 
 However, the models by \citet{Yaron2005}, who explored a large range
 of parameters, predict only relatively low ejection velocities
 for all RN and  there is no set of parameters
 that fits   a ivery fast RN like U Sco,
  with large ejection velocity and short decay
 times in optical and X-rays. 
 Thus, the specific characteristics of U Sco may be
 due to irregularly varying $\dot m$ and/or other peculiar conditions
 that are not examined in the work by \citet{Yaron2005}.
 The parameters of N LMC 2009, instead, 
 compared with \citet{Yaron2005}, are at least marginally
 consistent with their model of
 a very massive (1.4 M$_\odot$), initially hot WD 
 (probably recently formed) WD.

 Fig. 9 shows the comparison 
of the early {\sl Chandra} spectrum (day 18) of U Sco \citep{Orio2013},
 with the first spectrum (day 90) of N LMC2009.
 The apparent P-Cyg profile of the N VII H-like line (24.78
 \AA \ rest wavelength), and the
 N VI He-like resonance line with (28.78 \AA  \ rest
 wavelength)  is observed in N LMC 2009
 like in U Sco, with about the same redshift for the emission lines, and
 lower blueshift for the absorption.
 The definition ``pseudo P-Cyg'' in
 \citet[][]{Orio2013} was adopted, because there is 
 clear evidence that the absorption features of U Sco 
 in the first epoch of observation were in the Thomson-scattered radiation
 originally emitted near or on the WD surface. The
 emission lines originated in an outer region in the ejecta, at
 large distance from the WD.

 Apart from the similarity in the two strong nitrogen 
 features, the spectrum of N LMC 2009 at day 90 is much more
 intricate than that of U Sco, and  presents a ``forest'' of absorption
 and emission features that are not easily disentangled, also
 given the relatively poor S/N ratio.
 U Sco, unlike N LMC 2009a, never
 became much more X-ray luminous \citep[][]{Orio2013} and
 the following evolution was completely different from that of N LMC 2009,
 with strong emission lines and no more measurable absorption features.
 Only a portion of the predicted WD flux was detected in U Sco,  
and assuming a 12 kpc distance for U Sco \citep{Schaefer2010}, 
   N LMC 2009a was twice intrinsically more
 luminous already at day 90, ahead of the X-ray peak, 
 although above above 26 \AA \  the lower flux of U Sco
is due in large part to the much higher interstellar column density
 \citep[][]{Schaefer2010, Orio2013}, also apparent from the
 depth of the O I interstellar line.
 U Sco did not become more luminous in the following monitoring
 with the {\sl Swift} XRT \citep{Pagnotta2015}. 
Given the following increase in the N LMC 2009
 SSS flux, we suggest that in this nova we did not 
observe the supersoft flux only in a Thomson scattered
 corona in a  high inclination system, like in U Sco, but 
 probably there was a direct view of at
 least a large portion of the WD surface.

{\bf 2. Comparison with KT Eri.}
 A second comparison, shown in Fig. 10 and more
 relevant for the observations from day 165,
 is with unpublished archival
 observations of KT Eri, a Galactic nova with many
 aspects in common with N LMC 2009a, including
 the very short period modulation in X-rays  \citep[35 s, see][]{Beardmore2010,
 Ness2015}. Fig. 10 shows the spectrum
 of N LMC 2009 on day 165 overimposed on the KT Eri spectrum observed with
 {\sl Chandra} and the  Low Energy Transmission Grating (LETG) on day
 158 \citep{Ness2010, Orio2018}. 

The KT Eri GAIA parallax in DR2 does not have a large error and translates into a 
 distance of 3.69$^{+0.53}_{-0.33}$ kpc \footnote{From the GAIA database using ARI's
Gaia Services, see  \url{https://www2.mpia-hd.mpg.de/homes/calj/gdr2_distances/gdr2_distances.pdf}\label{fn:gaia}}.
 The total integrated absorbed flux in the RGS band was
 10$^{-8}$ erg s$^{-1}$ cm$^{-2}$, and the absorption 
 was low, comparable to that of the LMC (Pei et al. 2021)
implying that  
 N LMC 2009 was about 2.3 times more X-ray luminous than KT Eri,
 although this difference may be due  to the daily variability
 observed in both novae.
 \citet[][]{Bode2016} highlighted, among other similarities, a 
 similar X-ray light curve. However, there are some significant
 differences in the {\it Swift} XRT 
 data: the rise to maximum supersoft X-ray 
 luminosity was only about 60 days for KT Eri, and the decay started
 around day 180 from the outburst, shortly after the spectrum we
 show here. The theory predicts that
 the rise of the SSS is inversely proportional
 to the ejecta mass. The duration of the SSS is also
  a function of the ejected mass, which tends to be
 inversely proportional to both the WD mass and the mass accretion 
 rate before the outburst \citep[e.g.][]{Wolf2013}. 
 The values of T$_{\rm eff}$ for N LMC 2009a indicate a high
 mass WD and the time for accreting the envelope
 cannot have been long, with a low mass accretion rate,
 since it is a RN. The other factor contributing to accreting
 higher envelope mass is the WD effective temperature at the first
 onset of accretion \citep[][]{Yaron2005}. This would mean that the onset of
 accretion was recent in N LMC 2009a and its WD had a longer
 time to cool than KT Eri before accretion started. 

KT Eri was observed earlier after the outburst with
 {\sl Chandra} and the LETG, showing a much less luminous source
 on day 71 after the outburst,
 then a clear increase in luminosity on days 79 and following dates.
 The LETG spectrum of KT Eri appeared dominated by a WD continuum with
 strong absorption lines from the beginning, and in no observation did
 it resemble that of N LMC 2009a on day 90.
 Unfortunately, because of technical and visibility constraints, 
 the KT Eri observations ended at an earlier post-outburst phase 
 than the N LMC 2009a ones. However, some comparisons are possible. 
 Fig. 10 shows the comparison
 between the KT Eri spectrum of day 158 and the N LMC
 2009 spectrum of day 165. We did not observe flux shortwards
 of an absorption edge of O VII at 18.6718 \AA\ for KT Eri,
 but there is residual flux above this edge for N LMC 2009a,
 probably indicating a hotter source.
 We indicate the features assuming a blueshift velocity of 1400 km s$^{-1}$,
 more appropriate for KT Eri,
while the N LMC 2009 blueshift velocity is on average quite lower.
The figure also evidentiates that in N LMC 2009
 there may be an emission core superimposed on blueshifted absorption
 for the O VII triplet recombination He-like line
 with rest wavelength 21.6 \AA.

 N LMC 2009 does not share with KT Eri, or with other
 novae \citep[see]{Rauch2010, Orio2018}  a very deep
 absorption feature of N VIII at 24.78 \AA, which was almost saturated in
 KT Eri on days 78-84. It appears like a P Cyg profile 
 as shown in Fig. 4, and the absorption has   
 lower velocity of few hundred km s$^{-1}$.
 Despite many similarities in the two spectra, altogether
 the difference in the absorption lines
 depth is remarkable. The effective temperature upper limit derived by Pei et al.
 (2021 preprint,
 private communication) for KT Eri is 800,000 K, but the difference between
 the two novae is so large that it seems
 to be not only due to a hotter atmosphere (implying a  higher ionization parameter):
 we suggest that abundances and/or density must also be playing a role.   

{\bf 3. Comparison with Nova SMC 2016.}
 In Fig. 11 we show the comparison of the N LMC 2009 day 197
 spectrum with  the last high resolution X-ray
 spectrum, obtained on
 day 88, for N SMC 2016 \citep{aydi2018, Orio2018}.
 The evolution of N SMC 2016 was much more rapid, the 
 measured supersoft X-ray flux was larger by more than an order of
 magnitude, and
 the unabsorbed absolute luminosity was larger 
 by a factor of almost 30  \citep[see Table 2, and][]{Orio2018}. 
 The lines of N VII  (rest wavelength 24.78 \AA),  N VI recombination
 (28.79 \AA), and C VI (33.7342 \AA), 
 seem to be in emission for N LMC 2009a, but
 they are instead observed in absorption for N SMC 2016.
 The absorption features
 of N SMC 2016 are also significantly more blueshifted, by about 2000 km s$^{-1}$ in
 N SMC 2016. 
 The effective temperature in this nova was estimated by \citet{Orio2018} as 900,000 K 
 in the spectrum shown in the figure. The comparison shows that N LMC 2009 was
 probbaly hotter and/or had a denser photoionized plasma.
\begin{figure*}
\includegraphics[width=130mm]{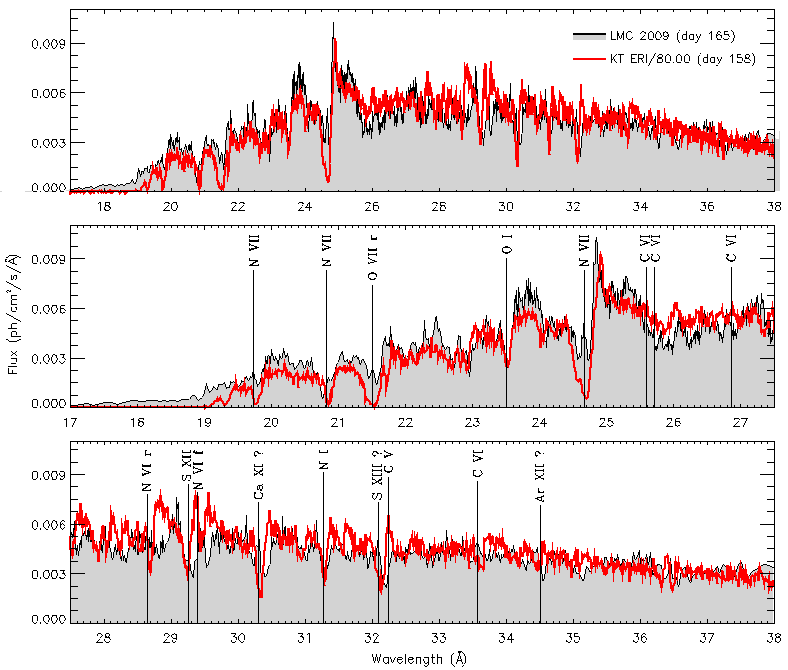}
\caption{Comparison  the spectrum
 of N LMC 2009 on day 165 with the spectrum of KT Eri on day 158,
 with the photon flux of N LMC 2009 divided 
 by a factor of 80. In the panel on the left we indicate 
 the lines position with a blueshift by 1410 km s$^{-1}$,
 which is a good fit for several absorption lines of KT Eri
 predicted by the atmospheric models
 (except for O I and N I, which are local ISM lines, observed at rest wavelength).
}
\end{figure*}

 {\bf 4.} Finally, another significant comparison can be made with
{\bf N LMC 2012} \citep[for which only one exposure was
 obtained , see][]{Schwarz2015}, 
 the only other MC nova for which an X-ray high
 resolution spectrum is available. 
  Despite the comparable count rate,
 the spectrum of N LMC 2012 shown by \citet{Schwarz2015}
 is completely different from those of N LMC 2009. It
 has a much harder portion of high continuum, that is not present in
 other novae, and the flux appears to be cut only by
 the O VIII absorption edge at 14.228 \AA.   
 We only note here that the continuum X-ray
 spectrum in this nova appears to
 span a much larger range than in most other novae, including N LMC 2009a. 
 We did attempt a preliminary
 fit with WD atmospheric models for this nova, and suggest 
 the possible presence in the spectrum
 of two separate zones at different temperature,
 possibly due to magnetic accretion onto polar caps,
 like suggested to explain the spectrum of V407 Lup \citep{aydi2018b}. 
\begin{figure*}
\includegraphics[width=130mm]{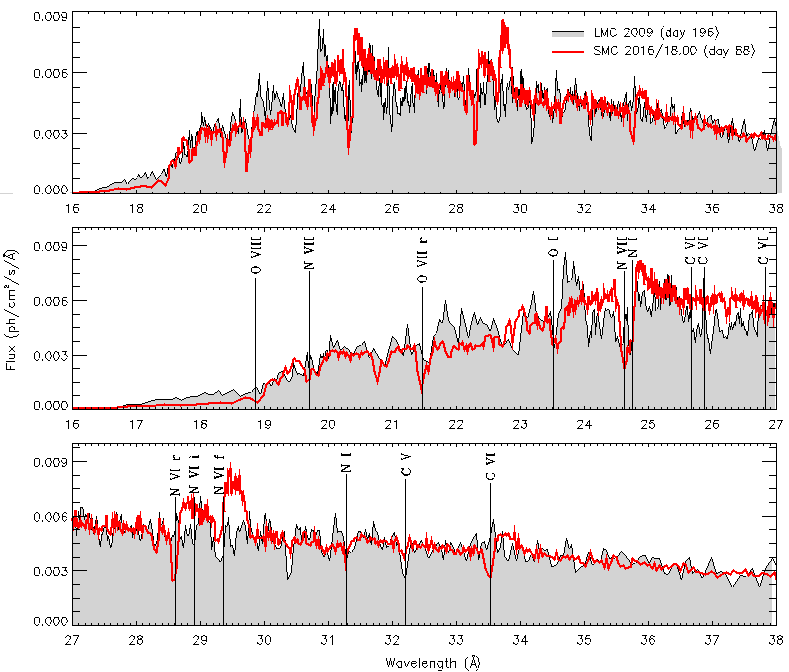}
\caption{Comparison of the spectrum 
 of N LMC 2009 on day 165 with the spectrum of N SMC 2016 on day 88,
 with the photon flux of N LMC 2009 multiplied by a factor of 18 for the
 comparison. 
}
\end{figure*}
\section{Timing analysis: revisiting the 33 s period}
\citet{Ness2015} performed a basic timing analysis for the  {\sl XMM-Newton}
 observations, detecting a significant modulation
 with a period around 33.3 s. The period  may have
 changed by a small amount between the dates of the observations.
 Here we explore this modulation more in detail,
 also with the aid of light curves' simulations.
 Such  short period modulations have been detected in other novae \citep[see][]{Ness2015,
Page2020} and non-nova SSS \citep{Trudo2008, Odendaal2014}. 
 \citet{Ness2015} attributed the modulations to non-radial
g-mode oscillations caused by the burning that induces gravity waves in the envelope 
(so called $\epsilon$ 
mechanism", but recently, detailed models seem to rule it out because the 
typical periods would
not exceed 10 s \citep{Wolf2018}.
 Other mechanisms invoked to explain the root cause of the
pulsations are connected with the WD rotation. If the WD accretes mass at high rate, the WD
may be spun to high rotation periods. In CAL 83 the oscillations have been attributed
 by \citet{Odendaal2017}
to ``dwarf nova oscillations'' (DNO) in an extreme ``low-inertia magnetic accretor''.
\subsection{Periodograms}
We compared periodograms of different instruments and exposures.
 Because the low count rate of the RGS resulted in low quality periodograms
 or absent signal, we focused on the EPIC cameras, despite some pile-up,
which effects especially the pn. 

\begin{figure}
\includegraphics[width=150mm,angle=-90]{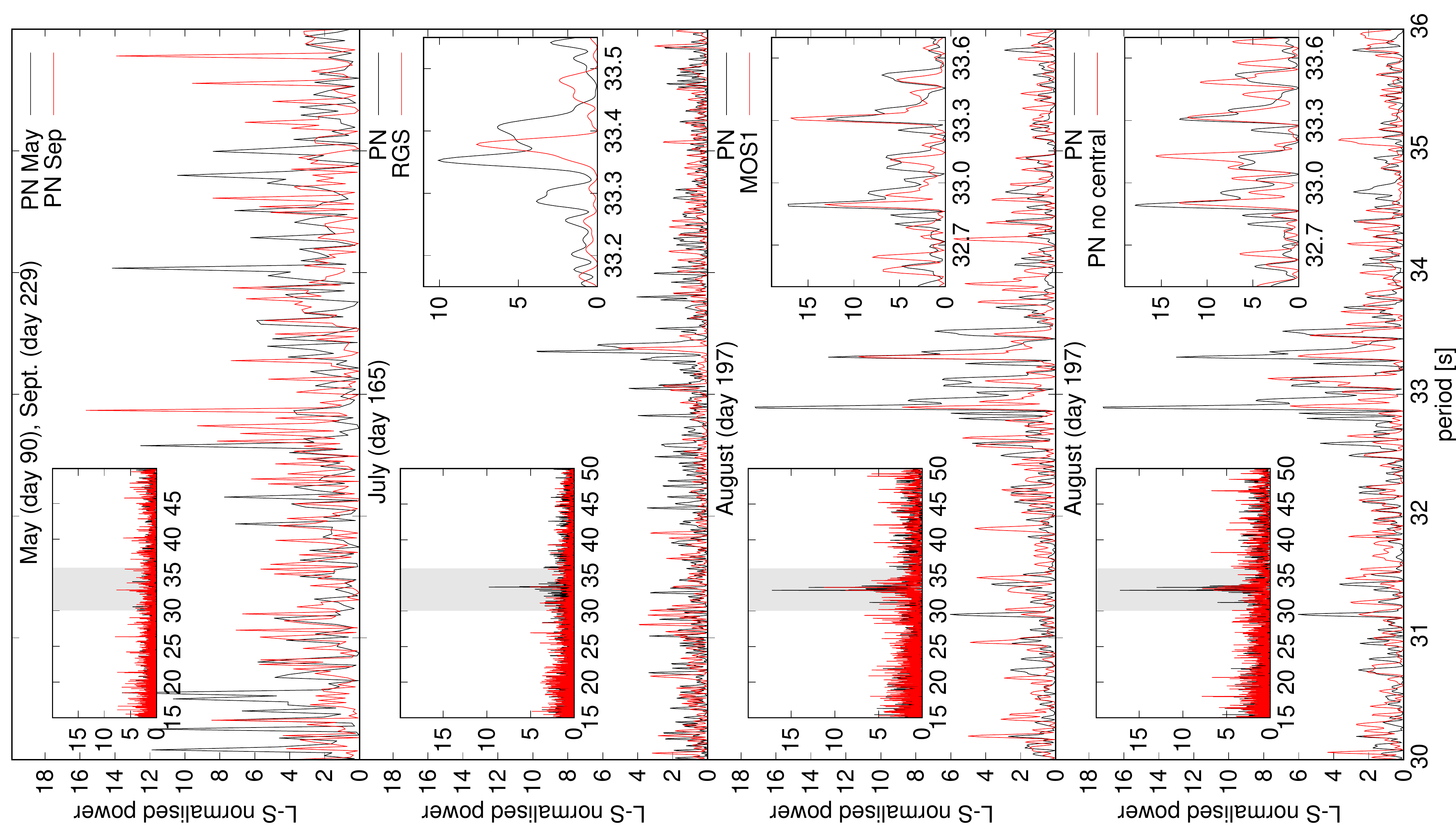}
\caption{Comparison of periodograms of different instruments and 
observations. 
The pile-up corrected pn  data (pn no central) were extracted after
 subtracting the central, most piled-up region of the source. 
The insets on the left
 show larger period intervals, and
 the shaded areas are the intervals plotted 
 in the main panels.
 The insets on the right show instead a narrower
 period interval than  the main panels, 
 with the August 2009 periodogram plotted
 in red.}
\label{pds_diff_instr}
\end{figure}

In Fig.~\ref{pds_diff_instr} we show and compare all the calculated
 periodograms. Even though the reading 
 times were uniform, we used the Lomb-Scargle (\citealt{scargle1982}) method,
 because it allows to better resolve
 the period than the Fourier transform, due to oversampling.
 We show a larger time interval in the left insets, while the most
 important features are shown in the main panels.
 Table~\ref{periods} summarizes the most significant periodicities with
 errors estimated from the half width at half maximum of the corresponding peaks.
\begin{table*}
\caption{Detected periodicities, in seconds, corresponding
 to the strongest signals
 shown  in Fig.~\ref{pds_diff_instr}. 
We include a possible periodicity in the September data of day 229,
 although the corresponding peak is not by any means as dominant 
as on days 165 and 197. 
In each row we report periods that are consistent 
 with each other within the statistical error.}
\begin{center}
\begin{tabular}{lcccccc}
\hline
\hline
label & pn & RGS & pn & pn no-pile-up & MOS1 & pn \\
& Day 164.9 & Day 164.9 & Day 196.17 & Day 196.17 & Day 196.17 & Day 228.93 \\
\hline
$P_1$ & -- & -- & $32.89 \pm 0.02$ & $32.90 \pm 0.02$ & $32.89 \pm 0.02$ & $32.89 \pm 0.01$\\
$P_2$ & -- & -- & -- & $33.13 \pm 0.02$ & -- & --\\
$P_3$ & $33.35 \pm 0.02$ & -- & $33.31 \pm 0.02$ & $33.32 \pm 0.02$ & $33.31 \pm 0.02$ & --\\
$P_4$ & -- & $33.38 \pm 0.01$ & -- & -- & -- & --\\
$P_5$ & $33.41 \pm 0.01$ & -- & -- & -- & -- & --\\
\hline
\end{tabular}
\end{center}
\label{periods}
\end{table*}
The periodograms for the light curves measured on days 90 and
 229 do not
 show an obviously dominant signal in the other observations
 (although, as Table 3 shows, we retrieved the period in
 the pn light curve on day 229, with lower significance),
so we did not investigate these data further. 
 It is very likely that the modulation started only
 in the plateau phase of the SSS and ceased around
 the time the final cooling started.
For the data of days 165 and 197 we found dominant peaks, especially 
 strong on day 197.
 The dominant signal is absent in the MOS 
 light curve of day 165, but was retrieved instead in the RGS data
 of the same date.
 Both periodograms suggest a single signal, with a double structure in the pn,
 that however may be an artifact. 
 Moreover, the periodicity measured
 with the RGS data on day 165 is the average of the two values
 measured with the pn on the same day. This may
 indicate that the amplitude in the pn light curve is not stable, generating
 a false beat like in V4743\,Sgr (\citealt{dobrotka2017}). The
beat causes
 us to measure a splitting of the peak, while the real 
signal is in between (so, it may be $P_4$  detected in the RGS data).

The pattern is more complex on day 197. The periodicity $P_3$ detected in July
 is retrieved in all the August light curves, suggesting that this feature is 
 real and stable. Moreover, the non pile-up
 corrected and the pile-up corrected pn periodograms, and the MOS-1 
 periodograms, for day 197 
 show the same patterns, implying that pile-up in the pn data does not
 strongly effect the timing results. We were interested in exploring
 whether the non pile-up corrected light curves,
 which have higher S/N,  also give reliable results.
 For the day 197 observation we compared the results using the
 pile-up corrected pn light curve, to
 those obtained for the MOS-1, 
 and found that the derived periodicities ($P_1$ and $P_3$) agree
 within the errors. Worth noting is the dominant peak $P_2$ in the non pile-up
 corrected pn data, measured 
also in the pile-up corrected light curve, but with rather low significance. 
 We note that also many of the low peaks and faint features are present 
in both periodograms, but with different power. This confirms that
 pile-up does not affect the signal detection, although 
 we cannot rule out that it affects its significance.
\subsection{Is the period variable?}
\begin{figure}
\resizebox{\hsize}{!}{\includegraphics[angle=-90]{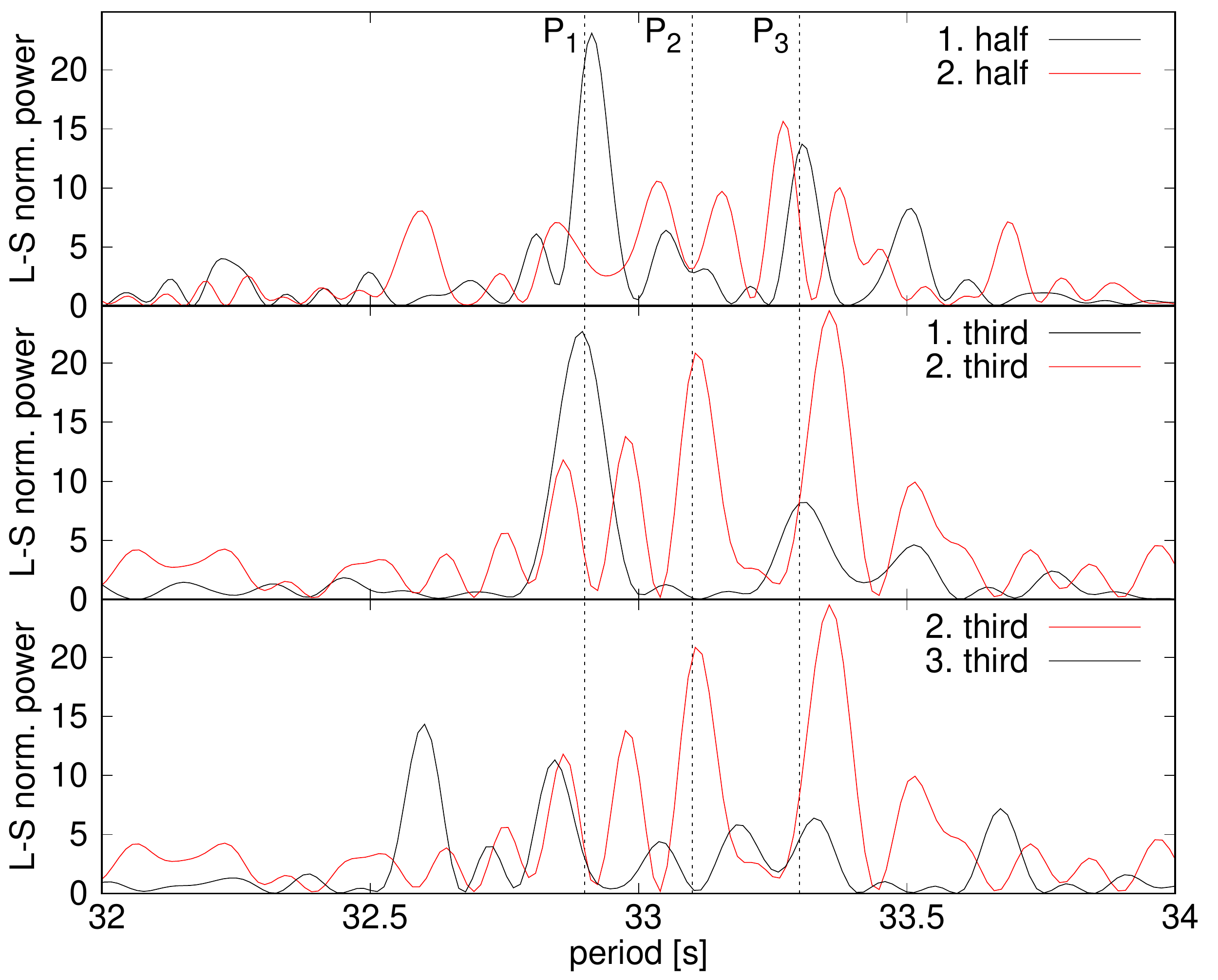}}
\caption{Periodograms for day 197, 
divided into halves and thirds. The vertical dashed lines indicate the periods
 $P_1$, $P_2$ and $P_3$ obtained by analyzing the whole exposure at once.}
\label{pds_aug_halves}
\end{figure}

The complex pattern discovered in the
 light curve of day 197 suggests that the period may
 have varied during the exposure. We 
 already mentioned the variable amplitude on day 165, and reminded that
 variable amplitude was discovered  
  in the case of V4743\,Sgr (\citealt{dobrotka2017}). In Appendix 1,
 we show how we used simulations to investigate the possibility
 of variable periodicity and amplitude. The conclusion
 that can be derived is that models with variable periodicity  
 match the data better than those with a constant period.
 The lingering question is whether
 different periodicities occur simultaneously, or whether instead the period
 of the modulation changes on short time scales, in the course of 
 each exposure. We reasoned that 
if the period varied during the exposures, 
 this must become evident by splitting the 
 original exposure in shorter intervals.  Therefore, we 
 experimented by dividing the
 day 197 light curve into two and three equally long segments.
 The corresponding periodograms are depicted in Fig.~\ref{pds_aug_halves}.
 Dividing the light curve into halves, shows that the first half comprises both dominant signals $P_1$ and $P_3$, while  
the second half is dominated only by $P_3$. 
 We tried a further subdivision in three parts, and found
 that the first interval is dominated by $P_1$, 
the second by $P_2$ and $P_3$ (with a slight offset), 
 and in the third portion, none of the $P_1$, $P_2$ and $P_3$ periods
 can be retrieved. This indicates that the period of the modulation
 was most likely variable during the exposure. This is probably the reason for which
 different periods of similar length were measured in the same
 periodogram by analyzing the light
 curve of the whole exposure time. We suggest
 that these periods did not coexist,
 but instead a single modulation occurred, whose
 slightly changed during time intervals  of minutes to hours.
  While \citet{Ness2015} found that the short-term periodicity may
 be a recurrent and transient phenomenon, here instead our interpretation is that the
 variations in amplitude and length of the period sometimes make
 it undetectable, but most likely it is always present.
	 \section{Discussion: N LMC 2009a}
Nova LMC 2009 is only one of three novae in the Clouds that could
 be observed with high resolution X-ray spectroscopy.
The others were N LMC 2012 \citep{Schwarz2015} and N SMC 2016 \citep{Orio2018}.
 All three novae are luminous and reached a range of effective
 temperature that can only be explained with the presence of a massive
 WD, close to the Chandrasekhar limit if we compare 
 the data with models by \citet{Yaron2005}. However, a word
 of warning should be given concerning the fact the models
 do not explain all the RN characteristics in a very
 consistent way. In fact, the ejection velocity
 inferred from the optical spectra exceeds the value predicted 
 by the models for a RN with a recurrence period of 38 years or
 shorter: this difficulty of modeling RN is a known problem. 
 A possible explanation is that
 $\dot m$ has been variable over the secular evolution of these novae: 
 if the recurrence time was longer, and $\dot m$ was
 lower in previous epochs, the closely spaced nova
 outbursts have only started very recently and
 the material in the burning layer may still be colder and more degenerate
 than it would be after many outburst with a very short recurrence time,
 thus causing a larger ejection velocity than in the models, which assume 
 that the accreted mass is accumulated on a hotter surface.

 N LMC 2009a was not as X-ray luminous as N LMC 2012 and N SMC 2016. 
 N SMC 2016 also remained a much more luminous SSS for many months.
 Like these and other novae observed with the gratings,
 LMC 2009a showed a hot continuum, compatible with a peak
 effective temperature of almost a million K, predicted by the models
 for a WD mass m(WD)$\geq$1.3 M$_\odot$. However, the absolute X-ray luminosity, 
 estimated by fitting the spectrum in a phase when it
 constitutes over 98\% of the bolometric luminosity, in N LMC 2009 is 
 only  a portion of the predicted
 Eddington luminosity. Because the X-ray flux of this nova was irregularly variable
 during all observations  over time scales of hours, our
 interpretation is that the filling factor in the outflow of the ejecta 
 varied, and was subject to instabilities even quite close to the  
 WD surface, never becoming completely
 optically thin to X-rays during the SSS phase.
 We suggest that
 the WD was observed through a large ``hole'' (or several ``holes'')
 of optically thin material that changed in size as the ejecta expanded
 clumped due to instabilities in the outflow,
 became shocked, and evolved. This is likely to have happened if there
 if the outflow, even at late epoch after maximum, was  not
 a continuous and smooth phenomenon. We note that  \citet{Aydi2020} explain
 the optical spectra of  novae as due to distinct outflow episodes.
 Also several emission and absorption features of this nova were not stable in
 the different exposures, varying significantly over timescales 
 of hours, but mostly without a clear correlation with the continuum level.

   It is remarkable that the nova does not show all the characteristic
 deep and broad absorption features of oxygen, nitrogen and carbon observed
 in other novae and attributed to the WD atmosphere. Some of these features 
   in this nova may be in emission and redshifted, 
 but an attempt to fit the spectra with a ``wind-atmosphere'' model
 by \citet{vanRossum2012} did not yield a result.   
 Perhaps the emission cores originate in the ejecta and are
not to be linked with the WD, as is the case in other novae.
 Although we identified and measured several absorption and
 emission features, we came to the conclusions that there are
 overlapping line systems produced in different regions and
 with different velocity, most of them originating
 in the ejected shell through
 which we observed only a portion of the WD luminous surface.

  Finally, the short period modulation observed in N LMC 2009 is
 intriguing, because it varies in amplitude and in period length,
 over time scales of few hours. The value obtained for the length of the period 
 is not compatible with against a non-radial g-mode oscillation due to the ``$\epsilon$''
 mechanism during nuclear burning, which is expected
 to have  shorter periods \citep{Wolf2018}. The non-stability of the period
 seems to rule out that it is due to the rotation of a WD that has been spun-up
 by accretion. Yet the very similar short term modulations
 observed in three other novae in the SSS phase and in CAL 83,
 a non-nova SSS, \citep[see][]{Ness2015}, suggests that the root cause
 has to do with the basic physics of nuclear burning WDs.  

\section{Archival X-ray exposures of other novae in the Magellanic Clouds} 
 Table 7 shows details of the pointed and serendipitous X-ray observations
 of the MC novae in outburst in the last 20 years.
 We observe only about two novae a year in the LMC and one every two years in the SMC \citep[see, among others,][]{Mroz2016},
 but thanks to the known distance, the low column density along the line of sight and
 the proximity the Magellanic Clouds provide very useful constraints to study the 
 SSS phase of novae. It is
 expected that in some cases even the emission of the ejecta, before the
 onset of the SSS, can
 be detected, albeit with low S/N. Since the advent of {\sl Chandra} and {\sl XMM-Newton}, 
 three novae, including N LMC 2009a (2009-02) described here, 
 have been observed with the X-ray gratings in high
 spectral resolution. However, the X-ray luminosity of novae in the Magellanic
 Clouds  appears to vary greatly, as Table 7 shows. 

 Recent {\sl Swift} observations of N LMC 2017-11a carried on 
 for 11 months did not yield any X-ray flux detection \citep{Aydi2019}. 
 N SMC 2019 was monitored
 for a few weeks and only a weak  X-ray source emerged after 4 months. The recurrent
 nova in outburst in the LMC in 2016 and in 2020 (previously known as 1968-12a and
1990b) was well followed with the XRT, but it was not X-ray luminous
 enough for high resolution spectroscopy \citep{Kuin2020, Page2020}.
 Two other X-ray-detected novae were not 
 luminous enough for the gratings:  N LMC 2000 \citep{Greiner2003}
 and N LMC 1995 \citep{Orio2003}. The latter had a long SSS phase between 5 and
 8  years. In fact, it was observed again by us in 2008 with {\sl XMM-Newton},
 and no longer detected. 

 {\sl XMM-Newton} and {\sl Chandra} typically offer much long exposures (several hours
 versus less than few tens of minutes for  
{\sl Swift} XRT). We explored the archival serendipitous X-ray observations 
 for MC novae of the last 20 years, to search for other possible detections. 
 This work was part of Sou Her senior thesis project in Wisconsin in 2016.
 Table 7  shows that we found 4 previously unknown serendipitous detections
 in {\sl XMM-Newton} deep exposures, among 13 novae that
 were observed with these satellites (only two were 
 serendipitously observed with {\sl Chandra}, and six more with
 with {\sl Swift} with shallow upper limits). 
 The upper limits on the luminosity were at least about 10$^{35}$ erg s$^{-1}$ for 
 the {XMM-Newton} exposures in Table 7. Observations of novae more than 10 years after
 the outburst were available in three cases: N LMC 2001-08 (11 and 16 years), N SMC 2002-10
 (10 and 15 years), N SMC 2005-08 (11 years) and yielded no detections. The most 
 advanced epoch for which we retrieved an X-ray detection was only 5 years and 3 months after
 maximum. 

 The measured count rates of the serendipitous targets are given in Table 7,
 and the broad band spectra with spectral fits with TMAP, yielding a reduced $\chi^2$ value of
 1.3 for N LMC 2008 and of about 1 for the other three novae, are shown in
Table 7 and Fig. 14.
 The SSS flux of the newly detected novae is much lower than predicted
 by the models and observed in the novae measured with the X-ray gratings.
 Even if the absolute luminosity value may vary significantly
 within the 90\% probability range and is poorly constrained, 
 we can rule out that these novae were near Eddington luminosity. 
 Rather than having observed only the final decline, which is predicted to 
 last for only a few weeks (Prialnik, private communication) and was indeed observed
 to take few weeks in RS Oph \citep{Nelson2008}, we suggest that we
 detected novae in which the WD was not fully visible and only a small
 region of the surface was observed, either because of obscuration
 by the ejecta, or by the accretion disk. Given the elapsed
post-outburst time, longer than a year, the latter is much
more likely. The value of the column density N(H) inferred in the spectral fits and shown in
 Table 7 is in fact consisting with no significant intrinsic absorption.
 With the higher column density in the Galaxy, such partially
 visible SSS may not be sufficiently luminous for detection: this
 is an important factor to take into account when constraining
 nova parameters on the basis of the SSS detection, duration and behavior.
\begin{table*}
\centering
\caption{Pointed and serendipitous observations of MC novae
 in outburst, satellite used,
 time after outburst in days (d), months (m) or years (yr),
number of exposures, X-ray detection (yes or no), whether the nova was a RN,
 and publications. The novae are identified by year and month of outburst, and the
 names of the (only) serendipitously observed ones are in boldface. 
The whole row is in boldface 
 if there was a detection.}
\begin{tabular}{lcccccccc}
\hline \\
 Name  &   Satellite & pointed? & when  & how many & X-ray on? &  RN?   & reference\\
       &             &          &       & times    &           &        &          \\
 \hline \\
 LMC 2020-07  & Swift   &  yes  & 3d-6m & $\simeq$80 & yes & 1968, 1990 & \cite{Page2020} \\
              &         &       &       &            &     & 2016,2020  & \\
     2016 outb.   & Swift   & yes   & 7d-11m  &  $\simeq$80     & yes & & \cite{Kuin2020} \\
 SMC 2019-7 &  Swift   & yes  & 10d-4m & 5 & yes(@4m) &  & \\
 LMC 2018-5  & eRosita & survey & 21m &  several  & yes  &  &  \cite{Ducci2020} \\
             &  Swift  & yes    & 22m-25m  &  16   &  yes &  &   \\
 LMC 2018-02 & Swift  & yes & 9d-5m    &  50   &  yes & yes, 1996  & \cite{Page2018} \\
 LMC 2017-11 & Swift &  yes & 14d-1yr   &  46   &  no  &            &  \cite{Bahramian2018} \\
 SMC 2016-10 & Swift &  yes &   6d-1yr & 111  &  yes  & perhaps     & \cite{Orio2018} \\
             &  XMM-Newton  &  yes &  1    & 75d       &       &          &       " \\
            &  Chandra  &  yes &  2     & 39d,88d       &       &          &       " \\
 LMC 2016-04 &  Swift   & yes  &  14d-48d &   29   &   yes & probably &  \\
 LMC 2015-03 & Swift    &  yes &  few days  &  3   &    no  &          &   \\
 LMC 2012-11  & Swift & yes & 1d         & 1      &    no   &          &  \\
 LMC 2012-10  & Swift & yes & $\simeq$7d &   1    &   no    &   &  \\
 {\bf SMC 2012-09}  & Swift & no  & $\geq$13m     &  several  &  no &   & \\
              & Chandra & no & 2,14m      & 2         &        &     &   & \\
 SMC 2012-03 & Swift &  yes  &  4m-14m  &   85  & yes  &   & \cite{Schwarz2012}    \\
             &       &       &            &       &      &   & \cite{Page2013a, Page2013b}  \\
 LMC 2012-03 & Swift & yes   &  1d-21 m &  72  &   yes &    & \cite{Schwarz2015} \\
             & Chandra & yes &            &     &   yes  &  &    "   \\
 {\bf SMC 2011-11} & {\bf Maxi}  & no  & 0d? & 1 & yes  &  & \cite{Li2012, 
Morii2013}         \\
                   & Swift       & yes &     &   & yes  &  &          \\ 
 {\bf LMC 2011-08} & Swift & no   & 1yr  & 1    & no &  & \\
 LMC 2009-05 & Swift & yes & 3yr  &  4 &  no  &   &  \\
 LMC 2009-02 &  Swift    &    yes   & 9d-1yr    & 82   & yes  & perhaps  & \cite{Bode2016} \\
             &  XMM-Newton & yes    &  90-229d  &  4   & yes  &          & this paper \\      
 {\bf SMC 2008-10} & {\bf XMM-Newton}  &  {\bf no} & {\bf 1yr}  & {\bf 1} & {\bf yes}  &     &        \\
             & Swift       &  no & 4-10yr & several &   no   &     &        \\
 SMC 2006-08 & XMM-Newton &  no  & $>$3yr & 3     & no  &  & \\
 {\bf LMC 2005-11} & Swift      & yes     & 1m, 3y, 4y & 7    &  no &   & \\
             & XMM-Newton & no &     5yr,7yr & 7      & no     &  & \\
             & Swift      & no &  $\geq$7yr & several & no & & \\
 {\bf LMC 2005-09} & {\bf XMM-Newton} & {\bf no} & {\bf 21m} & {\bf 1} & {\bf yes} &  &  &  \\
             & Swift, XMM & no & 3-13yr   & several & no & & \\
 {\bf SMC 2005-08} & XMM-Newton & y  &  6m      &  1      & no & & \\
             & Swift & no  & 1yr-11yr       &  7      &  no  &   & \\
             & XMM-Newton  & no & 4yr-11yr  &  8      & no   &   & \\
 {\bf LMC 2004-10} & {\bf XMM-Newton} & {\bf no} &  {\bf 2.1yr, 3.9yr}  &{\bf 1} & {\bf yes} &  &  \\
 {\bf SMC 2004-06} & {\bf XMM-Newton}  & {\bf no}  & {\bf 5.25yr} & {\bf 1}   & {\bf yes} & & \\
 {\bf LMC 2003-06} & XMM-Newton  & no & $>$10yr & 4    & no  &   &  \\
 {\bf SMC 2002-10} & XMM-Newton & no  & 10yr, 15yr & 2     &   no &  &  \\
{\bf SMC 2001-10}  & XMM-Newton & no & 4.5yr, 6yr, 8yr  & 3  &  no &  &  \\
 {\bf LMC 2001-08}  & Swift     &  no   &  5-10yr  & several  &   no &  &  \\
              & XMM-Newton & no & 11yr,16yr & 2 & no & & \\
 \hline
\end{tabular}
\end{table*}
\begin{figure*}
\includegraphics[width=85mm]{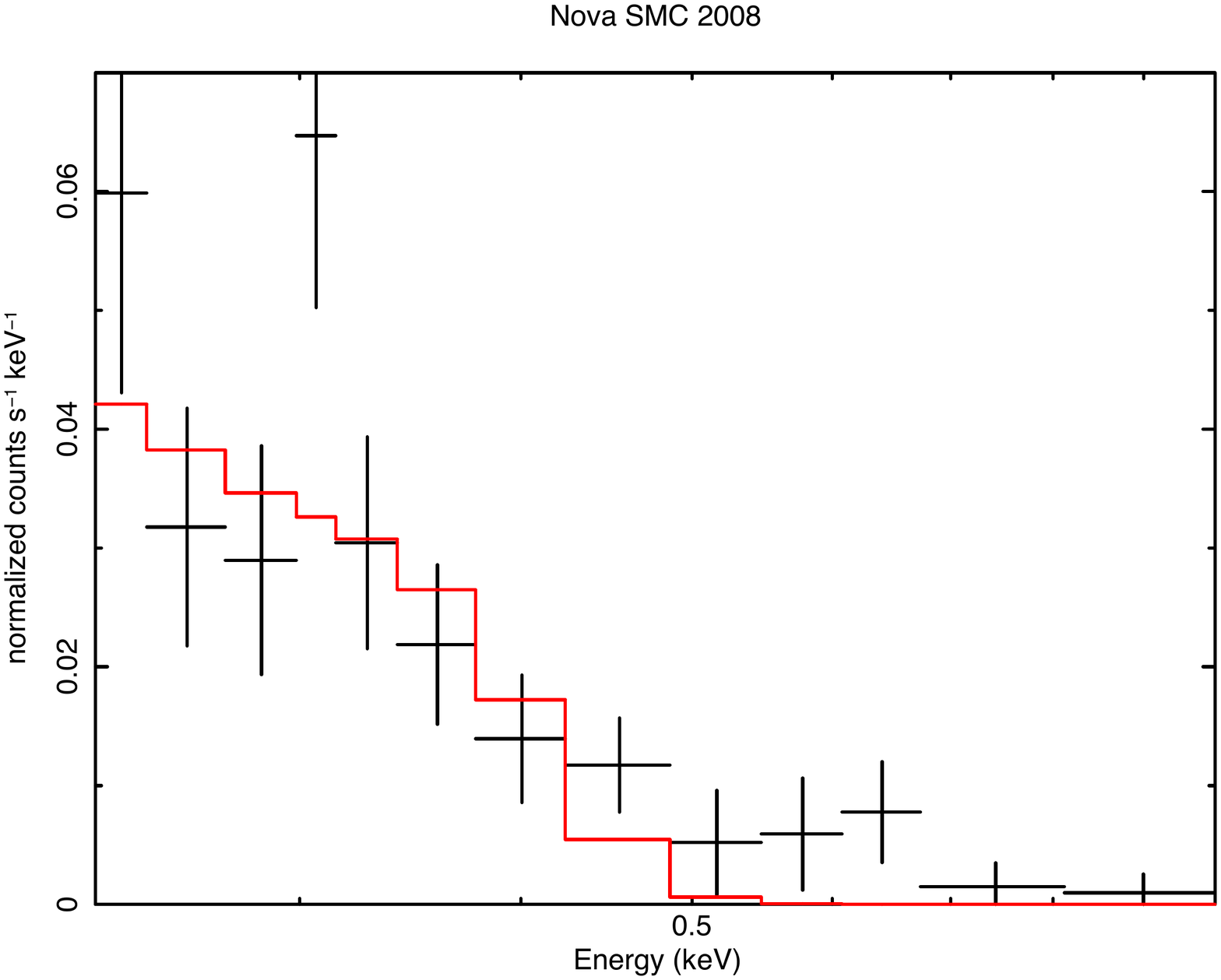}
\includegraphics[width=85mm]{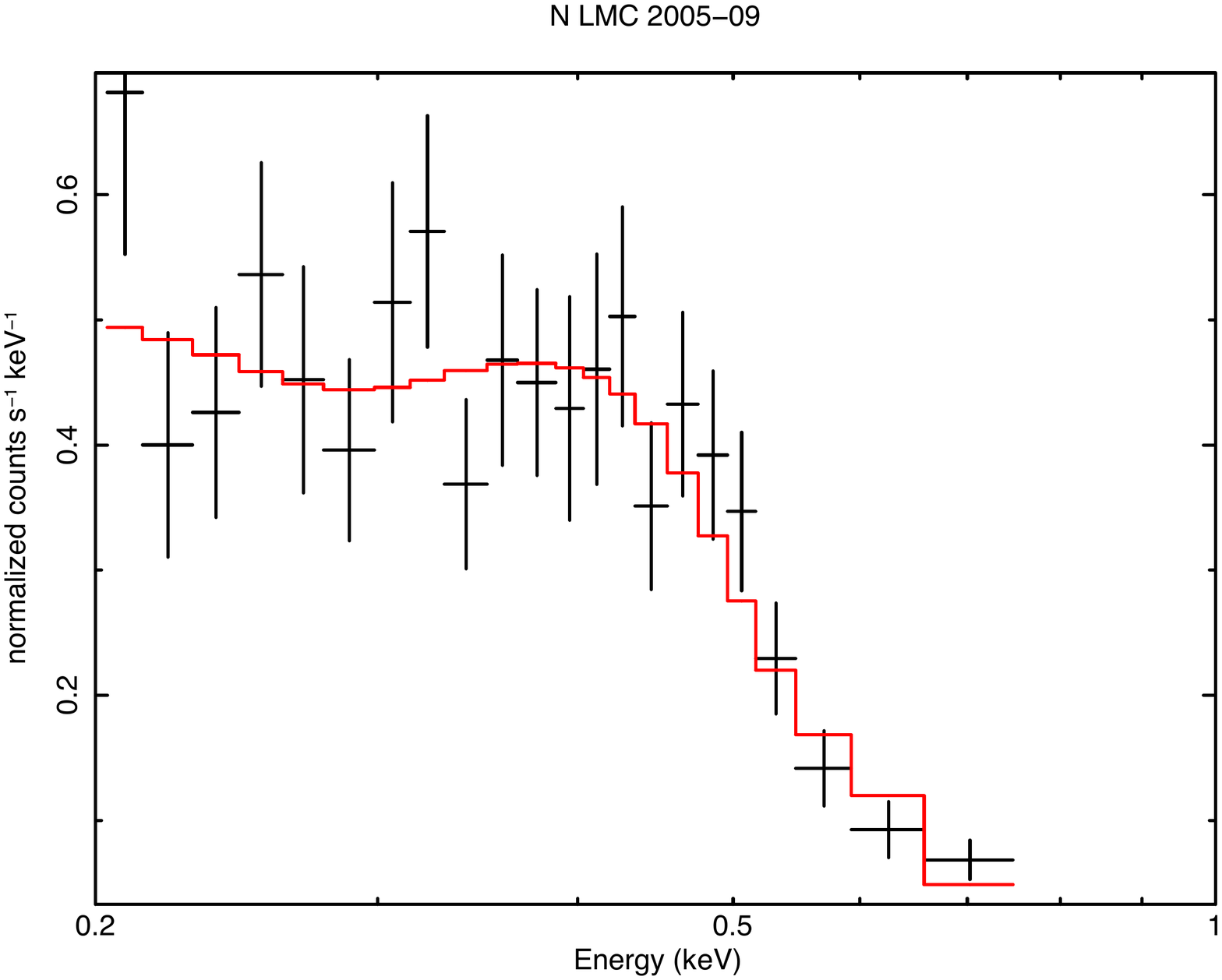}
\includegraphics[width=85mm]{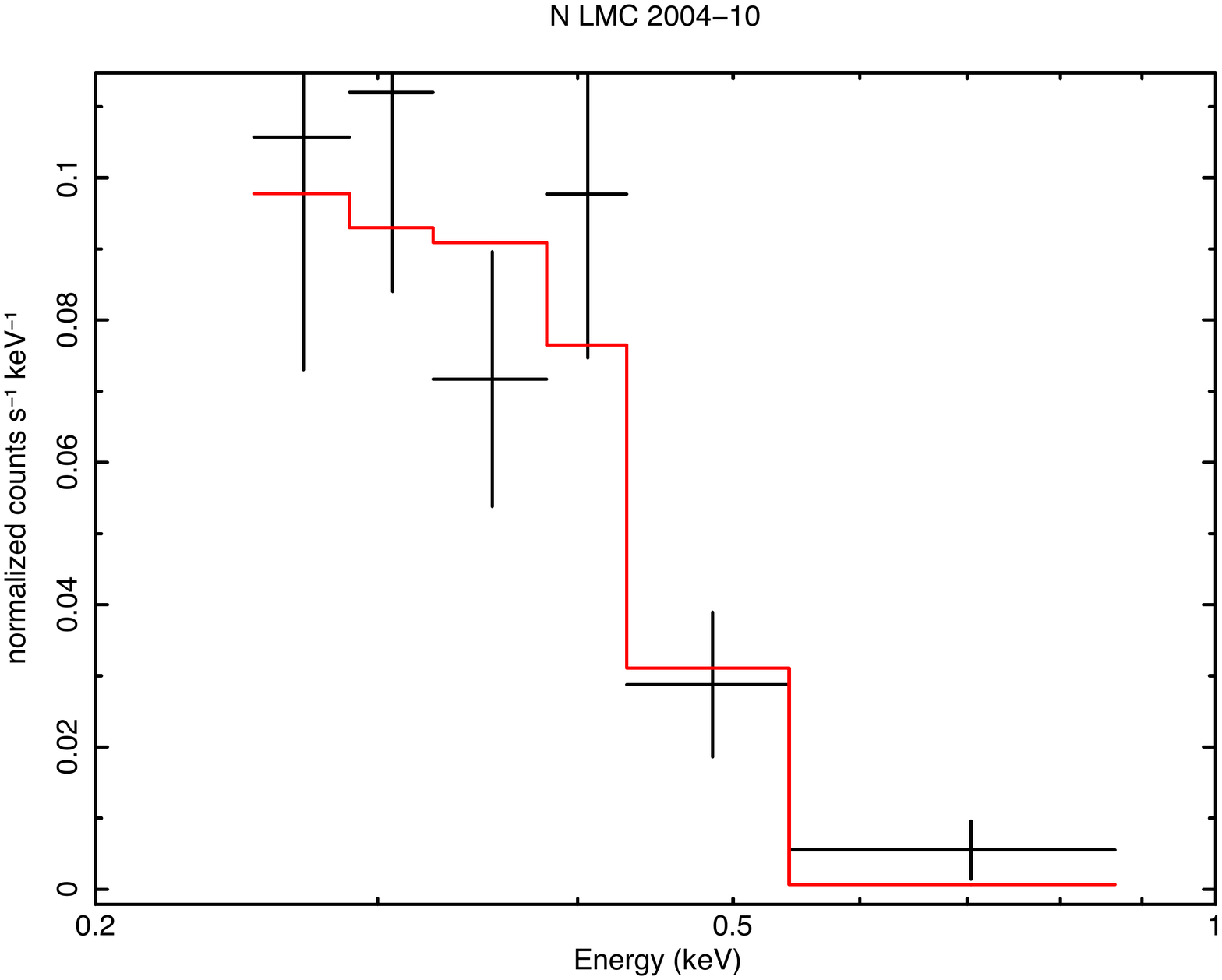}
\includegraphics[width=85mm]{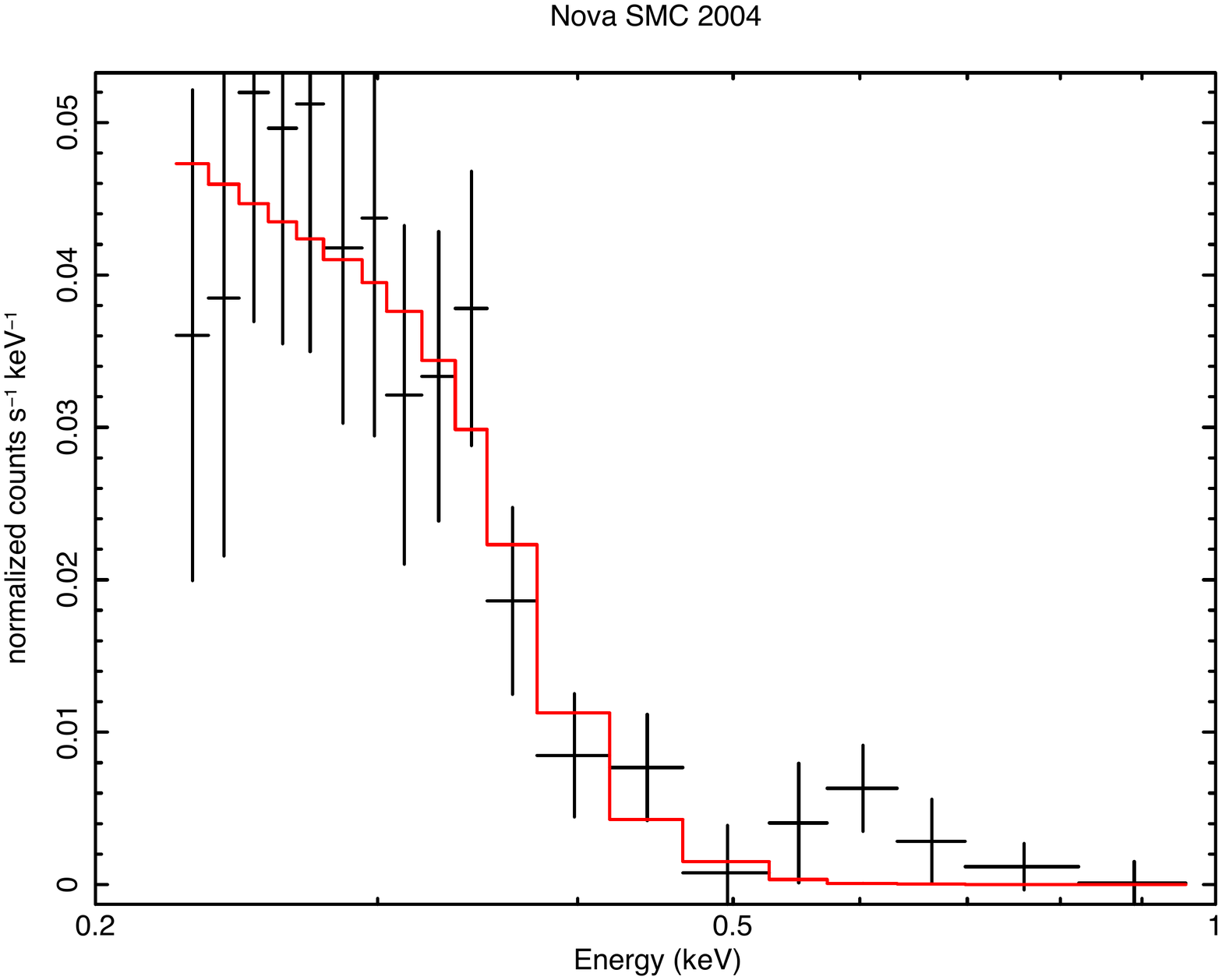}
\caption{TMAP atmospheric fit to the spectra of Novae SMC 2008-10, LMC 2005-11, LMC 2004-10, SMC 20046,
 at 1, 1.8, 3.9 and 5.25 years post-outburst, respectively.}
\end{figure*} 
\begin{table*}
\caption{Count rates and spectral fits
 with TMAP to the XMM-Newton observations of MC novae. For the best fit,
 we assumed a minimum N(H) of 3.5 $\times 10^{20}$ cm$^{-2}$}. 
\begin{center}
\begin{tabular}{rrrrrr}\hline\hline \noalign{\smallskip}
 Nova & Date & counts s$^{-1}$ & N(H) $\times 10^{20}$ cm$^{-2}$ & T$_{\rm eff}$ (K) & L$_{\rm x} \times 10^{35}$ erg s$^{-1}$ \\
 \hline
 \noalign{\smallskip}
LMC 2004-10 & 2006-12-2 & 0.0145$\pm$0.0033 & 3.5$^{+13}_{-3.5}$ & 476,000$\pm 148,000$
& 0.35$^{+1.14}_{-0.34}$ \\
            & 2008-09-03 & 0.0218$\pm$0.0028 & 3.5$^{+16}_{-3.5}$ & 671,000$\pm 112,000$
& 0.31$^{+0.56}_{-0.23}$ \\
SMC 2004-06  &  2009-09-13 & 0.0073$\pm$0.0007 & 8.8$^{+8.9}_{-5.6}$ &
 404,000$^{+49,000}_{-88,000}$ & 4.70$^{+5.30}_{-4.69}$ \\   
LMC 2005-09 & 2007-06-19  & 0.1645$\pm$0.0066 & 4.2$_{-1.8}^{+2.2}$ &
 715,000$_{-15,000}^{+24,000}$ & 2.40$_{-1.18}^{+1.72}$ \\
SMC 2008-10 & 2009-09-27 & 0.0081$\pm$0.0009 & 3.5$_{-3.5}^{+1.1}$ &
520,000$_{-50,000}^{+165,000}$ &  1.97$_{-1.87}^{+310.00}$ \\ 
\hline
 \noalign{\smallskip}
\end{tabular}
\end{center}
\end{table*}
\section{Conclusions}
 Our exploration of the N LMC 2009a X-ray high resolution spectra
 has highlighted their complexity and the likely superposition of different
 regions of emission.  We also searched the X-ray archives and
 discovered four more supersoft X-ray sources,
 albeit at lower luminosity than expected. These are our main conclusions.

$\bullet$ The continuum of the spectra of N LMC 2009 indicates the supersoft emission
 of the atmosphere of a WD with effective temperature peaking above 
 810,000 K (the PION blackbody source)
and likely to be around a million K (the TMAP model result),
 corresponding to a WD in the 1.2-1.4 M$_\odot$ range. 
 The continuum and its absorption edges can be well reproduced only assuming
 low oxygen abundance in  the intervening interstellar medium and, above all,
 enhanced nitrogen by a factor around 100 times solar in the nova atmosphere
 and/or residual wind near the surface, 
 where we assume that the absorption features originate. 

 $\bullet$ The blueshift of the absorption features is explained
 by a mass outflow rate of the order of 10$^{-8}$ M$_\odot$ yr$^{-1}$.
 Nova models usually assume that mass loss has ceased when the SSS emerges,
 but like in other  novae, this does not appear to be true: some residual
 mass loss is still occurring.

 $\bullet$ The absorption features in this nova were never as deep (or even saturated)
 as in other novae bserved in the Galaxy and in the Clouds, as the comparison
 with KT Eri and with N SMC 2016 shows clearly. While the peak temperature
 in N LMC 2009 may have been 100,000-200,000 K hotter, implying a higher
 ionization parameter, the difference is so large that it is 
 likely to have been due also to different abundances and density in the 
 medium in which the absorption features originated.  

$\bullet$ The ejecta of this nova emitted additional thermal X-ray flux that produced a complex
 emission line spectrum, most likely arising from multiple regions at different
 temperatures, with a contribution of both photoionization and shock ionization.

 $\bullet$ Nova KT Eri seems to have 
 a high mass WD and a partially evolved companion in 
 common with N LMC 2009. However, in N LMC 2009, 
 the turn-on time was longer, and the duration of the SSS was
 comparable, so the ejecta initially absorbing
 the SSS emission may have had higher mass, but the left-over envelope
 was of the same order of magnitude. 
 Since N LMC 2009a is a RN with an inter-outburst period
 of only 38 years, while no outburst of KT Eri
 have been found in tens of years
 before the eruption, N LMC 2009 did not have the time to accrete comparable or higher
 mass before the outburst unless it had  high $\dot m$, which
 in turn  would have resulted in ejection 
 after accretion of a small mass, unless the WD 
 before the outburst was still relatively cool \citep[see][]{Yaron2005}. 
 Most likely, this nova is at only the beginning of its RN cycle with
 short inter-outburst periods and the WD surface has not been heated
 yet by a repeated outburst cycle \citep[see][]{Yaron2005}.

$\bullet$ The 33s pulsation varied in amplitude and in length of the period,
 over time scales of hours.
 The search  for the physical root cause of this intriguing phenomenon,
 already observed also in other novae, must take 
 this fact into account.

 $\bullet$ N LMC 2009 was an order of magnitude less intrinsically luminous
 than the level compatible with emission arising from the whole surface.
 The predicted flux level was not observed in all
 SSS-novae, but it was indeed measured in N SMC 2016 as well
 as in other novae \citep[e.g. RS Oph,][]{Nelson2008}. 
 On the other hand, the portion of expected WD flux that is observed 
is much larger 
 than in U Sco, in which only Thomson scattered radiation was detected.
 We suggested that in N LMC 2009a we observed the WD directly along the line of sight,
 but either dense remaining clumps in the ejecta, or a non-disrupted accretion disk,
 blocked the whole of the whole WD atmosphere. 
 The clumps are a more likely explanation, because of the 
variability in the continuum level
 indicating varying visibility. The scenario we suggest 
 is one in which 
 part of the material, in an outflow in which instabilities occurred,
  was still opaque to X-rays.

 $\bullet$ We discovered that four other MC novae were serendipitously observed
 as SSS with low column density. There was no indication of  flux variation
 and the absolute SSS luminosity was only about a factor of a thousand less than
 predicted (and indeed observed  in the most luminous SSS-novae). 
 We hypothesize that partial obscuration of the SSS was due to
 an accretion disk at high inclination, since these novae were detected
 long after the end of mass loss and it is unlikely that the ejecta
 still obscured the WD. 

 $\bullet$  The statistics  of the MC novae of the last 20 years indicates
 that novae that are still SSS after 6 or more years from
the outburst are rare.
 We also found that half of all novae were detected as SSS 
 for at least several months
between few days and 5.5 years after the outburst. Since 
 the sampling was done mostly at sparse post-outburst times, and it never
 lasted for more than a few months with {\sl Swift},
or covered only one or two random
 epochs with {\sl XMM-Newton}, the percentage of novae detectable
 as SSS within 6 years from the outburst would actually be quite higher if an uninterrupted survey was possible.

\section*{Data Availability}
The data analyzed in this article are all available in the
HEASARC archive of NASA at the following URL: \url{https://
heasarc.gsfc.nasa.gov/db-perl/W3Browse/w3browse.pl}
\section*{Acknowledgements}
M. Orio was supported by a NASA grant for XMM-Newton data analysis. 
A.D. was supported by the Slovak grant VEGA 1/0408/20, and by the
Operational Programme Research and Innovation for the project:
``Scientific and Research Centre of Excellence SlovakION for Material and
Interdisciplinary Research'', project code ITMS2014+:313011W085,
co-financed by the European Regional Development Fund.


\bibliography{Final2} 

\begin{thebibliography}{}
\makeatletter
\relax
\def\mn@urlcharsother{\let\do\@makeother \do\$\do\&\do\#\do\^\do\_\do\%\do\~}
\def\mn@doi{\begingroup\mn@urlcharsother \@ifnextchar [ {\mn@doi@}
  {\mn@doi@[]}}
\def\mn@doi@[#1]#2{\def\@tempa{#1}\ifx\@tempa\@empty \href
  {http://dx.doi.org/#2} {doi:#2}\else \href {http://dx.doi.org/#2} {#1}\fi
  \endgroup}
\def\mn@eprint#1#2{\mn@eprint@#1:#2::\@nil}
\def\mn@eprint@arXiv#1{\href {http://arxiv.org/abs/#1} {{\tt arXiv:#1}}}
\def\mn@eprint@dblp#1{\href {http://dblp.uni-trier.de/rec/bibtex/#1.xml}
  {dblp:#1}}
\def\mn@eprint@#1:#2:#3:#4\@nil{\def\@tempa {#1}\def\@tempb {#2}\def\@tempc
  {#3}\ifx \@tempc \@empty \let \@tempc \@tempb \let \@tempb \@tempa \fi \ifx
  \@tempb \@empty \def\@tempb {arXiv}\fi \@ifundefined
  {mn@eprint@\@tempb}{\@tempb:\@tempc}{\expandafter \expandafter \csname
  mn@eprint@\@tempb\endcsname \expandafter{\@tempc}}}

\bibitem[\protect\citeauthoryear{{Aydi} et~al.,}{{Aydi}
  et~al.}{2018a}]{aydi2018}
{Aydi} E.,  et~al., 2018a, \mn@doi [\mnras] {10.1093/mnras/stx2678}, \href
  {https://ui.adsabs.harvard.edu/abs/2018MNRAS.474.2679A} {474, 2679}

\bibitem[\protect\citeauthoryear{{Aydi} et~al.,}{{Aydi}
  et~al.}{2018b}]{aydi2018b}
{Aydi} E.,  et~al., 2018b, \mn@doi [\mnras] {10.1093/mnras/sty1759}, \href
  {https://ui.adsabs.harvard.edu/abs/2018MNRAS.480..572A} {480, 572}

\bibitem[\protect\citeauthoryear{{Aydi} et~al.,}{{Aydi}
  et~al.}{2019}]{Aydi2019}
{Aydi} E.,  et~al., 2019, arXiv e-prints, \href
  {https://ui.adsabs.harvard.edu/abs/2019arXiv190309232A} {p. arXiv:1903.09232}

\bibitem[\protect\citeauthoryear{{Aydi} et~al.,}{{Aydi}
  et~al.}{2020}]{Aydi2020}
{Aydi} E.,  et~al., 2020, \mn@doi [\apj] {10.3847/1538-4357/abc3bb}, \href
  {https://ui.adsabs.harvard.edu/abs/2020ApJ...905...62A} {905, 62}

\bibitem[\protect\citeauthoryear{{Bahramian}, {Chomiuk}, {Strader}, {Kuin},
  {Darnley}  \& {Page}}{{Bahramian} et~al.}{2018}]{Bahramian2018}
{Bahramian} A.,  {Chomiuk} L.,  {Strader} J.,  {Kuin} P.,  {Darnley} M.~J.,
  {Page} K.,  2018, The Astronomer's Telegram, \href
  {https://ui.adsabs.harvard.edu/abs/2018ATel11301....1B} {11301, 1}

\bibitem[\protect\citeauthoryear{{Beardmore} et~al.,}{{Beardmore}
  et~al.}{2010}]{Beardmore2010}
{Beardmore} A.~P.,  et~al., 2010, The Astronomer's Telegram, \href
  {https://ui.adsabs.harvard.edu/abs/2010ATel.2423....1B} {2423, 1}

\bibitem[\protect\citeauthoryear{{Beardmore} et~al.,}{{Beardmore}
  et~al.}{2012}]{Beardmore2012}
{Beardmore} A.~P.,  et~al., 2012, \mn@doi [\aap] {10.1051/0004-6361/201219681},
  \href {http://adsabs.harvard.edu/abs/2012A%26A...545A.116B} {545, A116}

\bibitem[\protect\citeauthoryear{{Bode} et~al.,}{{Bode}
  et~al.}{2016}]{Bode2016}
{Bode} M.~F.,  et~al., 2016, \mn@doi [\apj] {10.3847/0004-637X/818/2/145},
  \href {http://adsabs.harvard.edu/abs/2016ApJ...818..145B} {818, 145}

\bibitem[\protect\citeauthoryear{{Bond}, {Walter}, {Cosgrove}, {Espinoza}  \&
  {Liller}}{{Bond} et~al.}{2009}]{Bond2009}
{Bond} H.~E.,  {Walter} F.~M.,  {Cosgrove} E.,  {Espinoza} J.,   {Liller} W.,
  2009, \iaucirc, \href {http://adsabs.harvard.edu/abs/2009IAUC.9019....2B}
  {9019}

\bibitem[\protect\citeauthoryear{{Cash}}{{Cash}}{1979}]{Cash1979}
{Cash} W.,  1979, \mn@doi [\apj] {10.1086/156922}, \href
  {https://ui.adsabs.harvard.edu/abs/1979ApJ...228..939C} {228, 939}

\bibitem[\protect\citeauthoryear{{Dobrotka} \& {Ness}}{{Dobrotka} \&
  {Ness}}{2017}]{dobrotka2017}
{Dobrotka} A.,  {Ness} J.-U.,  2017, \mn@doi [\mnras] {10.1093/mnras/stx442},
  \href {http://adsabs.harvard.edu/abs/2017MNRAS.467.4865D} {467, 4865}

\bibitem[\protect\citeauthoryear{{Dorman} \& {Arnaud}}{{Dorman} \&
  {Arnaud}}{2001}]{Dorman2001}
{Dorman} B.,  {Arnaud} K.~A.,  2001, in {Harnden} F.~R. J.,  {Primini} F.~A.,
  {Payne} H.~E.,  eds,  Astronomical Society of the Pacific Conference Series
  Vol. 238, Astronomical Data Analysis Software and Systems X. p.~415

\bibitem[\protect\citeauthoryear{{Drake} et~al.,}{{Drake}
  et~al.}{2003}]{Drake2003}
{Drake} J.~J.,  et~al., 2003, \mn@doi [ApJ] {10.1086/345534}, \href
  {https://ui.adsabs.harvard.edu/abs/2003ApJ...584..448D} {584, 448}

\bibitem[\protect\citeauthoryear{{Ducci} et~al.,}{{Ducci}
  et~al.}{2020}]{Ducci2020}
{Ducci} L.,  et~al., 2020, The Astronomer's Telegram, \href
  {https://ui.adsabs.harvard.edu/abs/2020ATel13545....1D} {13545, 1}

\bibitem[\protect\citeauthoryear{{Finzell}, {Chomiuk}, {Munari}  \&
  {Walter}}{{Finzell} et~al.}{2015}]{Finzell2015}
{Finzell} T.,  {Chomiuk} L.,  {Munari} U.,   {Walter} F.~M.,  2015, \mn@doi
  [\apj] {10.1088/0004-637X/809/2/160}, \href
  {http://adsabs.harvard.edu/abs/2015ApJ...809..160F} {809, 160}

\bibitem[\protect\citeauthoryear{{Heise}, {van Teeseling}  \&
  {Kahabka}}{{Heise} et~al.}{1994}]{Heise1994}
{Heise} J.,  {van Teeseling} A.,   {Kahabka} P.,  1994, \aap, \href
  {https://ui.adsabs.harvard.edu/abs/1994A&A...288L..45H} {288, L45}

\bibitem[\protect\citeauthoryear{{Jos{\'e}}, {Hernanz}  \&
  {Iliadis}}{{Jos{\'e}} et~al.}{2006}]{Jose2006}
{Jos{\'e}} J.,  {Hernanz} M.,   {Iliadis} C.,  2006, \mn@doi [\nphysa]
  {10.1016/j.nuclphysa.2005.02.121}, \href
  {https://ui.adsabs.harvard.edu/abs/2006NuPhA.777..550J} {777, 550}

\bibitem[\protect\citeauthoryear{{Kaastra}}{{Kaastra}}{2017}]{Kaastra2017}
{Kaastra} J.~S.,  2017, \mn@doi [\aap] {10.1051/0004-6361/201629319}, \href
  {https://ui.adsabs.harvard.edu/abs/2017A&A...605A..51K} {605, A51}

\bibitem[\protect\citeauthoryear{{Kaastra}, {Mewe}  \&
  {Nieuwenhuijzen}}{{Kaastra} et~al.}{1996}]{Kaastra1996}
{Kaastra} J.~S.,  {Mewe} R.,   {Nieuwenhuijzen} H.,  1996, in UV and X-ray
  Spectroscopy of Astrophysical and Laboratory Plasmas. pp 411--414

\bibitem[\protect\citeauthoryear{{Kuin} et~al.,}{{Kuin}
  et~al.}{2020}]{Kuin2020}
{Kuin} N.~P.~M.,  et~al., 2020, \mn@doi [\mnras] {10.1093/mnras/stz2960}, \href
  {https://ui.adsabs.harvard.edu/abs/2020MNRAS.491..655K} {491, 655}

\bibitem[\protect\citeauthoryear{{Li} et~al.,}{{Li} et~al.}{2012}]{Li2012}
{Li} K.~L.,  et~al., 2012, \mn@doi [Apj] {10.1088/0004-637X/761/2/99}, \href
  {https://ui.adsabs.harvard.edu/abs/2012ApJ...761...99L} {761, 99}

\bibitem[\protect\citeauthoryear{{Liller}}{{Liller}}{2009}]{Liller2009}
{Liller} W.,  2009, \iaucirc, \href
  {http://adsabs.harvard.edu/abs/2009IAUC.9019....1L} {9019}

\bibitem[\protect\citeauthoryear{{Mehdipour}, {Kaastra}  \&
  {Kallman}}{{Mehdipour} et~al.}{2016}]{Mehdipour2016}
{Mehdipour} M.,  {Kaastra} J.~S.,   {Kallman} T.,  2016, \mn@doi [\aap]
  {10.1051/0004-6361/201628721}, \href
  {https://ui.adsabs.harvard.edu/abs/2016A&A...596A..65M} {596, A65}

\bibitem[\protect\citeauthoryear{{Morii} et~al.,}{{Morii}
  et~al.}{2013}]{Morii2013}
{Morii} M.,  et~al., 2013, \mn@doi [Apj] {10.1088/0004-637X/779/2/118}, \href
  {https://ui.adsabs.harvard.edu/abs/2013ApJ...779..118M} {779, 118}

\bibitem[\protect\citeauthoryear{{Mr{\'o}z} et~al.,}{{Mr{\'o}z}
  et~al.}{2016}]{Mroz2016}
{Mr{\'o}z} P.,  et~al., 2016, \mn@doi [\apjs] {10.3847/0067-0049/222/1/9},
  \href {https://ui.adsabs.harvard.edu/abs/2016ApJS..222....9M} {222, 9}

\bibitem[\protect\citeauthoryear{{Nelson}, {Orio}, {Cassinelli}, {Still},
  {Leibowitz}  \& {Mucciarelli}}{{Nelson} et~al.}{2008}]{Nelson2008}
{Nelson} T.,  {Orio} M.,  {Cassinelli} J.~P.,  {Still} M.,  {Leibowitz} E.,
  {Mucciarelli} P.,  2008, \mn@doi [ApJ] {10.1086/524054}, \href
  {https://ui.adsabs.harvard.edu/abs/2008ApJ...673.1067N} {673, 1067}

\bibitem[\protect\citeauthoryear{{Ness}, {Starrfield}, {Jordan}, {Krautter}  \&
  {Schmitt}}{{Ness} et~al.}{2005}]{Ness2005}
{Ness} J.~U.,  {Starrfield} S.,  {Jordan} C.,  {Krautter} J.,   {Schmitt}
  J.~H.~M.~M.,  2005, \mn@doi [\mnras] {10.1111/j.1365-2966.2005.09664.x},
  \href {https://ui.adsabs.harvard.edu/abs/2005MNRAS.364.1015N} {364, 1015}

\bibitem[\protect\citeauthoryear{{Ness}, {Drake}, {Starrfield}, {Bode}, {Page},
  {Beardmore}, {Osborne}  \& {Schwarz}}{{Ness} et~al.}{2010}]{Ness2010}
{Ness} J.~U.,  {Drake} J.~J.,  {Starrfield} S.,  {Bode} M.,  {Page} K.,
  {Beardmore} A.,  {Osborne} J.~P.,   {Schwarz} G.,  2010, The Astronomer's
  Telegram, \href {https://ui.adsabs.harvard.edu/abs/2010ATel.2418....1N}
  {2418, 1}

\bibitem[\protect\citeauthoryear{{Ness} et~al.,}{{Ness}
  et~al.}{2011}]{Ness2011}
{Ness} J.~U.,  et~al., 2011, \mn@doi [Apj] {10.1088/0004-637X/733/1/70}, \href
  {https://ui.adsabs.harvard.edu/abs/2011ApJ...733...70N} {733, 70}

\bibitem[\protect\citeauthoryear{{Ness} et~al.,}{{Ness}
  et~al.}{2012}]{Ness2012}
{Ness} J.-U.,  et~al., 2012, \mn@doi [ApJ] {10.1088/0004-637X/745/1/43}, \href
  {http://adsabs.harvard.edu/abs/2012ApJ...745...43N} {745, 43}

\bibitem[\protect\citeauthoryear{{Ness} et~al.,}{{Ness}
  et~al.}{2013}]{Ness2013}
{Ness} J.~U.,  et~al., 2013, \mn@doi [\aap] {10.1051/0004-6361/201322415},
  \href {https://ui.adsabs.harvard.edu/abs/2013A&A...559A..50N} {559, A50}

\bibitem[\protect\citeauthoryear{{Ness} et~al.,}{{Ness}
  et~al.}{2015}]{Ness2015}
{Ness} J.~U.,  et~al., 2015, \mn@doi [\aap] {10.1051/0004-6361/201425178},
  \href {https://ui.adsabs.harvard.edu/abs/2015A&A...578A..39N} {578, A39}

\bibitem[\protect\citeauthoryear{{Odendaal} \& {Meintjes}}{{Odendaal} \&
  {Meintjes}}{2017}]{Odendaal2017}
{Odendaal} A.,  {Meintjes} P.~J.,  2017, \mn@doi [\mnras]
  {10.1093/mnras/stx233}, \href
  {https://ui.adsabs.harvard.edu/abs/2017MNRAS.467.2797O} {467, 2797}

\bibitem[\protect\citeauthoryear{{Odendaal}, {Meintjes}, {Charles}  \&
  {Rajoelimanana}}{{Odendaal} et~al.}{2014}]{Odendaal2014}
{Odendaal} A.,  {Meintjes} P.~J.,  {Charles} P.~A.,   {Rajoelimanana} A.~F.,
  2014, \mn@doi [\mnras] {10.1093/mnras/stt2111}, \href
  {https://ui.adsabs.harvard.edu/abs/2014MNRAS.437.2948O} {437, 2948}

\bibitem[\protect\citeauthoryear{{Orio}}{{Orio}}{2012}]{Orio2012}
{Orio} M.,  2012, Bulletin of the Astronomical Society of India, \href
  {https://ui.adsabs.harvard.edu/abs/2012BASI...40..333O} {40, 333}

\bibitem[\protect\citeauthoryear{{Orio}, {Covington}  \& {{\"O}gelman}}{{Orio}
  et~al.}{2001}]{Orio2001}
{Orio} M.,  {Covington} J.,   {{\"O}gelman} H.,  2001, \mn@doi [\aap]
  {10.1051/0004-6361:20010537}, \href
  {https://ui.adsabs.harvard.edu/abs/2001A&A...373..542O} {373, 542}

\bibitem[\protect\citeauthoryear{{Orio}, {Hartmann}, {Still}  \&
  {Greiner}}{{Orio} et~al.}{2003a}]{Greiner2003}
{Orio} M.,  {Hartmann} W.,  {Still} M.,   {Greiner} J.,  2003a, \mn@doi [The
  Astrophysical Journal] {10.1086/376828}, \href
  {https://ui.adsabs.harvard.edu/abs/2003ApJ...594..435O} {594, 435}

\bibitem[\protect\citeauthoryear{{Orio}, {Hartmann}, {Still}  \&
  {Greiner}}{{Orio} et~al.}{2003b}]{Orio2003}
{Orio} M.,  {Hartmann} W.,  {Still} M.,   {Greiner} J.,  2003b, \mn@doi [The
  Astrophysical Journal] {10.1086/376828}, \href
  {https://ui.adsabs.harvard.edu/abs/2003ApJ...594..435O} {594, 435}

\bibitem[\protect\citeauthoryear{{Orio}, {Mason}, {Gallagher}  \&
  {Abbott}}{{Orio} et~al.}{2009}]{Orio2009}
{Orio} M.,  {Mason} E.,  {Gallagher} J.,   {Abbott} T.,  2009, The Astronomer's
  Telegram, \href {http://adsabs.harvard.edu/abs/2009ATel.1930....1O} {1930}

\bibitem[\protect\citeauthoryear{{Orio}, {Behar}, {Gallagher}, {Bianchini},
  {Chiosi}, {Luna}, {Nelson}  \& {Rauch}}{{Orio} et~al.}{2013a}]{OrioIAU}
{Orio} M.,  {Behar} E.,  {Gallagher} J.,  {Bianchini} A.,  {Chiosi} E.,  {Luna}
  J.,  {Nelson} T.,   {Rauch} T.,  2013a, in {Di Stefano} R.,  {Orio} M.,
  {Moe} M.,  eds,  Vol. 281, Binary Paths to Type Ia Supernovae Explosions. pp
  181--185, \mn@doi{10.1017/S1743921312014950}

\bibitem[\protect\citeauthoryear{{Orio} et~al.,}{{Orio}
  et~al.}{2013b}]{Orio2013}
{Orio} M.,  et~al., 2013b, \mn@doi [\mnras] {10.1093/mnras/sts421}, \href
  {http://adsabs.harvard.edu/abs/2013MNRAS.429.1342O} {429, 1342}

\bibitem[\protect\citeauthoryear{{Orio}, {Henze}  \& {Ness}}{{Orio}
  et~al.}{2017}]{Orio2017}
{Orio} M.,  {Henze} M.,   {Ness} J.,  2017, in {Ness} J.-U.,  {Migliari} S.,
  eds, The X-ray Universe 2017. p.~165

\bibitem[\protect\citeauthoryear{{Orio} et~al.,}{{Orio}
  et~al.}{2018}]{Orio2018}
{Orio} M.,  et~al., 2018, \mn@doi [ApJ] {10.3847/1538-4357/aacf06}, \href
  {https://ui.adsabs.harvard.edu/abs/2018ApJ...862..164O} {862, 164}

\bibitem[\protect\citeauthoryear{{Orio} et~al.,}{{Orio}
  et~al.}{2020}]{Orio2020}
{Orio} M.,  et~al., 2020, \mn@doi [\apj] {10.3847/1538-4357/ab8c4d}, \href
  {https://ui.adsabs.harvard.edu/abs/2020ApJ...895...80O} {895, 80}

\bibitem[\protect\citeauthoryear{{Page}, {Walter}, {Schwarz}  \&
  {Osborne}}{{Page} et~al.}{2013a}]{Page2013a}
{Page} K.~L.,  {Walter} F.~M.,  {Schwarz} G.~J.,   {Osborne} J.~P.,  2013a, The
  Astronomer's Telegram, \href
  {https://ui.adsabs.harvard.edu/abs/2013ATel.4853....1P} {4853, 1}

\bibitem[\protect\citeauthoryear{{Page}, {Osborne}, {Beardmore}  \&
  {Schwarz}}{{Page} et~al.}{2013b}]{Page2013b}
{Page} K.~L.,  {Osborne} J.~P.,  {Beardmore} A.~P.,   {Schwarz} G.~J.,  2013b,
  The Astronomer's Telegram, \href
  {https://ui.adsabs.harvard.edu/abs/2013ATel.4920....1P} {4920, 1}

\bibitem[\protect\citeauthoryear{{Page}, {Kuin}  \& {Henze}}{{Page}
  et~al.}{2018}]{Page2018}
{Page} K.~L.,  {Kuin} N.~P.~M.,   {Henze} M.,  2018, The Astronomer's Telegram,
  \href {https://ui.adsabs.harvard.edu/abs/2018ATel11410....1P} {11410, 1}

\bibitem[\protect\citeauthoryear{{Page}, {Beardmore}  \& {Osborne}}{{Page}
  et~al.}{2020}]{Page2020}
{Page} K.~L.,  {Beardmore} A.~P.,   {Osborne} J.~P.,  2020, \mn@doi [Advances
  in Space Research] {10.1016/j.asr.2019.08.003}, \href
  {https://ui.adsabs.harvard.edu/abs/2020AdSpR..66.1169P} {66, 1169}

\bibitem[\protect\citeauthoryear{{Pagnotta} et~al.,}{{Pagnotta}
  et~al.}{2015}]{Pagnotta2015}
{Pagnotta} A.,  et~al., 2015, \mn@doi [\apj] {10.1088/0004-637X/811/1/32},
  \href {https://ui.adsabs.harvard.edu/abs/2015ApJ...811...32P} {811, 32}

\bibitem[\protect\citeauthoryear{{Payne-Gaposchkin}}{{Payne-Gaposchkin}}{1964}]{PayneG1964}
{Payne-Gaposchkin} C.~H.,  1964, {The galactic novae}

\bibitem[\protect\citeauthoryear{{Peretz}, {Orio}, {Behar}, {Bianchini},
  {Gallagher}, {Rauch}, {Tofflemire}  \& {Zemko}}{{Peretz}
  et~al.}{2016}]{Peretz2016}
{Peretz} U.,  {Orio} M.,  {Behar} E.,  {Bianchini} A.,  {Gallagher} J.,
  {Rauch} T.,  {Tofflemire} B.,   {Zemko} P.,  2016, \mn@doi [ApJ]
  {10.3847/0004-637X/829/1/2}, \href
  {https://ui.adsabs.harvard.edu/abs/2016ApJ...829....2P} {829, 2}

\bibitem[\protect\citeauthoryear{{Pietrzy{\'n}ski} et~al.,}{{Pietrzy{\'n}ski}
  et~al.}{2019}]{Pietrzynski2019}
{Pietrzy{\'n}ski} G.,  et~al., 2019, \mn@doi [\nat]
  {10.1038/s41586-019-0999-4}, \href
  {https://ui.adsabs.harvard.edu/abs/2019Natur.567..200P} {567, 200}

\bibitem[\protect\citeauthoryear{{Pinto}, {Ness}, {Verbunt}, {Kaastra},
  {Costantini}  \& {Detmers}}{{Pinto} et~al.}{2012}]{Pinto2012}
{Pinto} C.,  {Ness} J.~U.,  {Verbunt} F.,  {Kaastra} J.~S.,  {Costantini} E.,
  {Detmers} R.~G.,  2012, \mn@doi [\aap] {10.1051/0004-6361/201117835}, \href
  {https://ui.adsabs.harvard.edu/abs/2012A&A...543A.134P} {543, A134}

\bibitem[\protect\citeauthoryear{{Prialnik}}{{Prialnik}}{1986}]{Prialnik1986}
{Prialnik} D.,  1986, \mn@doi [Apj] {10.1086/164677}, \href
  {https://ui.adsabs.harvard.edu/abs/1986ApJ...310..222P} {310, 222}

\bibitem[\protect\citeauthoryear{{Rauch}, {Orio}, {Gonzales-Riestra}, {Nelson},
  {Still}, {Werner}  \& {Wilms}}{{Rauch} et~al.}{2010}]{Rauch2010}
{Rauch} T.,  {Orio} M.,  {Gonzales-Riestra} R.,  {Nelson} T.,  {Still} M.,
  {Werner} K.,   {Wilms} J.,  2010, \mn@doi [ApJ]
  {10.1088/0004-637X/717/1/363}, \href
  {https://ui.adsabs.harvard.edu/abs/2010ApJ...717..363R} {717, 363}

\bibitem[\protect\citeauthoryear{{Rohrbach}, {Ness}  \&
  {Starrfield}}{{Rohrbach} et~al.}{2009}]{Rohrbach2009}
{Rohrbach} J.~G.,  {Ness} J.~U.,   {Starrfield} S.,  2009, \mn@doi [\aj]
  {10.1088/0004-6256/137/6/4627}, \href
  {https://ui.adsabs.harvard.edu/abs/2009AJ....137.4627R} {137, 4627}

\bibitem[\protect\citeauthoryear{{Scargle}}{{Scargle}}{1982}]{scargle1982}
{Scargle} J.~D.,  1982, \mn@doi [\apj] {10.1086/160554}, \href
  {http://adsabs.harvard.edu/abs/1982ApJ...263..835S} {263, 835}

\bibitem[\protect\citeauthoryear{{Schaefer} et~al.,}{{Schaefer}
  et~al.}{2010}]{Schaefer2010}
{Schaefer} B.~E.,  et~al., 2010, \mn@doi [\aj] {10.1088/0004-6256/140/4/925},
  \href {http://adsabs.harvard.edu/abs/2010AJ....140..925S} {140, 925}

\bibitem[\protect\citeauthoryear{{Schwarz} et~al.,}{{Schwarz}
  et~al.}{2011}]{Schwarz2011}
{Schwarz} G.~J.,  et~al., 2011, \mn@doi [\apjs] {10.1088/0067-0049/197/2/31},
  \href {https://ui.adsabs.harvard.edu/abs/2011ApJS..197...31S} {197, 31}

\bibitem[\protect\citeauthoryear{{Schwarz}, {Osborne}, {Page}, {Walter}  \&
  {Starrfield}}{{Schwarz} et~al.}{2012}]{Schwarz2012}
{Schwarz} G.~J.,  {Osborne} J.~P.,  {Page} K.,  {Walter} F.~M.,   {Starrfield}
  S.,  2012, The Astronomer's Telegram, \href
  {https://ui.adsabs.harvard.edu/abs/2012ATel.4501....1S} {4501, 1}

\bibitem[\protect\citeauthoryear{{Schwarz} et~al.,}{{Schwarz}
  et~al.}{2015}]{Schwarz2015}
{Schwarz} G.~J.,  et~al., 2015, \mn@doi [Apj] {10.1088/0004-6256/149/3/95},
  \href {https://ui.adsabs.harvard.edu/abs/2015AJ....149...95S} {149, 95}

\bibitem[\protect\citeauthoryear{{Smith}, {Brickhouse}, {Liedahl}  \&
  {Raymond}}{{Smith} et~al.}{2001}]{Smith2001}
{Smith} R.~K.,  {Brickhouse} N.~S.,  {Liedahl} D.~A.,   {Raymond} J.~C.,  2001,
  \mn@doi [\apjl] {10.1086/322992}, \href
  {https://ui.adsabs.harvard.edu/abs/2001ApJ...556L..91S} {556, L91}

\bibitem[\protect\citeauthoryear{{Starrfield}, {Timmes}, {Iliadis}, {Hix},
  {Arnett}, {Meakin}  \& {Sparks}}{{Starrfield} et~al.}{2012}]{Starrfield2012}
{Starrfield} S.,  {Timmes} F.~X.,  {Iliadis} C.,  {Hix} W.~R.,  {Arnett} W.~D.,
   {Meakin} C.,   {Sparks} W.~M.,  2012, \mn@doi [Baltic Astronomy]
  {10.1515/astro-2017-0361}, \href
  {https://ui.adsabs.harvard.edu/abs/2012BaltA..21...76S} {21, 76}

\bibitem[\protect\citeauthoryear{{Strope}, {Schaefer}  \& {Henden}}{{Strope}
  et~al.}{2010}]{Strope2010}
{Strope} R.~J.,  {Schaefer} B.~E.,   {Henden} A.~A.,  2010, \mn@doi [\aj]
  {10.1088/0004-6256/140/1/34}, \href
  {http://adsabs.harvard.edu/abs/2010AJ....140...34S} {140, 34}

\bibitem[\protect\citeauthoryear{{Tofflemire}, {Orio}, {Page}, {Osborne},
  {Ciroi}, {Cracco}, {Di Mille}  \& {Maxwell}}{{Tofflemire}
  et~al.}{2013}]{Tofflemi2013}
{Tofflemire} B.~M.,  {Orio} M.,  {Page} K.~L.,  {Osborne} J.~P.,  {Ciroi} S.,
  {Cracco} V.,  {Di Mille} F.,   {Maxwell} M.,  2013, \mn@doi [ApJ]
  {10.1088/0004-637X/779/1/22}, \href
  {https://ui.adsabs.harvard.edu/abs/2013ApJ...779...22T} {779, 22}

\bibitem[\protect\citeauthoryear{{Trudolyubov} \& {Priedhorsky}}{{Trudolyubov}
  \& {Priedhorsky}}{2008}]{Trudo2008}
{Trudolyubov} S.~P.,  {Priedhorsky} W.~C.,  2008, \mn@doi [\apj]
  {10.1086/526397}, \href
  {https://ui.adsabs.harvard.edu/abs/2008ApJ...676.1218T} {676, 1218}

\bibitem[\protect\citeauthoryear{{Williams}}{{Williams}}{1992}]{Williams1992}
{Williams} R.~E.,  1992, \mn@doi [\aj] {10.1086/116268}, \href
  {http://adsabs.harvard.edu/abs/1992AJ....104..725W} {104, 725}

\bibitem[\protect\citeauthoryear{{Wilms}, {Allen}  \& {McCray}}{{Wilms}
  et~al.}{2000}]{Wilms2000}
{Wilms} J.,  {Allen} A.,   {McCray} R.,  2000, \mn@doi [\apj] {10.1086/317016},
  \href {https://ui.adsabs.harvard.edu/abs/2000ApJ...542..914W} {542, 914}

\bibitem[\protect\citeauthoryear{{Wolf}, {Bildsten}, {Brooks}  \&
  {Paxton}}{{Wolf} et~al.}{2013}]{Wolf2013}
{Wolf} W.~M.,  {Bildsten} L.,  {Brooks} J.,   {Paxton} B.,  2013, \mn@doi [ApJ]
  {10.1088/0004-637X/777/2/136}, \href
  {https://ui.adsabs.harvard.edu/abs/2013ApJ...777..136W} {777, 136}

\bibitem[\protect\citeauthoryear{{Wolf}, {Townsend}  \& {Bildsten}}{{Wolf}
  et~al.}{2018}]{Wolf2018}
{Wolf} W.~M.,  {Townsend} R. H.~D.,   {Bildsten} L.,  2018, \mn@doi [ApJ]
  {10.3847/1538-4357/aaad05}, \href
  {https://ui.adsabs.harvard.edu/abs/2018ApJ...855..127W} {855, 127}

\bibitem[\protect\citeauthoryear{{Yaron}, {Prialnik}, {Shara}  \&
  {Kovetz}}{{Yaron} et~al.}{2005}]{Yaron2005}
{Yaron} O.,  {Prialnik} D.,  {Shara} M.~M.,   {Kovetz} A.,  2005, \mn@doi [AJ]
  {10.1086/428435}, \href
  {https://ui.adsabs.harvard.edu/abs/2005ApJ...623..398Y} {623, 398}

\bibitem[\protect\citeauthoryear{{den Herder} et~al.,}{{den Herder}
  et~al.}{2001}]{denHerder2001}
{den Herder} J.~W.,  et~al., 2001, \mn@doi [\aap] {10.1051/0004-6361:20000058},
  \href {https://ui.adsabs.harvard.edu/abs/2001A&A...365L...7D} {365, L7}

\bibitem[\protect\citeauthoryear{{van Rossum}}{{van
  Rossum}}{2012}]{vanRossum2012}
{van Rossum} D.~R.,  2012, \mn@doi [Apj] {10.1088/0004-637X/756/1/43}, \href
  {https://ui.adsabs.harvard.edu/abs/2012ApJ...756...43V} {756, 43}

\makeatother
\end{thebibliography}
\bibliographystyle{mnras}
\bsp    
\appendix
\section{Simulations of the August 2009 light curve}
\begin{figure}
\resizebox{\hsize}{!}{\includegraphics[angle=-90]{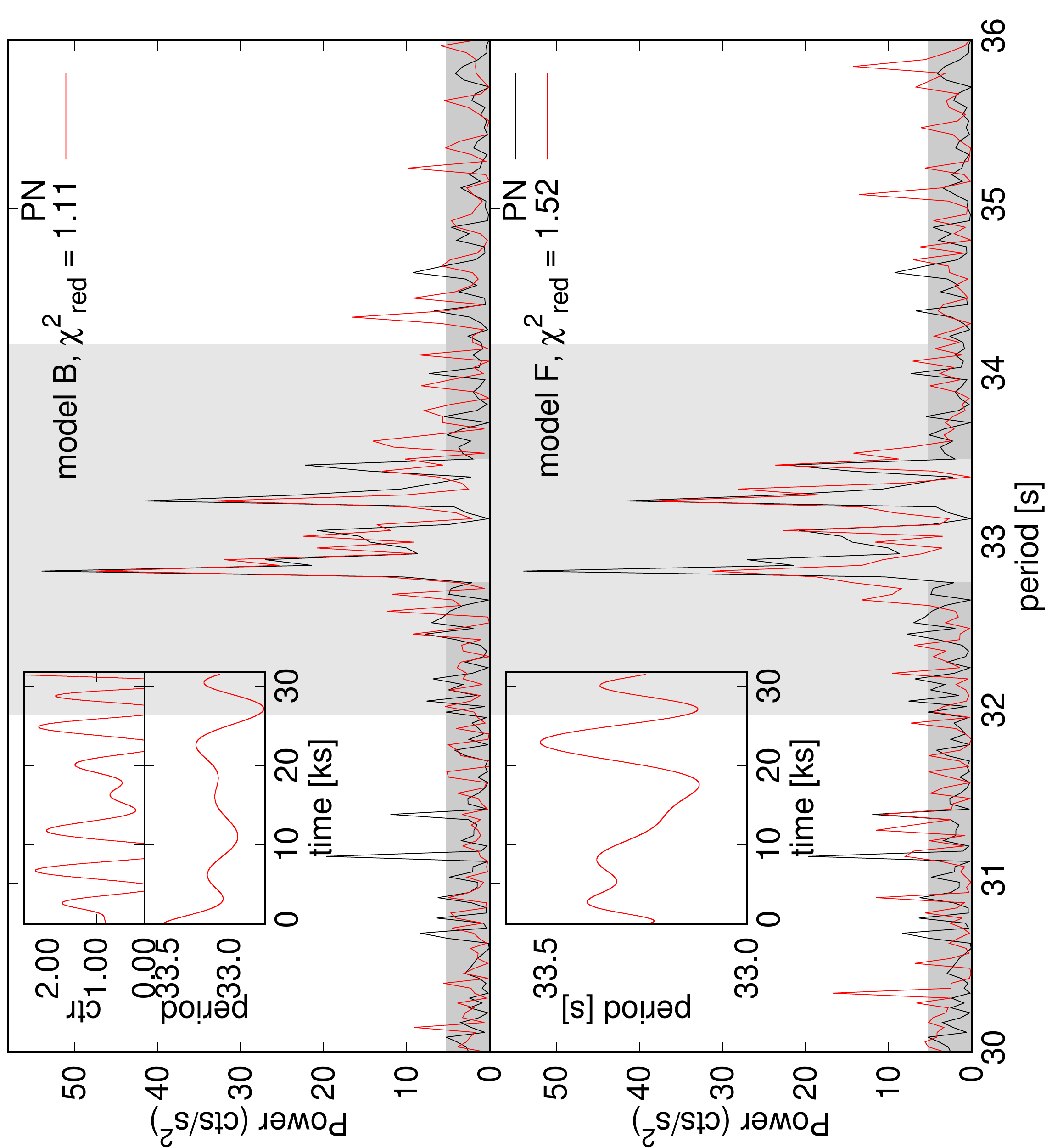}}
\caption{Comparison of the observed data in August and the best of 10000 simulated periodograms for models B and F, as in Table~\ref{var_param}. The insets show the evolution of the simulated amplitude (in count rate) and periodicity. The light shaded area is the frequency interval over which the $\chi^2_{\rm red}$ is calculated. The darker area
represents the interval over which we calculated the power uncertainty $\sigma$ (equal to the vertical extent of the shaded area).}
\label{pds_pvar}
\end{figure}

\begin{figure}
\resizebox{\hsize}{!}{\includegraphics[angle=-90]{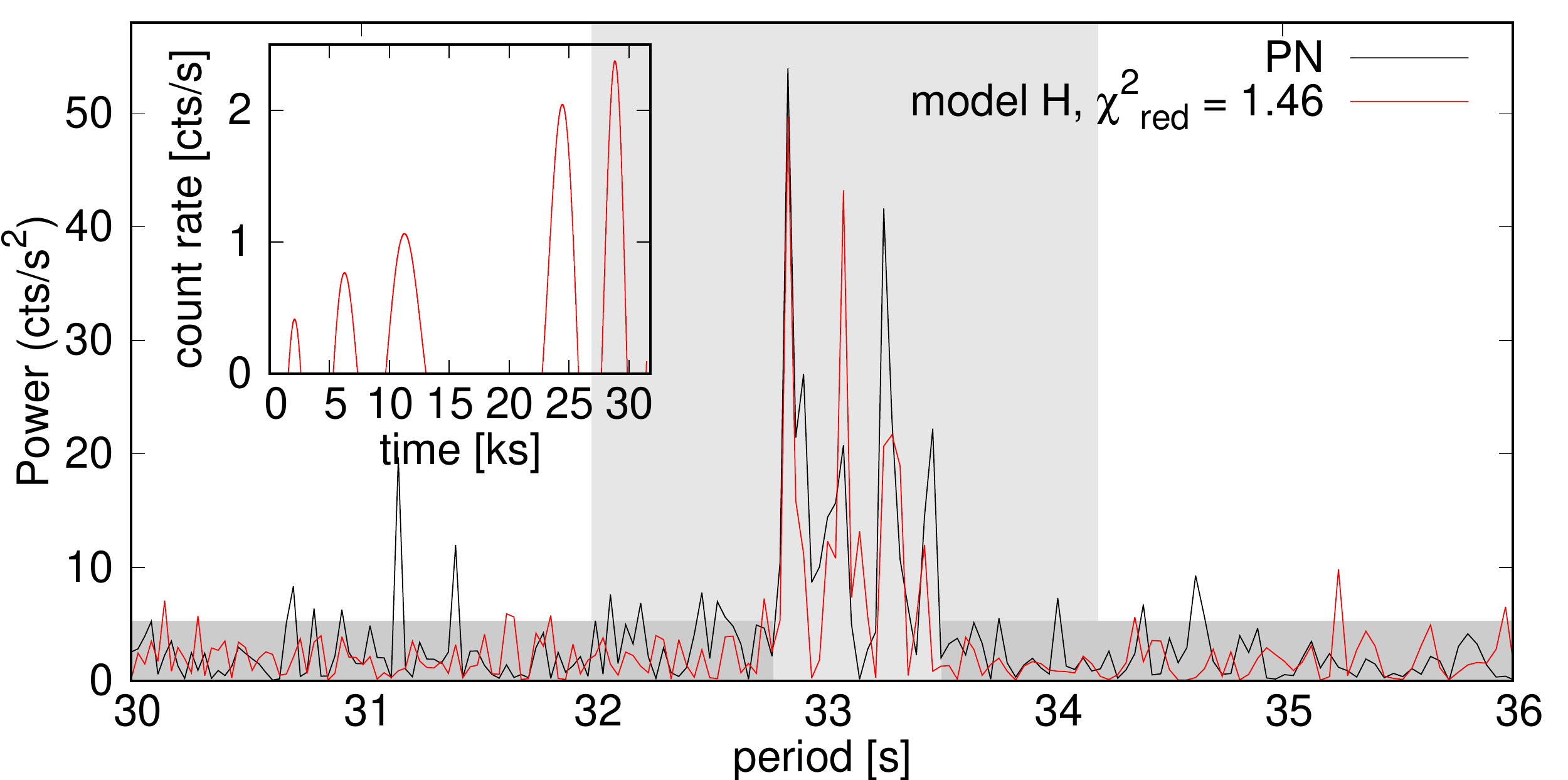}}
\caption{The same as Fig.~\ref{pds_pvar} but for model H.}
\label{pds_pconst}
\end{figure}

We based this simulation on the August pn data,
 because they yielded the highest count rate and the most complex features.
 We first fitted the pn data with a polynomial (25th order, P$_{25}$),
 in order to model the long-term trend. and  on P$_{25}$
 we added a sinusoidal function and Gaussian noise G.
 The latter represents the residual scatter
 of flux after subtracting P$_{25}$
 from the observed data. \footnote{It yields a very similar flux distribution to the 
 simple Poisson noise.}
 The flux $\psi$ thus can be expressed with this equation:
\begin{equation}
\psi = {\rm P}_{25} + a\,{\rm sin}(2\,\pi\,t/p) + {\rm G}
\end{equation}
We performed simulations with constant and with variable periodicity
 $p$ and amplitude $a$.
 Table~\ref{var_param} summarizes the parameters $p$ and $a$.
 For the variable periodicity
 simulations we needed a smooth function describing the variability of
 amplitude or/and periodicity, so we generated random points over the
 time interval
 of the exposure\footnote{Also a short
 time before and after the exposure, to avoid boundary fitting effects,
 and using a step of 200\,s.}, with a Gaussian distribution in which $p$ and $a$ are the mean values, and d$p$ and d$a$ are the
corresponding variances. We fitted these points with a polynomial (P$_a$ or P$_p$), the required smooth function used as input for the sine function;
\begin{equation}
\psi = {\rm P}_{25} + {\rm P}_a\,{\rm sin}(2\,\pi\,t/{\rm P}_p) + {\rm G}
\label{model}
\end{equation}
If the amplitude drops at or below zero,  the modulation
 is considered to be absent ($a = 0$).
 As input values for the mean periodicity $p$ we used the most
 significant periodicities $P_1$, $P_2$ and $P_3$ from Table~\ref{periods}, and we varied the amplitude until we obtained the best match.
\begin{table}
\caption{Sine parameters of the simulated signal, amplitude ($a$) and periodicity ($p$). d$a$ and d$p$ represent the rate of variability of the corresponding parameter (see text for details). No d$a$ or no d$p$ values means (constant) P$_a = a$ or P$_p = p$ in equation~(\ref{model}), respectively.}
\begin{center}
\begin{tabular}{lccccr}
\hline
\hline
model & $p$ & ${\rm d}p$ & $a$ & ${\rm d}a$ & best $\chi^2_{\rm red}$\\
 & (s) & (s) & (cts/s) & (cts/s)\\
\hline
A & 32.9 & 0.5 & 1.0 & 2.5 & 1.87\\
B & 33.1 & 0.5 & 1.0 & 2.5 & 1.11\\
C & 33.3 & 0.5 & 1.0 & 2.5 & 1.99\\
D & 32.9 & 0.5 & 1.0 & -- & 2.02\\
E & 33.1 & 0.5 & 1.0 & -- & 1.74\\
F & 33.3 & 0.5 & 1.0 & -- & 1.52\\
G & 32.9 & -- & 0.0 & 2.5 & 1.86\\
H & 33.1 & -- & 0.0 & 2.5 & 1.46\\
I & 33.3 & -- & 0.0 & 2.5 & 1.93\\
\hline
\end{tabular}
\end{center}
\label{var_param}
\end{table}
For every model, we run 10000 simulations and calculated the periodogram using
a Fast Fourier transform (it
 allows a much faster calculation
 than the Lomb-Scargle method).
 We selected a best case, using the sum of the residual squares
 $\Sigma (o - s)^2$ calculated over a given frequency interval,
 where $o$ and $s$ are the observed and simulated powers at a given
 periodicity in the periodogram. The gray shaded area in the main panels
of Figs.~\ref{pds_pvar} and \ref{pds_pconst} shows the frequency interval
 we used. The sums of residual squares are relative numbers,
 and the minimum indicates the best value. However,
 it is important to evaluate also the real goodness of the model;
\begin{equation}
\chi^2_{\rm red} = \frac{1}{N} \Sigma \frac{(o - s)^2}{\sigma^2},
\end{equation}
where $N$ is the number of degrees of freedom (number of periodogram points
over which the $\chi^2_{\rm red}$ is calculated). In order
 to estimate 
 $\sigma$, we assumed that the periodogram outside the main
pattern (marked as darker shaded area in Figs.~\ref{pds_pvar} and
\ref{pds_pconst}) contains random features attributed to noise.
We assumed that the noise power is exponentially distributed,
and as $\sigma$ we chose the power at the
 90\% level in the cumulative histogram (5.27), excluding one relatively
 significant peak at 31.2\,s, which is of unknown origin.
Fig.~\ref{pds_pvar} shows the comparison of the observed periodogram
 with the best
 simulated one with variable amplitude and period.
In the upper panel we show a case in which both parameters are variable,
and in the lower panel a case with constant amplitude and variable period.
The best case is model B, in which both parameters are variable.
The simulated periodogram describes almost all the observed features,
supporting the interpretation that the signal is variable and the
periodicity oscillates around $P_2$. The period variability is
depicted in the lower inset. Model F, with constant amplitude and
variable period, also reproduces all the observed features,
but the amplitudes are significantly different and the periodicity
oscillates around $P_3$, matching the dominant peaks in all the observed
 light curves, except the non pile-up-corrected August pn light curve.
We remind that \citet{dobrotka2017} explained the double peak periodogram feature in V4743\,Sgr as a false beat in the observed data. Two close periodicities  result in a beat with low frequency, but if a single signal changes its amplitude or disappears for some time, this effect mimics a beat, so and the numerical method "interprets" it as due to two close frequencies. The same principle can explain the non-single peak in V2491 Cyg \citep{Ness2011}.

 Fig.~\ref{pds_pconst} shows the best case with constant periodicity but variable amplitude.
However, $a$ was set to zero, yielding the amplitude oscillate around zero, and disappearing
more frequently, mimicking a false beat. Larger values of $a$ yield only a single dominant peak, which do not describe the observed feature at all. The best case reproduces all the observed features with $\chi^2_{\rm red}$ even better than model F. Models G and I yield single peak solutions and do not describe the complex observed pattern.
Model H
 is acceptable, and it supports a constant periodicity $P_2$ with variable
 amplitude. However, model B describes the observed periodogram much better,
 and indicates a variable signal. With model H we were not able to
 retrieve the observed power configuration, i.e. $P_2$ having lower power
 than $P_1$ and $P_3$\footnote{We repeated the 10000 simulation
 process several times.}.
 Any periodicity introduced
 in the model produces the highest, or a very significant peak at the
 given period,
 while the models with variable periodicity are more flexible,
 and appear to be more realistic.
 We note here
 that interesting periodicity evolution is
shown in the upper inset of Fig.~\ref{pds_pvar}. Although the
 figure shows only a  solution out of many
 possible ones, and an exact match is not expected, the
 significant decrease of the period length towards the end of the
 exposure is in agreement with the periodogram of the third
 time interval obtained when we split the exposure in three portions
 (see text).
\label{lastpage}
\end{document}